\documentclass[11pt]{article}

\usepackage[margin=1in]{geometry}
\usepackage{amsmath,amssymb,amsfonts,amsthm,mathtools}
\usepackage{bm}
\usepackage{graphicx}
\usepackage{xcolor}
\usepackage{booktabs}
\usepackage{multirow}
\usepackage{hyperref}
\usepackage[numbers,sort&compress]{natbib}
\usepackage{setspace}
\usepackage{enumitem}
\usepackage{microtype}
\usepackage[font=small]{caption}
\usepackage{float}

\hypersetup{colorlinks=true,citecolor=blue,linkcolor=blue,urlcolor=blue}
\newcommand{\R}{\mathbb{R}}

\newcommand{\norm}[1]{\left\lVert #1 \right\rVert}
\newcommand{\thetaVec}{\bm{\theta}}
\newcommand{\xVec}{\bm{x}}
\newcommand{\IMap}{\mathcal{I}}
\newcommand{\Mset}{\mathcal{M}}

\title{Diffusion learning reveals viable parameter manifolds and compensation geometry in biological dynamical systems}
\author{Ruilin Zhang$^{1}$, Louis Tao$^{2,3}$, Zhuo-Cheng Xiao$^{4,5}$}
\date{}

\begin{document}
\maketitle

\begin{center}
\footnotesize
$^{1}$Peking University--Tsinghua University--National Institute of Biological Sciences Joint Graduate Program,\\
Academy for Advanced Interdisciplinary Studies, Peking University, Beijing 100871, China\\
$^{2}$Center for Bioinformatics, School of Life Sciences,\\
Peking University, Beijing 100871, China\\
$^{3}$Center for Quantitative Biology, Academy for Advanced Interdisciplinary Studies, Peking University, Beijing 100871, China\\
$^{4}$NYU-ECNU Institute of Mathematical Sciences, New York University Shanghai, Shanghai 200124, China\\
$^{5}$NYU-ECNU Institute of Brain and Cognitive Science, New York University Shanghai, Shanghai 200124, China
\end{center}

\begin{center}
\textbf{Classification:} Applied Mathematics; Dynamical Systems; Computational Neuroscience
\end{center}

\section*{Significance Statement}
Biological dynamical systems often admit many parameter combinations that produce the same observed behavior. We show that these non-unique inverse solutions form viable sets with locally manifold-like components. The geometrical features, including dimension, tangent directions, curvature, and multimodality, quantify how observables constrain the underlying dynamics. Conditional diffusion models learn these manifolds directly from simulation data in Lorenz, Izhikevich, and neuronal network models. The learned geometry reveals compensation laws, hidden parameter couplings, and the number of effective degrees of freedom that remain after the function is fixed. By turning robustness from a qualitative observation into a measurable geometric object, this framework provides a practical route to analyzing degeneracy and mechanism in complex biological models.

\begin{abstract}
Models of complex systems often have many parameters, yet are constrained by far fewer experimentally accessible observables; consequently, similar activity can emerge from coordinated parameter changes. We formalize these compatible parameter sets as \emph{viable parameter manifolds}: the inverse images of target dynamical features under a parameter-to-feature map. The relevant codimension is not the number of reported features, but the effective rank of that map at the target scale. Locally redundant features lower the effective codimension, while poor conditioning, high curvature, or regime mixing degrade learnability. We train conditional score-based diffusion models on simulated parameter--feature pairs and use them as amortized samplers of prior-weighted viable sets. In the Lorenz system, scalar trajectory statistics generate thin viable sheets, and a finite-tolerance conditioning localizes a transition-adjacent corridor. In the Izhikevich neuron model, four firing descriptors lie close to a nearly two-dimensional family of features, and the learned inverse images reveal distinct regular and irregular compensation geometries. In a deterministic ODE reduction of finite spiking networks \cite{Chang2025dsODE}, the same framework reveals excitatory--inhibitory compensation, timescale--coupling tradeoffs, and viable manifolds across 4--12 parameter dimensions. In this view, robustness, compensation, and hidden parameter dependencies are
organized as inverse geometry, with diffusion models providing practical tools
for sampling, visualizing, and interrogating that geometry.
\end{abstract}

\noindent\textbf{Keywords:} diffusion models; viable parameter manifold; robustness; compensation geometry; computational neuroscience; dynamical systems

\section{Introduction}

A central lesson from biological dynamics is that function can be reproducible even when the parameters that generate it are not. Neurons with different conductance combinations can fire in similar ways, networks with different synaptic couplings can produce comparable population activity, and systems biology models often preserve output while many microscopic parameters vary over wide ranges \citep{Prinz2004,MarderGoaillard2006,Marder2011,OLeary2013,Drion2015,Gutenkunst2007}. Across fields, this phenomenon appears as sloppiness, model-manifold compression, structural non-identifiability, dynamical compensation, and biological degeneracy \citep{BrownSethna2003,Gutenkunst2007,Machta2013,Transtrum2015Sloppiness,BellmanAstrom1970,Sontag2017,Karin2016,Prinz2004}. This degeneracy is not merely a nuisance for parameter estimation. It is one mechanism by which biological systems remain robust, adaptable, and evolvable \citep{Daniels2008,Wagner2008,Ciliberti2007,MarderGoaillard2006,Drion2015}. Yet most inverse analyses still report isolated parameter fits, local sensitivities, or low-dimensional sweeps, leaving the full shape of the compatible parameter set largely unseen.

We propose that the natural object of study is the \textit{viable parameter manifold}:
the set of parameters that realizes a specified dynamical behavior. If \(\thetaVec\in\Theta\subset\R^k\) denotes model parameters and \(\IMap(\thetaVec)\) denotes a low-dimensional vector of statistics extracted from simulated dynamics, then a target behavior \(y\) defines
\begin{equation}
\Mset_y=\{\thetaVec\in\Theta:\IMap(\thetaVec)=y\}.
\label{eq:intro_viable_set_revised}
\end{equation}
Within a fixed dynamical regime, the statistics \(\IMap\) often vary smoothly with the parameters, so \(\Mset_y\) is locally a level set. This simple observation changes the inverse problem. Instead of asking which parameter configuration is best, we study the viable parameter manifold as the inverse image of the parameter-to-feature map, $\Mset_y= \IMap^{-1}(y)$. Then we inquire the geometrical features of $\Mset_y$, namely the number of (local) dimensions, how it bends, and whether it splits into multiple mechanistic branches.


This geometric viewpoint gives direct scientific meaning to quantities that are otherwise treated as technical details. Intrinsic dimension estimates the number of effective degrees of freedom that remain after fixing a target behavior. Tangent spaces encode compensation laws, i.e., infinitesimal parameter co-variations that preserve the same output. Curvature measures how these laws change across the viable set. Nearby level sets show how changes in the desired behavior moves through parameter space. Disconnected components or irregular folds suggest that the chosen observables may be mixing distinct dynamical regimes, often near bifurcations or attractor transitions. 
Thus, the viable-manifold language used here gives an inverse-geometric way to connect parameter nonuniqueness to prior work on robustness, adaptability, and evolvability \citep{Daniels2008,Wagner2008,Ciliberti2007}.

The computational challenge is that these manifolds are embedded in parameter spaces too large for exhaustive sweeps, whereas pointwise optimization returns only isolated solutions. We first sample the ambient parameter domain, simulate the associated features, and train a conditional score-based diffusion model on the resulting parameter--feature pairs; the target fiber is neither known nor sampled directly when this library is constructed \citep{Ho2020DDPM,Song2021SDE}. Existing diffusion theory gives intrinsic-dimension-adaptive rates when the training observations are already drawn from the manifold-supported distribution to be learned \citep{TangYang2024}. That result is an informative oracle benchmark for learning variation along a single inverse fiber, but it is not an end-to-end rate for our ambiently sampled library. Conditional diffusion theory based on global joint samples and smoothly varying manifold-supported conditional laws is closer to our design \citep{TangLinYang2025}; however, it imposes uniform regularity directly on the conditional supports and densities and controls error averaged over target values. We therefore use diffusion here as an empirically validated, amortized proposal mechanism for inverse geometry, leaving a minimax guarantee for the simulator-induced inverse fibers to future studies.

We develop this view across three dynamical systems of increasing complexity and biological relevance. In the Lorenz system \citep{Lorenz1963}, scalar trajectory statistics generate thin viable sheets and provide a controlled calibration of the geometry. In the Izhikevich neuron model \citep{Izhikevich2003,Izhikevich2004}, target firing statistics reveal parameter manifolds whose tangent directions express compensatory changes among excitability, reset, and adaptation parameters. In dsODE reductions of finite spiking networks \citep{Chang2025dsODE}, viable manifolds expose how excitatory--inhibitory balance extends beyond two-dimensional sweeps to hidden dependencies among synaptic strengths, timescales, connection probabilities, and external drives. Across these systems, the central message is the same: compatible parameters are not a cloud of arbitrary alternatives. They are organized geometric objects, and learning their geometry turns robustness from a qualitative observation into a measurable structure.

\section{Results}

\subsection{Conditional diffusion samples viable parameter sets}

\begin{figure}[htbp]
\centering
\includegraphics[width=\linewidth,page=1]{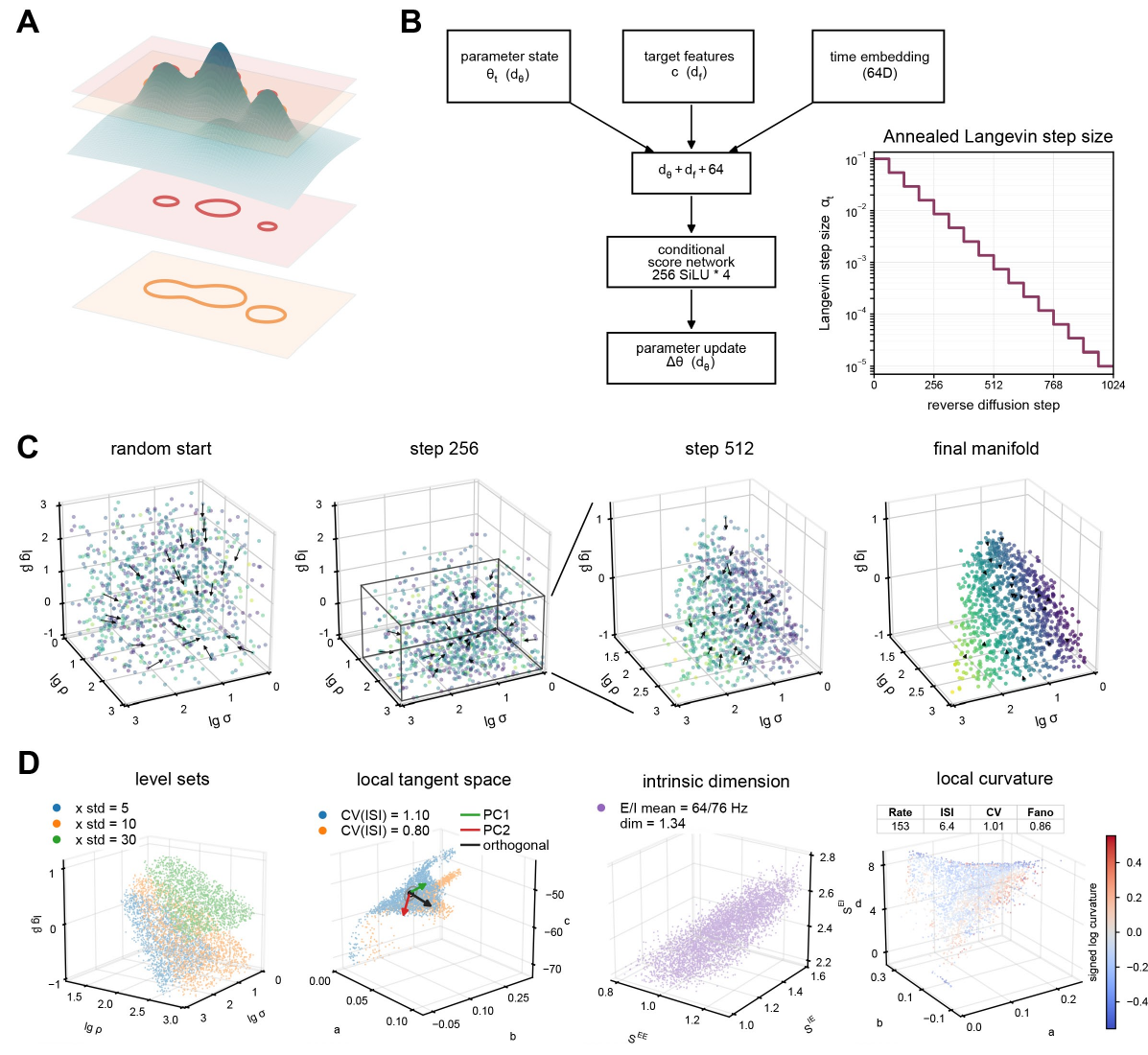}
\caption{Conditional learning and geometric analysis frameworks of viable parameter sets. (A) Target fibers in parameter space. (B) Conditional score model and sampling procedure. (C) Reverse diffusion for a Lorenz target. (D) Neighboring target sets and the empirical dimension, tangent, and curvature measurements used in the main text.}
\label{fig:overview}
\end{figure}

We seek the set of parameters compatible with a prescribed dynamical behavior, rather than a single fitted parameter vector. Consider a dynamical system with parameters \(\thetaVec\in\Theta\subset\R^k\),
\begin{equation}
\dot{\xVec}=f(\xVec,\thetaVec),
\end{equation}
and let \(\IMap:\Theta\to\R^m\) collect trajectory statistics. For a target \(y\), the exact viable set and its finite-tolerance counterpart are
\begin{equation}
\Mset_y=\{\thetaVec\in\Theta:\IMap(\thetaVec)=y\},
\qquad
\Mset_{y,\varepsilon}=\{\thetaVec\in\Theta:\norm{\IMap(\thetaVec)-y}\leq\varepsilon\}.
\label{eq:viable_set_main}
\end{equation}
The feature discrepancy and tolerance may be feature- and target-scaled; model-specific definitions are given in Materials and Methods.
A target may select one connected set or several components in parameter space (Fig.~\ref{fig:overview}A). On a differentiable dynamical stratum, at a regular point where \(D\IMap\) has locally constant rank \(r\), the constant-rank theorem gives a local fiber of dimension \(k-r\), with tangent space
\begin{equation}
T_{\thetaVec}\Mset_y=\ker D\IMap(\thetaVec).
\label{eq:tangent_space_main}
\end{equation}
Thus the effective number of constraints is the local rank of the parameter-to-feature map, not the nominal number of reported features, and its null directions describe first-order parameter compensation.

Throughout the main text, geometric quantities are empirical measurements of finite generated point clouds: participation ratio summarizes effective dimension, local principal components summarize directions, and local quadratic fits summarize curvature. Their definitions and detailed comparisons with fitted forward-Jacobian ranks and nullspaces are given in the SI Appendix.

We construct an ambient library by sampling \(\thetaVec\), simulating the model, and pairing each parameter vector with \(\IMap(\thetaVec)\). A conditional score model trained on these pairs transforms a random parameter cloud into target-conditioned proposals (Fig.~\ref{fig:overview}B,C). The learned proposal law is prior-dependent and aims to approximate the conditional law induced by the parameter prior and feature map; direct re-simulation then identifies points lying in the stated target-tolerance tube. Unless stated otherwise, exploratory geometry is measured on proposal clouds and target fidelity on re-simulated or tolerance-accepted points. Acceptance quantifies proposal precision under the chosen tolerances; component-coverage analyses quantify recall and relative component mass.

\subsection{Scalar targets select Lorenz sheets}

\begin{figure}[htbp]
\centering
\includegraphics[width=.97\linewidth,page=1]{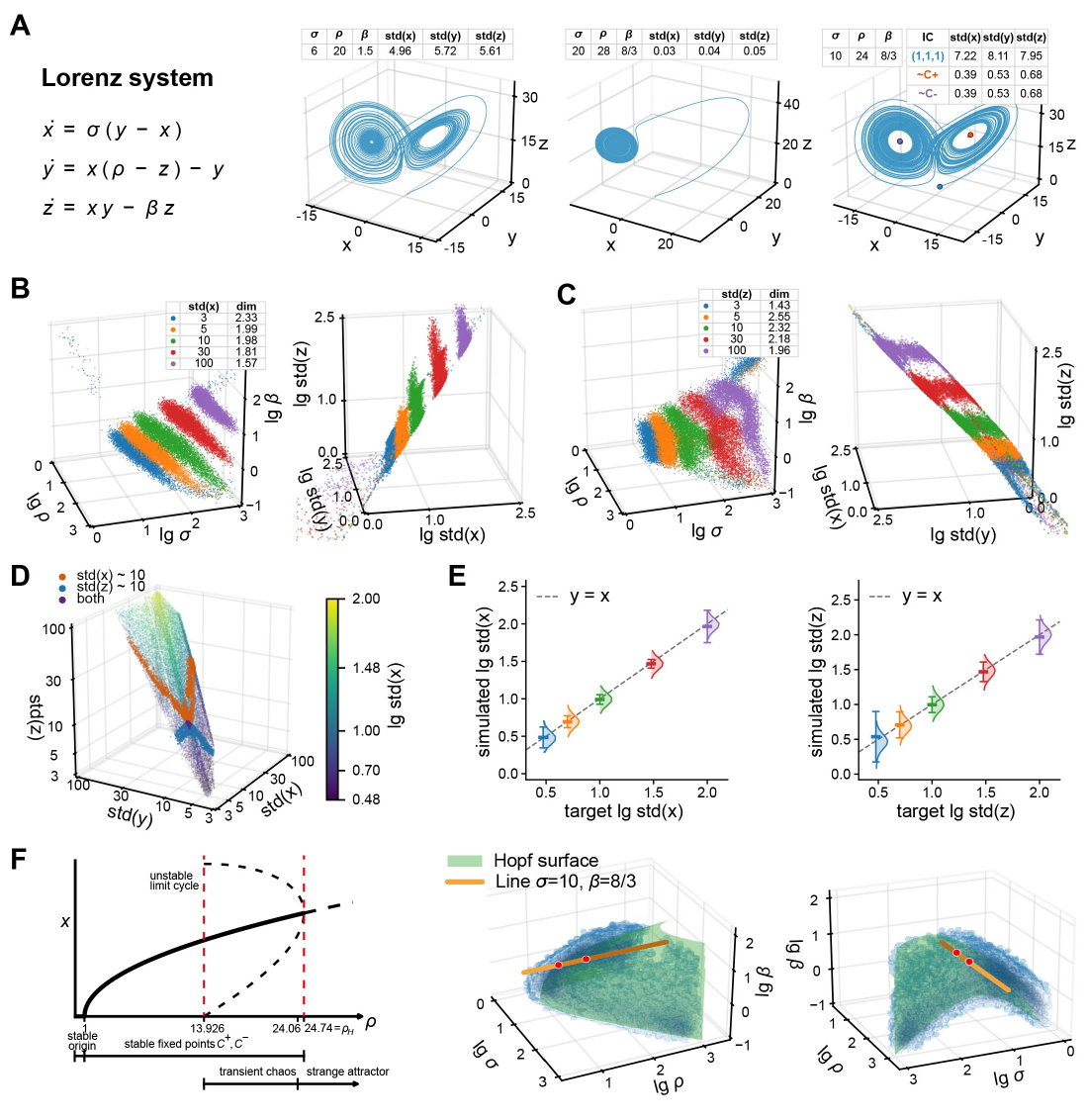}
\caption{Lorenz viable sets. (A) Model, sampled parameter domain, and representative trajectories. (B,C) Generated proposal clouds conditioned on scalar \(\log_{10}\operatorname{std}(x)\) and \(\log_{10}\operatorname{std}(z)\) targets. (D) Simulated feature cloud and scalar- or joint-target subsets. (E) Re-simulation of scalar-target proposals. (F) Joint initial-condition-dependent conditioning and comparison with the classical homoclinic and Hopf thresholds.}
\label{fig:lorenz}
\end{figure}

We first calibrated the approach on the Lorenz system,
\begin{equation}
\dot x=\sigma(y-x),\qquad
\dot y=x(\rho-z)-y,\qquad
\dot z=xy-\beta z,
\end{equation}
whose attractors and transitions are well characterized \citep{Lorenz1963,Sparrow1982Lorenz}. We used coordinate-wise trajectory standard deviations as scalar summaries of attractor size (Fig.~\ref{fig:lorenz}A). Conditioning on \(\log_{10}\operatorname{std}(x)\) generated thin, similarly oriented sheets in \((\log_{10}\sigma,\log_{10}\rho,\log_{10}\beta)\), whereas \(\log_{10}\operatorname{std}(z)\) selected a differently oriented family (Fig.~\ref{fig:lorenz}B,C). Their participation ratios clustered around 2, corresponding to the codimension-1 inverse mapping. Hence the sets are often sheet-like in the three-dimensional parameter space. In feature space, the two scalar guides selected different regions of the simulated response cloud (Fig.~\ref{fig:lorenz}D), and direct re-simulation preserved the ordering of the requested targets (Fig.~\ref{fig:lorenz}E).

Joint conditioning further localized the compatible set for transient chaos. Combining a large-variability criterion from a generic initial state with a low-variability criterion near the nonzero equilibria selected a finite-width corridor adjacent to the classical transient-chaos interval (Fig.~\ref{fig:lorenz}F). On the standard one-dimensional slice \((\sigma,\beta)=(10,8/3)\), the displayed corridor intersects the interval between the known homoclinic and Hopf thresholds. Moreover, across the full parameter domain, the upper edge of the corridor qualitatively tracks the analytic Hopf surface, which separates the transient chaos and chaotic regimes. The construction therefore localizes a transition-adjacent dynamical regime from trajectory summaries. Simulation windows, initial conditions, tolerances, and boundary comparisons are reported in the SI Appendix.

\subsection{Firing statistics reveal neuronal compensation}

\begin{figure}[htbp]
\centering
\includegraphics[width=0.94\linewidth,page=1]{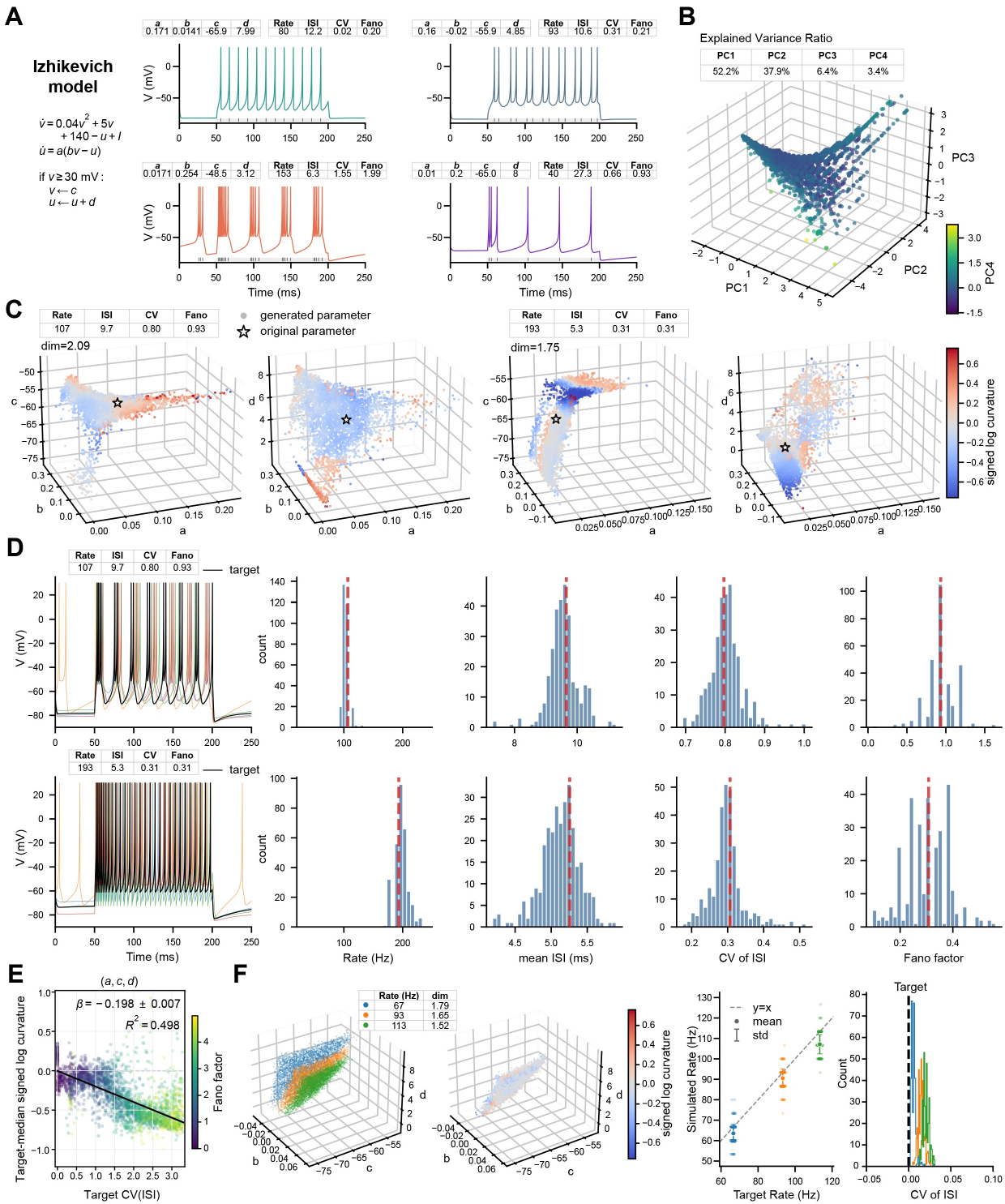}
\caption{Izhikevich viable sets. (A) Model, parameter domain, and representative current-step responses. (B) Principal components of the four-feature response library. (C) Generated proposal clouds in two three-parameter projections; stars mark held-out anchors, colors show signed-log empirical curvature, and participation ratios are annotated. (D) Re-simulated traces and feature distributions. (E) Target-median signed-log empirical curvature in the \((a,c,d)\) projection versus target CV(ISI), colored by target Fano factor; the line and band show the fit and 95\% target-bootstrap interval. (F) Regular-spiking target families, a 113-Hz example, and validation of firing rate and CV(ISI).}
\label{fig:izhikevich}
\end{figure}

We then applied the framework to the Izhikevich neuron model,
\begin{equation}
\dot v=0.04v^2+5v+140-u+I,
\qquad
\dot u=a(bv-u),
\end{equation}
with the reset \(v\mapsto c\), \(u\mapsto u+d\) at threshold \citep{Izhikevich2003,Izhikevich2004}. Under a fixed current step, the four parameters \((a,b,c,d)\) control recovery, subthreshold coupling, voltage reset, and post-spike adaptation. We summarized each response by firing rate, mean interspike interval (ISI), CV(ISI), and a temporal-bin Fano factor (Fig.~\ref{fig:izhikevich}A).

The four descriptors were strongly correlated in the analyzed response library: its first two principal components explained \(52.2\%\) and \(37.9\%\) of the variance, respectively (Fig.~\ref{fig:izhikevich}B). This is because of our design: mean firing rate is closely related to the inverse of mean ISI, and both CV(ISI) and Fano factor quantify the irregularity of the firing patterns. Consistent with this redundancy, four-feature conditioning produced extended, approximately two-dimensional proposal clouds rather than isolated points (zero-dimension) (Fig.~\ref{fig:izhikevich}C). The displayed targets were defined by held-out parameter--feature pairs, and direct re-simulation recovered their firing-rate and spike-timing statistics (Fig.~\ref{fig:izhikevich}D). Dominant local cloud directions nominate coordinated variation among recovery, excitability, reset, and adaptation parameters that may preserve the measured response, providing a population-of-models representation of neuronal compensation \citep{Prinz2004,MarderGoaillard2006,Marder2011,OLeary2013,Drion2015,Fyon2024}.

The shape of this compensation depended on firing irregularity. Across the target screen, the median signed-log empirical curvature became more negative as target CV(ISI) increased (Fig.~\ref{fig:izhikevich}E). The displayed regular-spiking targets instead produced visually near-affine families across firing rates, with accurate rate recovery and low CV(ISI) (Fig.~\ref{fig:izhikevich}F). A reduced spike-to-spike return-map analysis offers a hypothesis for this contrast: regular firing is organized by a stable periodic recovery state, whereas irregular targets require joint control of the average reset--adaptation load and recovery-map stability. Those two requirements play antagonistic roles across \((a,b,c,d)\), yielding the negative curvature observed in the irregular-target projections. The derivation and quantitative curvature analysis are given in the SI Appendix.

\subsection{Rate moments separate dsODE branches}
\label{sec:main-dsODE-4D}

\begin{figure}[htbp]
\centering
\includegraphics[width=\linewidth]{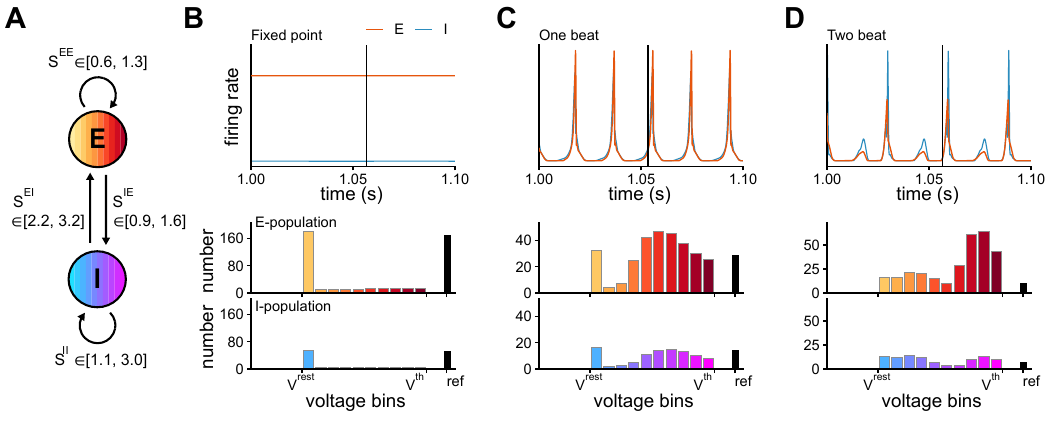}
\caption{dsODE model and population activity. (A) Recurrent excitatory--inhibitory network and sampled coupling ranges. (B--D) E/I firing-rate traces and voltage-bin population states for a fixed point, a one-beat oscillation, and a two-beat oscillation; vertical lines mark the displayed time points.}
\label{fig:dsode_eq}
\end{figure}

We then turned to dsODE, a deterministic reduction of finite, structurally homogeneous spiking networks that represents excitatory and inhibitory populations \citep{Chang2025dsODE}. dsODE counts the number of neurons in discretized voltage-bin and traces the moments of pending-synaptic-drive variables. It faithfully inherits physiological and network parameters from spiking networks, including recurrent strengths, synaptic and neuronal timescales, connection probabilities, and external drives. The model not only reproduces dynamical statistics, but also accurately captures the bifurcations between the stationary, one-beat, and multibeat regimes of the corresponding spiking networks (Fig.~\ref{fig:dsode_eq}). At the same time, dsODE remains efficient enough for ambient library construction.

\begin{figure}[H]
\centering
\includegraphics[width=\linewidth]{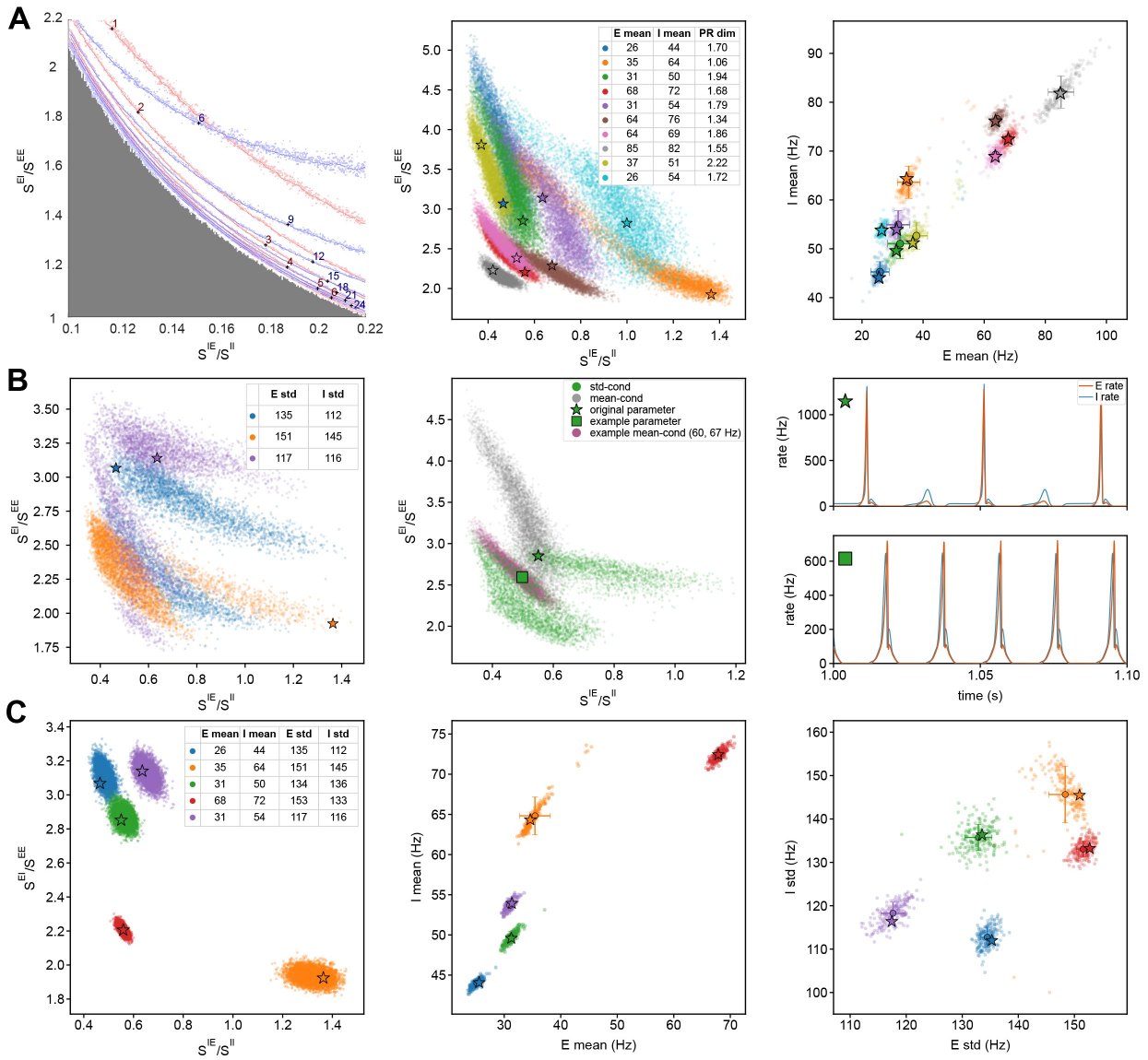}
\caption{Four-dimensional dsODE synaptic viable sets. (A) E/I-mean-conditioned proposal clouds in the inhibition plane, participation-ratio estimates, and re-simulated mean rates. (B) E/I-standard-deviation-conditioned proposal clouds, their overlap with a mean-conditioned cloud near held-out parameters, and representative one- and two-beat traces. (C) Joint mean--standard-deviation-conditioned proposal clouds and re-simulated recovery of all four target moments.}
\label{fig:dsode_4d}
\end{figure}

We first varied the four recurrent strengths \((S_{EE},S_{IE},S_{EI},S_{II})\). In the background regime, E and I mean rates co-vary and the compatible parameters form near-codimension-one strips in the inhibition-plane coordinates \((S_{IE}/S_{II},S_{EI}/S_{EE})\) \citep{Xiao2021}. Their organization by the product \((S_{IE}/S_{II})(S_{EI}/S_{EE})\) recovers the balance relation found in earlier network analyses (SI Appendix) \citep{Xiao2021,Chang2025dsODE} (Fig.~\ref{fig:dsode_4d}A left). In the more strongly driven regime studied here, conditioning jointly on the E and I mean rates still produced strip-like sets on the inhibition-plane coordinate, but E and I firing rates covary less hence the viable sets are closer to codimension-2 (Fig.~\ref{fig:dsode_4d}A middle). Direct re-simulations confirm the close alignment with the target E and I mean rates (Fig.~\ref{fig:dsode_4d}A right). 

The broadened and split mean-conditioned sets reflected dynamics that the means did not distinguish. Conditioning on E/I rate standard deviations produced a complementary organization, and re-simulation of overlapping mean- and standard-deviation-conditioned regions revealed distinct one-beat and two-beat trajectories (Fig.~\ref{fig:dsode_4d}B). Conditioning on all four moments \((\mu_E,\mu_I,\sigma_E,\sigma_I)\) contracted the displayed sets to compact neighborhoods and recovered both pairs of target statistics (Fig.~\ref{fig:dsode_4d}C). Within this driven beat regime, means constrain population activity levels, whereas standard deviations add oscillation-amplitude information; together they separate the displayed dynamical families more effectively than either pair alone. Nevertheless, when only targeting mean firing rates, the temporal dynamical information can hide in the thickness of strips on the inhibition plane. This is confirmed by the tight relation between mean firing rate ratio $(\mu_I/\mu_E)$ and set transverse width $\sigma_{\perp}$ (see SI Fig.~S2A). Together, our results are consistent with previous studies showing temporal-branch structure that is invisible to the mean-rate guide \citep{WallaceEtAl2011}. The SI Appendix derives the information retained by these four moments on periodic beat families and quantifies the branch analysis.

\subsection{Strength--timescale coordinates organize high-dimensional dsODE sets}

\begin{figure}[htbp]
\centering
\includegraphics[width=\linewidth,page=1]{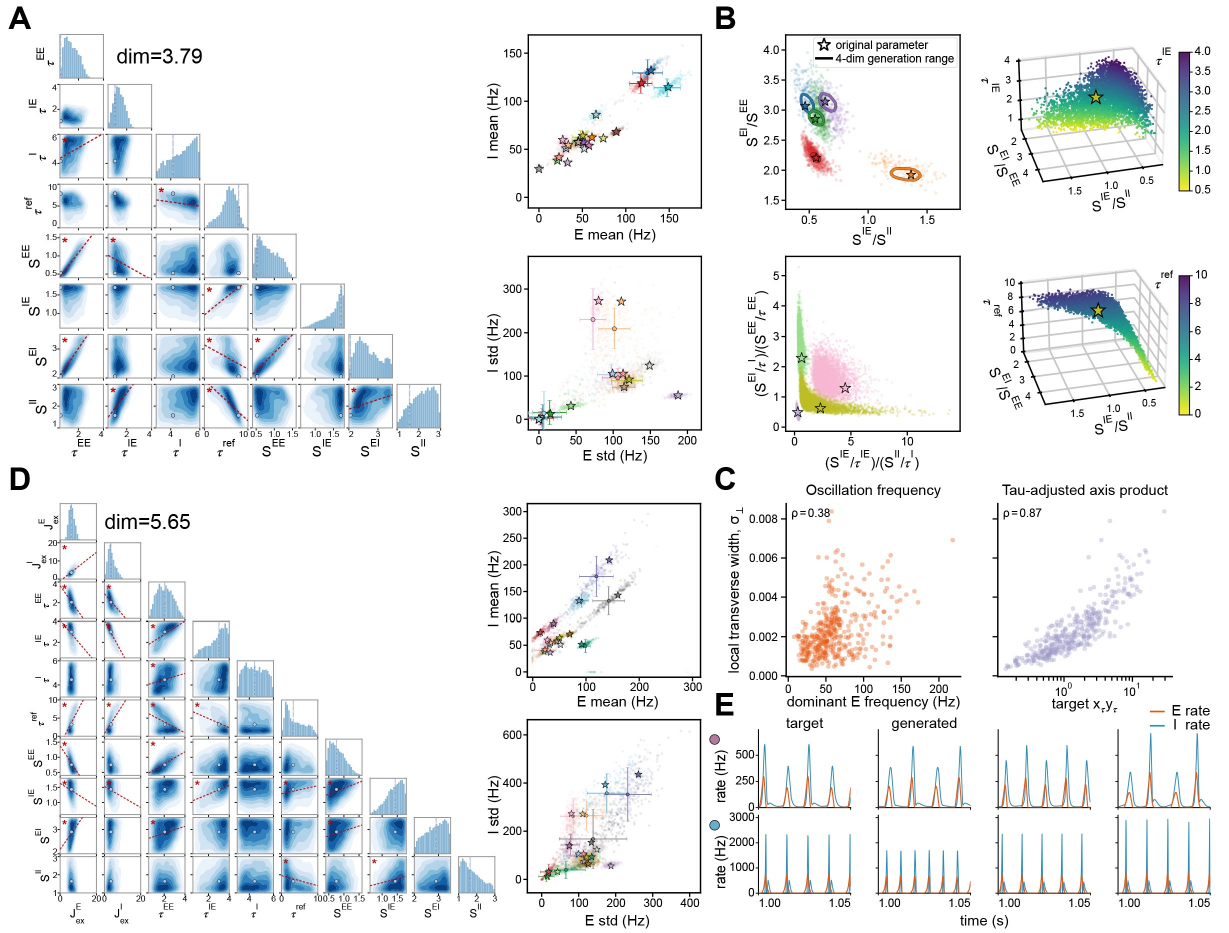}
\caption{High-dimensional dsODE viable sets. (A) Pairwise projections of an eight-dimensional strength--timescale proposal cloud and re-simulated E/I means and standard deviations; the annotated participation ratio is 3.79. (B) Strength-ratio and timescale-adjusted projections, colored by \(\tau_{IE}\) or \(\tau_{\mathrm{ref}}\). (C) Local transverse width \(\sigma_\perp\) versus dominant E-population frequency and \(\mathcal R_\tau=x_\tau y_\tau\) across 400 oscillatory target-conditioned clouds; \(\rho\), Spearman correlation. (D) Pairwise projections and target recovery for a ten-dimensional proposal cloud that also varies external drives; the annotated participation ratio is 5.65. (E) Target E/I rate traces and corresponding generated examples.}
\label{fig:dsode_highdim}
\end{figure}

We expanded the dsODE parameterization to an eight-dimensional space by varying the four recurrent strengths together with \(\tau_{EE}\), \(\tau_{IE}\), \(\tau_I\), and \(\tau_{\mathrm{ref}}\), while retaining the four-moment guide. The resulting eight-dimensional sets remained structured: a representative cloud had a participation ratio of 3.79, close to codimension-4 (Fig.~\ref{fig:dsode_highdim}A left). Direct re-simulation recovered the target moments, and strength and timescale parameters showed coordinated variation (Fig.~\ref{fig:dsode_highdim}A right). Empirically, the intersection between the eight-dimensional sets and the 4D synaptic parameter space shows consistent occupancy (Fig.~\ref{fig:dsode_highdim}B upper left), though with more diffused spans compared to Fig.~\ref{fig:dsode_4d}C left. This is expected due to the sparser training data in the 8D parameter space.

Projecting the parameter sets to 2D planes of parameter pairs reveals a systematic dependence between synaptic strengths ($S$) and timescale parameters ($\tau$) (Fig.~\ref{fig:dsode_highdim}A left). In particular, the proposal clouds appeared more sharply organized in projections that combined each strength with its channel timescale (Fig.~\ref{fig:dsode_highdim}B). We summarize this organization by
\begin{equation}
x_\tau=\frac{S_{IE}/\tau_{IE}}{S_{II}/\tau_I},
\qquad
y_\tau=\frac{S_{EI}/\tau_I}{S_{EE}/\tau_{EE}},
\qquad
\mathcal R_\tau=x_\tau y_\tau,
\label{eq:tau_loop_index}
\end{equation}
where \(\mathcal R_\tau\) compares the timescale-adjusted reciprocal E--I loop with the two within-population feedback channels. It is the high-frequency limit of a finite-frequency loop-gain ratio when connection-probability and susceptibility factors are shared or cancel to leading order (SI Appendix). Many target-conditioned clouds form thin ribbons in the \((x_\tau,y_\tau)\) plane, extending the strength-ratio organization found in four dimensions. The transverse width \(\sigma_\perp\) of clouds on the timescale-adjusted inhibition plane is significantly lower than the width projected on the unadjusted inhibition plane (see SI Fig.~S2B).

The ribbon width varied systematically across oscillatory targets. Across 400 oscillatory target-conditioned clouds, the transverse width \(\sigma_\perp\), defined as the median local minor-axis standard deviation in the normalized \((x_\tau,y_\tau)\) plane, was positively associated with dominant E-population frequency (Spearman \(\rho=0.38\)) and more strongly with \(\mathcal R_\tau\) (\(\rho=0.87\); Fig.~\ref{fig:dsode_highdim}C). At fixed feature tolerance, transverse width increases as the four-moment guide becomes less sensitive to displacement across this plane. 
Generally, a larger \(\mathcal R_\tau\) shifts control toward the reciprocal E--I timing loop, allowing additional variation in frequency, E--I phase, harmonic allocation, and beat shape at nearly fixed marginal moments. This reduced transverse sensitivity offers a mechanistic explanation for the broader high-frequency, high-\(\mathcal R_\tau\) ribbons and predicts that frequency, phase, or cross-spectral targets will preferentially contract them.

Furthermore, added parameters generated distinct compensation directions. Refractory time showed visible associations with \(S_{IE}\), \(S_{EI}\), and \(S_{II}\) in the displayed projections, consistent with recurrent changes offsetting reduced neuronal availability during longer refractory periods (Fig.~\ref{fig:dsode_highdim}B). When the two external drives were added, the ten-dimensional proposal clouds remained structured, with a representative participation ratio of 5.65; re-simulation recovered the target moments and produced trajectories representative of the target regimes (Fig.~\ref{fig:dsode_highdim}D,E). A local mean-input balance calculation provides a qualitative interpretation of the displayed drive--strength relations. Transfer-function, refractory-timing, and external-drive calculations, together with robustness analyses for the width statistics, are provided in the SI Appendix. A twelve-dimensional proposal cloud is also provided in the SI Appendix.


\subsection{Amortized inference and ambient-library complexity}
\label{sec:2.6-cost}
Once a shared parameter--feature library has been constructed, conditional generation amortizes inverse inference across targets: the generator is trained once, and each new target requires only proposal generation and final re-simulation rather than a new simulator-driven search. In the aggregate Lorenz and Izhikevich benchmarks, diffusion required \(1.13\)--\(4.59\) validation simulations per accepted proposal, compared with \(132\)--\(542\) for MCMC and \(511\)--\(883\) for local continuation; uniform rejection also had substantially lower acceptance. Flow matching produced similar target-conditioned clouds. These comparisons isolate per-query efficiency, while the shared library, training costs, and benchmark protocols are reported in the SI Appendix.

Library construction is a different statistical problem because parameters are sampled before any target fiber is known. Let \(\IMap:\Theta\subset\mathbb R^k\rightarrow\mathcal Y\), draw \(\thetaVec_i\sim\pi\), and set \(Y_i=\IMap(\thetaVec_i)\). If \(D\IMap\) has rank \(r\) near an interior regular target \(y^\ast\), the fiber has local dimension
\[
d=k-r.
\]
If the pushforward distribution of \(Y=\IMap(\thetaVec)\) has a continuous density \(p_Y\) with respect to the local \(r\)-dimensional volume measure on the feature image, then
\begin{equation}
\mathbb E\,N_h
=
N\Pr\!\left(\|Y-y^\ast\|\leq h\right)
=
N v_r p_Y(y^\ast)h^r+o(Nh^r),
\label{eq:ambient_localization}
\end{equation}
where \(N_h\) counts nearby library points and \(v_r\) is the \(r\)-dimensional unit-ball volume. Thus \(d\) controls variation along the fiber, whereas \(r=k-d\) and \(p_Y(y^\ast)\) control how often ambient sampling encounters it. A lower-dimensional fiber may be simpler to represent once found but harder to reach because its codimension is larger. This occupancy law is not a complete learning rate, because a regular conditional estimator can share information across nearby targets.

Two existing results provide complementary benchmarks. Given \(n\) samples already supported on a regular \(d\)-dimensional manifold with an \(\alpha\)-smooth density, manifold-adaptive diffusion estimation can attain, up to logarithmic factors,
\begin{equation}
\mathbb E W_1(\widehat\mu,\mu)
=
\widetilde{\mathcal O}\!\left(
n^{-1/2}\vee n^{-(\alpha+1)/(2\alpha+d)}
\right)
=
\widetilde{\mathcal O}\!\left(n^{-\gamma(d,\alpha)}\right),
\qquad
\gamma(d,\alpha)
=
\min\!\left\{\frac12,\frac{\alpha+1}{2\alpha+d}\right\},
\label{eq:oracle_manifold_w1}
\end{equation}
under quantitative regularity assumptions \citep{TangYang2024}. This oracle rate describes learning along a fiber but omits its acquisition from the ambient library. Conditional-distribution theory instead starts from global paired samples. Identifying the conditioning and conditional-support dimensions with \(r\) and \(d\), respectively, its target-averaged benchmark is
\begin{equation}
\mathbb E_Y W_1(\widehat\mu_Y,\mu_Y)
=
\widetilde{\mathcal O}\!\left[
N^{-\frac{\alpha_y}{2\alpha_y+r}}
+
N^{-\frac{\alpha_\parallel+1}
        {2\alpha_\parallel+d+
        (\alpha_\parallel/\alpha_y)r}}
\right],
\label{eq:conditional_w1}
\end{equation}
where \(\alpha_y\) and \(\alpha_\parallel\) describe regularity across targets and along fibers \citep{TangLinYang2025}. Broader minimax theory for moving supports adds the term
\[
N^{-1/(d/\beta_\parallel+r/\beta_y)},
\]
where \(\beta_\parallel\) and \(\beta_y\) quantify support regularity along and across fibers \citep{TangYang2025DistributionRegression}. These results show that target dimension, fiber dimension, and regularity enter jointly, but they control target-averaged risks for uniformly regular conditional families.

Regularity can therefore mitigate the curse of dimensionality. For fixed \(d\) and \(r\), increasing smoothness improves the exponents; an arbitrarily smooth, uniformly regular conditional family formally approaches \(\widetilde{\mathcal O}(N^{-1/2})\). Thus dimension need not enter exponentially through the accuracy exponent, although it can remain in constants, conditioning, and approximation complexity. Smoothness of \(\IMap\) alone does not ensure this uniform regularity.

These benchmarks do not yet solve our inverse problem. We require not only an accurate global parameter-to-feature relation or small target-averaged conditional error, but accurate recovery of the viable set at a specified target. Near a bifurcation, a globally accurate learner can still infer the wrong fiber: small forward or distributional errors may displace it, merge or omit components, or misidentify its dimension as the nonzero singular values of \(D\IMap\) shrink or its rank changes. A complete theory is therefore needed for how pointwise viable-set error scales with library size \(N\), ambient and intrinsic dimensions, regularity, target rarity and inverse conditioning, and generative-network size, and for how distributional error propagates to support, component, tangent-space, and curvature recovery.

\subsection{Independent targets contract viable sets and resolve branches}

\begin{figure}[htbp]
\centering
\includegraphics[width=\linewidth,page=1]{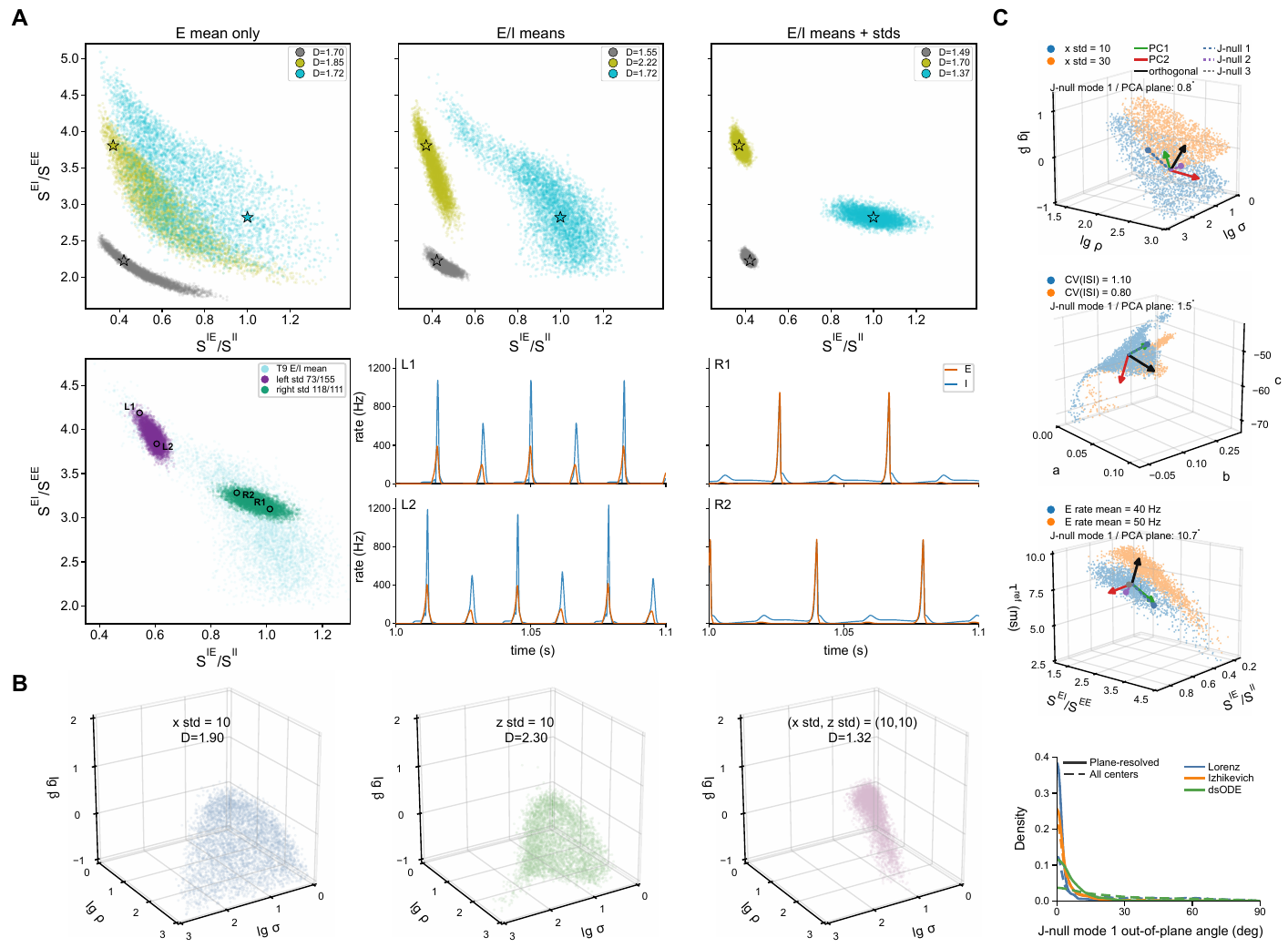}
\caption{Effects of target choice and local geometric comparisons. (A) Top: four-dimensional dsODE proposal clouds conditioned successively on the E mean, both population means, and both means and standard deviations. Bottom: overlap of mean- and standard-deviation-conditioned proposal clouds and representative one- and two-beat traces. (B) Lorenz proposal clouds conditioned on \(\log_{10}\operatorname{std}(x)\), \(\log_{10}\operatorname{std}(z)\), or both. (C) Representative cloud-PCA axes and projected fitted-Jacobian null directions for Lorenz, Izhikevich, and dsODE sets; lower curves show the leading null direction's out-of-plane angle for all eligible centers and for the subset whose null directions are resolved in the displayed plane. Eligibility and angle definitions are given in the SI Appendix.}
\label{fig:geometry_summary}
\end{figure}

Adding a target contracts a viable set locally only when it contributes an independent constraint. For a current guide \(G\) and an added scalar feature \(g\), the feature is nonredundant at \(\thetaVec\) when \(Dg(\thetaVec)\notin\operatorname{row}DG(\thetaVec)\). At a regular point satisfying both target conditions, adding \(g\) raises the local rank by one and lowers the exact fiber dimension by one; a redundant feature leaves that dimension unchanged. The displayed finite-tolerance clouds show the corresponding projected pattern (Fig.~\ref{fig:geometry_summary}A,B): the full four-moment dsODE guide has smaller projected extent than E-mean guidance, and the joint \((x\text{-std},z\text{-std})\) Lorenz cloud is narrower than either scalar-guided cloud. These projected changes measure added target information at the cloud scale; the exact local dimensional statement is determined by incremental Jacobian rank.

Branch separation is a different consequence of adding targets. A coarse guide can take the same value on dynamically distinct branches, so one inverse set may contain several components even when each component is locally regular. For the displayed dsODE targets, E/I-mean guidance pools regions whose re-simulated trajectories show one- and two-beat dynamics; their E/I standard deviations differ, and adding those statistics selects branch-specific neighborhoods from the mean-matched set (Fig.~\ref{fig:geometry_summary}A). Thus an informative target can either remove a free direction within a branch or distinguish branches merged by a coarser summary. The latter exposes family structure consistent with the known dsODE bifurcation organization without locating the intervening bifurcation \citep{Chang2025dsODE,wu2023multi}.

\section{Discussion}

Inverse modeling often ends when one acceptable parameter vector is found. Here we instead characterize the structured families of parameters compatible with a target behavior. This changes the central question from whether a model can reproduce the target to how much parameter freedom remains and which covariations preserve that behavior. Across the three systems, the conditioned point clouds exhibit lower-dimensional organization on locally regular components. For the target families tested, direct re-simulation confirms that substantial but target-dependent fractions satisfy the prescribed feature tolerances. This inverse-set perspective connects model-manifold and sloppiness concepts in prediction space with identifiability and dynamical compensation in parameter space, informing questions of generalization, explanation, and mechanistic interpretation \citep{BrownSethna2003,Gutenkunst2007,Transtrum2011Geometry,Machta2013,Transtrum2015Sloppiness}.

Across the case studies, compatible parameter sets remain structured as model complexity increases. In Lorenz, fixing a scalar trajectory statistic selects thin, approximately two-dimensional sheets within a three-dimensional parameter box. In the Izhikevich model, four co-varying firing descriptors select approximately two-dimensional proposal clouds embedded across all four parameter coordinates; local geometric estimates nominate candidate compensation directions, and the projected empirical curvature proxy becomes more negative as target spiking irregularity increases. In dsODE, mean-rate targets define excitatory--inhibitory balance-like strips, whereas adding rate-standard-deviation targets separates one- and two-beat families merged by mean rates alone. Extensions to higher-dimensional parameter spaces reveal tradeoffs among recurrent strengths, filtering timescales, refractory effects, and external drives. These structures are consistent with population-of-models accounts of neural degeneracy and provide candidate covariations through which hypotheses of homeostatic compensation can be tested \citep{Prinz2004,MarderGoaillard2006,Marder2011,OLeary2013,Drion2015,Fyon2024}. 

It is the geometry of viable parameter sets that makes that degeneracy operational. Complementary point-cloud and local forward-Jacobian analyses nominate candidate compensation and sensitivity axes (Fig.~\ref{fig:geometry_summary}C). Neighboring target-conditioned sets describe local functional variation, projected bending indicates where a single linear compensation law may fail, and component structure can flag candidate mechanisms or regimes. Additional observables contract a locally regular viable set only when they contribute independent active constraints; branch-discriminating observables can also separate dynamical families merged by coarser summaries. These geometric readouts can therefore guide experimental design and model reduction.

Conditional generation provides reusable access to these structured sets. Once trained, a generator supplies structured proposals for many targets without repeating a complete target-specific search. Compared to previous studies on amortized neural posterior estimation of inferred parameter distributions \citep{Goncalves2020,Bittner2021}, we used conditional generation to efficiently benchmark parameter geometry, including inverse-set dimension, compensation directions, curvature, and branch structure across targets, linking reusable inference to candidate mechanisms of dynamical robustness. The comparable preliminary flow-matching results suggest that this amortization is a broader property of conditional generative inference rather than one unique to diffusion. The generated clouds are finite-tolerance proposals for a prior-weighted conditional law; direct simulation remains the membership test, and uniform surface-area sampling of a level set is a separate problem. Section~\ref{sec:2.6-cost} distinguishes this post-training query efficiency from the one-time cost of constructing the ambient library. The missing theory is pointwise: small target-averaged distributional error need not guarantee accurate recovery of rare, critical, or multicomponent viable sets and their geometry.

The manifold description is local to regular dynamical regimes. Bifurcations, coexisting attractors, transient chaos, and branch switching can match the same low-order statistics through different mechanisms \citep{GuckenheimerHolmes1983,Sparrow1982Lorenz}; irregular or disconnected conditioned sets can therefore flag regime mixing, insufficient observables, or nearby bifurcation structure and should motivate targeted re-simulation, continuation, or componentwise analysis rather than be pooled into one smooth manifold. High-dimensional bifurcation discovery will require regime-aware extensions: componentwise atlases, active refinement near transitions, and diffusion-assisted continuation along codimension-one boundaries. The framework thereby connects inverse-set geometry, biological robustness and degeneracy, and generative sampling of complicated embedded sets.

Across the models studied here, robust behavior occupies a structured parameter family rather than one privileged vector. Learning that family makes residual freedom measurable, candidate compensatory covariations visible, and hidden dependencies experimentally testable. The resulting shift is from selecting one parameter answer to identifying, testing, and distinguishing the parameter families and local mechanisms that preserve function amid variability.
\section{Materials and Methods}

\subsection{Ambient simulation libraries and conditional generation}

For each model, we first sampled parameters \(\thetaVec_i\) from a prescribed distribution on the full parameter domain \(\Theta\), simulated the corresponding dynamics, and computed a feature vector \(y_i=\IMap(\thetaVec_i)\). Thus, the training data were ambient parameter--feature pairs; samples from any target-specific viable set were not assumed to be available. Parameters and features were transformed and normalized as described below and in the SI Appendix.

We trained a conditional score model \(s_\phi(\thetaVec_t,t,y)\) by denoising score matching on these pairs \citep{Hyvarinen2005,Vincent2011,Ho2020DDPM,Song2021SDE}. The network concatenated the noisy parameter vector, the conditioning features, and a 64-dimensional Fourier embedding of the noise level, followed by four 256-unit hidden layers with SiLU activations and a linear \(d_\theta\)-dimensional output layer. Samples were generated by annealed Langevin dynamics using 16 noise levels and 64 updates per level. The same architecture was used across systems; only the parameter and conditioning dimensions changed. Model-specific library sizes, optimization settings, and the complete noise schedule are reported in the SI Appendix.

The sampling and validation stages produce distinct objects. An ambient sample is a parameter--feature pair used to train the generator. A proposal is a parameter drawn from the learned conditional model at a requested target. Because the ambient library is sampled from a prior on \(\Theta\), the ideal conditional law on a regular fiber is prior- and coarea-weighted rather than uniform. Diffusion approximation and finite training resolution may further broaden the proposal distribution. We therefore re-simulated proposals and called a proposal accepted only when its recomputed features lay within the stated target tolerance. The accepted cloud is a proposal-weighted sample of the finite-tolerance viable set; component-coverage analyses separately assess the recovery and relative mass of its branches.

Unless a caption states otherwise, cloud-shape analyses use generated proposals to retain the sampled extent, whereas target-recovery statistics use re-simulated proposals or accepted points. Where iterative regeneration was used, accepted points together with small perturbations around them initialized another pass of the trained sampler before re-simulation. The SI Appendix gives the acceptance criteria, perturbation rule, target list, proposal counts, and three-round acceptance table; library construction and network training are accounted for separately.

\subsection{Empirical geometry of conditioned clouds}

We used the cloud-based geometric quantities introduced in the Results throughout the main text. The SI Appendix gives the participation-ratio, local-PCA, and empirical-curvature estimators; their neighborhood, scaling, and uncertainty analyses; and detailed comparisons with fitted forward-Jacobian ranks and nullspaces.

\subsection{Lorenz system}

We simulated
\begin{equation}
\dot x=\sigma(y-x),\qquad
\dot y=x(\rho-z)-y,\qquad
\dot z=xy-\beta z
\end{equation}
over the logarithmic parameter box
\begin{equation}
\log_{10}\sigma\in[0,3],\qquad
\log_{10}\rho\in[0,3],\qquad
\log_{10}\beta\in[-1,3].
\end{equation}
Parameters were sampled in these log coordinates and then mapped linearly to \([-1,1]^3\) for training. Each Lorenz conditional generator used 200{,}000 matched parameter--feature pairs. Coordinate-wise trajectory standard deviations were computed over \(t\in[50,150]\), after discarding \(t<50\). Scalar experiments conditioned on \(\log_{10}\operatorname{std}(x)\) or \(\log_{10}\operatorname{std}(z)\); joint experiments conditioned on both. The transition-adjacent experiment compared a fixed distant initial condition with initial conditions perturbed near the nonzero equilibria. Its construction, analytic Hopf surface, and boundary-offset calculation are described in the SI Appendix.

For direct feature validation, we re-simulated 5{,}000 proposals per scalar target. A Lorenz proposal was accepted when the maximum absolute error among the conditioned log-standard-deviation features was at most \(0.10\). The SI Appendix reports joint-target proposal counts, exclusions, and regeneration results.

\subsection{Izhikevich neuron model}

The neuron simulations used
\begin{equation}
\dot v=0.04v^2+5v+140-u+I,
\qquad
\dot u=a(bv-u),
\end{equation}
with the reset \(v\leftarrow c\) and \(u\leftarrow u+d\) whenever \(v\geq30\) mV \citep{Izhikevich2003,Izhikevich2004}. The sampled ranges were
\begin{equation}
a\in[0.001,0.25],\quad
b\in[-0.15,0.35],\quad
c\in[-75,-40],\quad
d\in[-1,9].
\end{equation}
A current step \(I=20\) was applied from 50 to 200 ms. From spikes in this interval we computed firing rate, mean interspike interval (ISI), CV(ISI), and a temporal-bin spike-count Fano statistic obtained from 10-ms bins. We use ``Fano factor'' below for this within-window dispersion statistic. For training, \(a\) was log-transformed; all four parameters were then normalized. The conditioning features were scaled by 200 Hz, 0.05 s, 5, and 5, respectively.

The generator was trained on 900{,}000 parameter--feature pairs, with a separate 5{,}000-pair held-out library used to select the displayed reference targets. We re-simulated 300 generated neurons per displayed target. A proposal was accepted when
\begin{equation}
\max_j
\frac{|f_j(\mathrm{sim})-f_j(\mathrm{target})|}
{\max\{|f_j(\mathrm{target})|,q_j\}}
\leq0.25,
\qquad
q=(10\ {\rm Hz},0.001\ {\rm s},0.05,0.10).
\end{equation}
The SI Appendix reports the sampling distribution within the parameter box, integration protocol, target-selection rules, held-out recovery statistics, and acceptance results.

\subsection{dsODE network model}

We used the deterministic spiking-network reduction of \citet{Chang2025dsODE}, whose state variables describe coarse-grained population activity and pending synaptic events. All simulations used \(n_E=300\), \(n_I=100\), a time step of 0.1 ms, and a duration of 5 s. E- and I-population rate means and standard deviations were computed over \([1,5]\) s after a 1-s burn-in and divided by 200 Hz for conditioning. In the 4D, 8D, and 10D experiments, all four connection probabilities were fixed at \(P_{QR}=0.8\). Although we refer to the complete dsODE equations in \citet{Chang2025dsODE}, fixed parameters, state initialization, and numerical implementation are provided in the SI Appendix.

The 4D experiments varied only the recurrent strengths,
\begin{equation}
(S_{EE},S_{IE},S_{EI},S_{II})
\in[0.60,1.30]\times[0.90,1.60]\times[2.20,3.20]\times[1.10,3.00].
\end{equation}
The 8D experiments varied
\begin{equation}
\begin{aligned}
&(\tau_{EE},\tau_{IE},\tau_I,\tau_{\mathrm{ref}},S_{EE},S_{IE},S_{EI},S_{II}),\\
&(\tau_{EE},\tau_{IE},\tau_I,\tau_{\mathrm{ref}})
\in[0.5,4.0]^2\times[3.0,6.0]\times[0,10],\\
&(S_{EE},S_{IE},S_{EI},S_{II})
\in[0.40,1.60]\times[0.60,1.80]\times[2.00,3.50]\times[1.00,3.00].
\end{aligned}
\end{equation}
The 10D experiments added external drives
\begin{equation}
(J_E^{\mathrm{ext}},J_I^{\mathrm{ext}})\in[0,20]^2
\end{equation}
to this 8D box. The 12D probability-varying extension and its parameter domain are reported in the SI Appendix. All varied parameters were linearly mapped to \([-1,1]\) before training.

For target-recovery analyses, we directly re-simulated generated dsODE parameters; displayed validation summaries used 200 proposals per target unless otherwise indicated. The principal acceptance rule required a maximum target-normalized feature error of 0.10, and a sensitivity analysis used a maximum absolute rate-feature error of 10 Hz. The SI Appendix specifies the normalization rule and its feature-specific floors.

Beat classification, dominant-frequency extraction, and the transverse-width statistic in the \(S/\tau\) plane were computed from post-burn-in trajectories and target-conditioned clouds as detailed in the SI Appendix. Correlations between cloud width and target properties were evaluated across target-conditioned clouds using Spearman's \(\rho\).

\subsection{Validation, robustness, and reporting}

Direct simulation determined whether the generated parameters reproduced the requested target. We recorded feature-error distributions in addition to thresholded acceptance fractions and inspected representative trajectories to verify their dynamical regime. The SI tables report the number of targets and proposals, acceptance counts over regeneration rounds, and sensitivity to alternative tolerances.

All post-training comparisons with uniform rejection, continuation, MCMC, and flow matching, including the definition of their success denominators and simulator-call accounting, are reported in the SI Appendix. Training-library construction and neural optimization are accounted for separately from post-training proposal generation.

\section*{Acknowledgments}
R.Z. and L.T. were partially supported by the National Natural Science and Technology Innovation STI2030-Major Project No.~2022ZD0204600. 
Z.-C.X. was supported by the Shanghai Municipal Education Commission through talent programs and NYU Shanghai boost fund. We thank for the useful discussion with Professor Lai-Sang Young and Professor Aaditya Rangan at New York University.

Author contributions: Z.-C.X. and L.T. conceived the study and jointly supervised the research; Z.-C.X. and R.Z. analyzed the data and derived the parameter compensations in the Izhkevich and dsODE models; R.Z. developed the software and prepared all figures; Z.-C.X. wrote the original draft; the authors acknowledge the use of ChatGPT in polishing the language and improving the readability of this manuscript, which was reviewed and edited by each of the authors. 

The author reports no conflict of interest.

\section*{Data, code, and materials availability}
\hypersetup{urlcolor=black}
The data and code available for public release, including the simulation, training, and figure-generation scripts, can be found at \url{https://github.com/JerryZ-PKU/Diffusion_Learning_Parameter_Manifold}.

\bibliographystyle{unsrtnat}
\bibliography{pnas_viable_manifolds_refs_updated_v3}

\end{document}


\maketitle

\section{Overview}
\label{sec:si_overview}
This SI Appendix collects the mathematical conventions, numerical protocols, and model-specific analyses supporting the main text. Section~\ref{sec:si_regularity} defines viable sets, states the result of the regular-level-set, and gives the empirical point-cloud estimators used for dimension, directions, and bending. Section~\ref{sec:si_diffusion} documents the conditional generator, resimulation, acceptance, and regeneration procedures. Sections~\ref{sec:si_lorenz}--\ref{sec:si_dsode} give the complete Lorenz, Izhikevich, and dsODE protocols and the reduced mechanistic calculations omitted from the Results. Section~\ref{sec:si_cost_scaling} separates ambient library construction, conditional-law learning, target-specific inverse recovery, and geometric support recovery. 

\section{Viable parameter manifolds and shared geometric conventions}
\label{sec:si_regularity}
Let
\begin{equation}
\dot{\xVec}=f(\xVec,\thetaVec),\qquad \xVec\in\R^n,\quad \thetaVec\in\Theta\subset\R^k,
\end{equation}
with $f$ smooth in both arguments. Let $\IMap(\thetaVec)\in\R^m$ denote a simulation-based feature map. Within an open parameter set that does not cross a bifurcation boundary, hyperbolic equilibria and hyperbolic periodic orbits vary smoothly with parameters, and any smooth functional of those invariant sets also varies smoothly. In a differentiable stratum, where
\begin{equation}
\mathrm{rank}\,D\IMap(\thetaVec)=r
\end{equation}
is locally constant, the constant-rank theorem gives the level set
\begin{equation}
\Mset_y=\{\thetaVec\in\Theta: \IMap(\thetaVec)=y\}
\end{equation}
as a $(k-r)$-dimensional embedded submanifold in local image coordinates; full row rank is the special case \(r=m\). In computation, we work with a tolerance-thickened viable set
\begin{equation}
\Mset_{y,\varepsilon}=\{\thetaVec\in\Theta:\norm{\IMap(\thetaVec)-y}\le \varepsilon\},
\end{equation}
which is the finite-tolerance object identified by re-simulating proposals and applying the stated acceptance rule.

\subsection{Intrinsic dimension and local compensation geometry}
\label{sec:si_geometry_estimators}
At regular points, the tangent space of $\Mset_y$ is the nullspace of the Jacobian:
\begin{equation}
T_{\thetaVec}\Mset_y = \{\delta\thetaVec : D\IMap(\thetaVec)\,\delta\thetaVec = 0\}.
\end{equation}
Thus the intrinsic dimension is
\begin{equation}
d_{\mathrm{int}}(\thetaVec)=k-\mathrm{rank}\,D\IMap(\thetaVec).
\end{equation}
All main-text geometric quantities are empirical measurements of generated point clouds in normalized parameter coordinates. We use the participation ratio as a continuous measure of covariance-spectrum concentration,
\begin{equation}
d_{\mathrm{PR}}=\frac{\left(\sum_i \lambda_i\right)^2}{\sum_i \lambda_i^2},
\end{equation}
where $\lambda_i$ are the covariance eigenvalues of a target-conditioned cloud. We use local PCA to describe directions and finite-scale spread within a generated cloud, a locally fitted forward Jacobian to estimate first-order level-set null directions, and a local quadratic patch to summarize projected bending. These estimators answer different finite-sample questions and need not agree exactly for a curved, tolerance-thickened, truncated, or nonuniformly sampled set. Throughout the main text, ``dimension,'' ``tangent direction,'' and ``curvature'' refer to these empirical quantities; comparisons with Jacobian nullity, PCA rank, and estimator sensitivity are reported in this SI.

At each generated center $\theta_i$, we selected the 2{,}000 nearest training parameters in diffusion-scaled coordinates and paired them with their simulated guide features. Parameters and guide coordinates were standardized within this neighborhood. We then fit $g_{\mathrm{std}}\approx A_{\mathrm{std}}\theta_{\mathrm{std}}+b$ by Gaussian-weighted ridge least squares, using the 75th percentile of local training-neighbor distances as the bandwidth, a minimum weight of $10^{-6}$, and a ridge penalty of $10^{-4}\operatorname{trace}(X^\top W X)/d_\theta$. Singular values satisfying $s\ge 0.05s_{\max}$ defined the active local rank. The remaining right singular vectors of $A_{\mathrm{std}}$ defined the fitted forward-map nullspace and were converted back to native scaled-parameter increments.

The 420 nearest generated samples to each center supplied the local PCA basis, eigenvalues, 95\% local PCA rank, and local participation ratio. The fitted-Jacobian nullspace estimates first-order parameter changes that preserve the included feature vector, whereas local PCA describes the dominant directions of the sampled, finite-tolerance cloud. Figure~7C therefore compares a tangent-side quantity: the leading projected Jacobian-null compensation mode with the empirical PC1--PC2 plane. A small out-of-plane angle indicates that this projected compensation direction lies close to the sampled plane. The exact projection, angle, eligibility filter, and system-level distributions are given in Section~\ref{sec:si_fig7c_geometry}.

We tested the sensitivity of the comparison to the neighborhood sizes used for the Jacobian fit and local PCA and to the numerical-rank threshold. The plane-resolved angle distributions remained concentrated at low values across these choices. Detailed results are reported in Section~\ref{sec:si_fig7c_geometry}.

A single Jacobian contains only first-order information. We therefore estimated bending separately, at centers passing the specified regularity screen. For the systematic Izhikevich analysis, displayed coordinates were linearly rescaled to comparable ranges. We used a radius-0.24 neighborhood, falling back to the nearest 30 samples when fewer than 30 points lay inside the radius. In local PCA coordinates $(u,v)$ and a leading normal coordinate $w$, we fit a quadratic Monge patch
\begin{equation}
w=\frac12\bigl(\kappa_1 u^2 + 2\kappa_{12}uv + \kappa_2 v^2\bigr)+\ell_1 u+\ell_2 v+\varepsilon.
\end{equation}
We oriented the fitted normal away from the global cloud center before assigning the sign and summarized the fitted mean-curvature proxy $H$ by $\operatorname{sign}(H)\log_{10}(1+|H|)$. This is a coordinate-scaled local bending diagnostic. 

\subsection{Ideal conditional law and finite-tolerance acceptance}
\label{sec:si_prior_weighted}
If the training parameters are drawn from a prior density \(p_\Theta(\thetaVec)\), ideal noiseless conditioning weights a regular level set by the co-area Jacobian over the locally independent feature directions:
\begin{equation}
J_{r_y}(\thetaVec)=\prod_{j=1}^{r_y}\sigma_j(D\IMap(\thetaVec)),
\qquad
\dd\mu_y(\thetaVec)\propto
\frac{p_\Theta(\thetaVec)}{J_{r_y}(\thetaVec)}\,\dd\Haus^{k-r_y}(\thetaVec).
\label{eq:coarea_main_revised}
\end{equation}
More precisely, if the feature map is restricted to local image coordinates of dimension $r_y$, the pushforward density and inverse conditional law are
\begin{align}
p_Y(y)&=\int_{\Mset_y}\frac{p_\Theta(\thetaVec)}{J_{r_y}(\thetaVec)}\,\dd\Haus^{k-r_y}(\thetaVec),\\
\dd\mu_y(\thetaVec)&=
\frac{p_\Theta(\thetaVec)}{p_Y(y)J_{r_y}(\thetaVec)}\,
\dd\Haus^{k-r_y}(\thetaVec).
\end{align}
The same formulas apply to a level set truncated by the boundary of $\Theta$, with integration over $\Mset_y\cap\Theta$; corners and tangencies to the box boundary require one-sided local charts. Conditional generation therefore approximates a prior- and coarea-weighted law rather than a uniform measure on $\Mset_y$.
Finite libraries and diffusion smoothing broaden the learned conditional law. Direct re-simulation and explicit feature tolerances then restrict the proposal law to a stated tube around the target; the resulting accepted cloud remains proposal-weighted within that tube.

Throughout the manuscript we therefore interpret the generated clouds as empirical atlases of prior-weighted, tolerance-thickened viable sets. 

\section{Conditional diffusion model}
\label{sec:si_diffusion}
\subsection{Training objective}
We train a conditional score-based diffusion model on parameter--feature pairs $(\thetaVec_i,y_i)$. Let $q_t(\thetaVec_t\mid \thetaVec_0)$ denote the forward noising process and $s_\phi(\thetaVec_t,t,y)$ the conditional score network. The objective is a denoising-score-matching loss of the form
\begin{equation}
\mathcal L(\phi)=\mathbb E_{t,\thetaVec_0,\thetaVec_t}\big\| s_\phi(\thetaVec_t,t,y)-\nabla_{\thetaVec_t}\log q_t(\thetaVec_t\mid \thetaVec_0)\big\|^2,
\end{equation}
implemented with Gaussian perturbations of the parameter vector \citep{Ho2020DDPM,Song2021SDE,Vincent2011,Hyvarinen2005}.

\subsection{Network architecture and sampling schedule}
Across the final Lorenz, Izhikevich, and dsODE experiments, we use the same conditional score-network architecture; only the parameter dimension $d_\theta$ and feature dimension $d_f$ change between applications. The network inputs are the current parameter state, target feature vector, and a Fourier embedding of the diffusion noise level:
\begin{equation}
[\theta_t,\; y,\; e(\sigma_t)]\in\R^{d_\theta+d_f+64},
\end{equation}
where the 64-dimensional embedding is
\begin{equation}
e(\sigma)=\bigl[\sin(\log \sigma\, B),\;\cos(\log \sigma\, B)\bigr].
\end{equation}
Here $B\in\R^{32}$ represents 32 sin-cosine pairs fixed at initialization from $\mathcal N(0,16^2)$ and kept non-trainable. They encode the current Gaussian noise scale rather than the numerical sampling-step index. $B$ is stored in the model checkpoint to make sure that training and sampling use the same frequencies.

The score network applies
\begin{equation}
\mathrm{Linear}(d_\theta+d_f+64,256)
\rightarrow \mathrm{SiLU}
\rightarrow \bigl[\mathrm{Linear}(256,256)\rightarrow \mathrm{SiLU}\bigr]^{\times3}
\rightarrow \mathrm{Linear}(256,d_\theta).
\end{equation}
The backbone contains four 256-unit affine hidden layers, each followed by a SiLU activation, and a final linear layer mapping 256 units to $d_\theta$ outputs. The final layer is initialized with zero weights and biases. Its output estimates the conditional score $s_\phi(\theta_t,y,\sigma_t)$, which supplies the deterministic direction in the reverse sampler.
The score networks were optimized using AdamW with zero weight decay, gradient-norm clipping at 1, and a constant learning rate without scheduling. The Lorenz and dsODE generators used minibatches of 256 with a learning rate of $10^{-4}$, whereas the Izhikevich generator used minibatches of 4096 with a learning rate of $3\times10^{-3}$.

Sampling uses annealed Langevin dynamics with 16 geometrically spaced noise levels and 64 updates per level, for 1024 reverse-diffusion steps. At noise level $\sigma_t$, the step size is
\begin{equation}
\alpha_t=0.1\,\sigma_t^2.
\label{eq:ald-step-size}
\end{equation}
Across the 16-level schedule, $\alpha_t$ therefore decreases geometrically from $10^{-1}$ to $10^{-5}$, as shown in main-text Fig.~1B. One update is
\begin{equation}
\theta_{t+1}
=
\theta_t+\alpha_t s_\phi(\theta_t,y,\sigma_t)+\sqrt{2\alpha_t}\,z,
\qquad z\sim\mathcal N(0,I).
\end{equation}
The fixed noise schedule was $\sigma_j=10^{-2j/15}$ for $j=0,\ldots,15$, giving 16 equally spaced values in $\log\sigma$ from $\sigma_{\max}=1$ to $\sigma_{\min}=0.01$. During training, one index $j$ was drawn uniformly and independently for each example, and the perturbed parameter was $\theta_j=\theta_0+\sigma_j\epsilon$, where $\epsilon\sim\mathcal N(0,I)$. The regression target was $-\epsilon/\sigma_j$, and the per-example score MSE was weighted by $\sigma_j^2$. During sampling, the levels were traversed from largest to smallest. Sampling was initialized uniformly in $[-1,1]^{d_\theta}$, a fresh independent Gaussian vector was drawn at every Langevin update, and each updated coordinate was clipped to $[-1,1]$.

\subsection{Resimulation-first validation}
\label{sec:si_sampling_validation}
At test time, the model is conditioned on a target feature vector $y$ and generates parameter proposals $\tilde\thetaVec$. Every displayed trajectory and feature-validation statistic is obtained by simulating a generated parameter. Unless a panel is explicitly labeled ``accepted,'' geometric point clouds show re-simulated proposals before feature-tolerance filtering. For Lorenz validation histograms we simulated 5{,}000 generated parameters per scalar target. For Izhikevich validation traces and histograms we simulated 300 generated neurons per target. For dsODE validation planes, dots and error bars summarize 200 generated parameters per target. Acceptance-filtered and regenerated clouds are used for the three-round precision analysis below.

\heading{Acceptance standard and resimulation} For Lorenz, a proposal was accepted when
\begin{equation}
\max_j\left|\log_{10}\operatorname{std}_j(\mathrm{sim})-\log_{10}\operatorname{std}_j(\mathrm{target})\right|\le 0.10.
\end{equation}
For Izhikevich, with features ordered as firing rate, mean ISI, CV(ISI), and Fano factor, a proposal was accepted when
\begin{equation}
\max_j\frac{|f_j(\mathrm{sim})-f_j(\mathrm{target})|}{\max(|f_j(\mathrm{target})|,q_j)}\le0.25,
\qquad q=(10\ \mathrm{Hz},0.001\ \mathrm{s},0.05,0.10).
\end{equation}
For dsODE, with $f_j$ denoting the conditioned E/I firing-rate means and/or standard deviations, a proposal was accepted under the main normalized rule when
\begin{equation}
\max_j\frac{|f_j(\mathrm{sim})-f_j(\mathrm{target})|}{\max(|f_j(\mathrm{target})|,10\ \mathrm{Hz})}\le0.10.
\end{equation}
An absolute-error sensitivity branch used a maximum error of 10 Hz. Because both strict criteria retained relatively few 10D proposals, an additional 10D branch used a relaxed maximum target-normalized error of 0.35.

After each round, accepted points and small jitter-fill starts were passed through the trained diffusion sampler, and the resulting proposals were re-simulated before the next acceptance decision. The recorded jitter scales in normalized parameter coordinates were 0.15 for Lorenz, 0.10 for Izhikevich, and 0.05 for dsODE. The jitter was Gaussian and clipped to the normalized parameter box; Lorenz and Izhikevich scaled the local parent covariance, whereas 0.05 was the coordinatewise standard deviation for dsODE. The same 1{,}024-update sampler schedule was used in each regeneration round. Across rounds, most collections changed only modestly, whereas the acceptance fraction for the 4D mean-only collection increased from 0.9435 in Round 0 to 0.9955 in Round 2 under the target-normalized rule. For the 10D collection, acceptance was approximately 11--12\% under the target-normalized 0.10 rule and 15\% under the absolute-error rule, and increased to approximately 59\% when the target-normalized threshold was relaxed to 0.35.

\begin{table}[htbp]
\centering
\scriptsize
\setlength{\tabcolsep}{3pt}
\caption{Acceptance under direct re-simulation across three diffusion-seeded regeneration rounds. Counts are accepted proposals; fractions are shown before counts in parentheses.}
\label{tab:si_acceptance_rounds}
\begin{tabular}{@{}lp{0.27\linewidth}lrrrr@{}}
\toprule
System & Parameterization / guide & Error rule & Proposals & Round 0 & Round 1 & Round 2 \\
\midrule
Lorenz & $x$-std & abs. $0.10$ & 5{,}000 & 0.8862 (4{,}431) & 0.8894 (4{,}447) & 0.8834 (4{,}417) \\
Lorenz & $z$-std & abs. $0.10$ & 5{,}000 & 0.6978 (3{,}489) & 0.6918 (3{,}459) & 0.6938 (3{,}469) \\
Lorenz & joint $x/z$-std & abs. $0.10$ & 4{,}000 & 0.1948 (779) & 0.1920 (768) & 0.2023 (809) \\
\addlinespace[3pt]
Izhikevich & full four-feature examples & target norm $0.25$ & 4{,}000 & 0.6875 (2{,}750) & 0.6928 (2{,}771) & 0.6925 (2{,}770) \\
Izhikevich & regular examples & target norm $0.25$ & 4{,}000 & 0.5193 (2{,}077) & 0.5198 (2{,}079) & 0.5223 (2{,}089) \\
\addlinespace[3pt]
dsODE & 4D $S$, E/I mean & target norm $0.10$ & 2{,}000 & 0.9435 (1{,}887) & 0.9950 (1{,}990) & 0.9955 (1{,}991) \\
dsODE & 4D $S$, E/I mean & abs. $10$ Hz & 2{,}000 & 0.9820 (1{,}964) & 0.9980 (1{,}996) & 0.9985 (1{,}997) \\
dsODE & 4D $S$, E/I std & target norm $0.10$ & 2{,}000 & 0.8810 (1{,}762) & 0.8840 (1{,}768) & 0.8945 (1{,}789) \\
dsODE & 4D $S$, E/I std & abs. $10$ Hz & 2{,}000 & 0.8630 (1{,}726) & 0.8630 (1{,}726) & 0.8790 (1{,}758) \\
dsODE & 4D $S$, E/I mean+std & target norm $0.10$ & 2{,}000 & 0.9830 (1{,}966) & 0.9845 (1{,}969) & 0.9790 (1{,}958) \\
dsODE & 4D $S$, E/I mean+std & abs. $10$ Hz & 2{,}000 & 0.9780 (1{,}956) & 0.9795 (1{,}959) & 0.9765 (1{,}953) \\
dsODE & 8D $\tau+S$, E/I mean+std & target norm $0.10$ & 4{,}000 & 0.6385 (2{,}554) & 0.6430 (2{,}572) & 0.6430 (2{,}572) \\
dsODE & 8D $\tau+S$, E/I mean+std & abs. $10$ Hz & 4{,}000 & 0.6455 (2{,}582) & 0.6583 (2{,}633) & 0.6498 (2{,}599) \\
dsODE & 10D $J_{\mathrm{ex}}+\tau+S$, E/I mean+std & target norm $0.10$ & 4{,}000 & 0.1248 (499) & 0.1113 (445) & 0.1145 (458) \\
dsODE & 10D $J_{\mathrm{ex}}+\tau+S$, E/I mean+std & abs. $10$ Hz & 4{,}000 & 0.1538 (615) & 0.1460 (584) & 0.1495 (598) \\
dsODE & 10D $J_{\mathrm{ex}}+\tau+S$, E/I mean+std & target norm $0.35$ & 4{,}000 & 0.5910 (2{,}364) & 0.5903 (2{,}361) & 0.5943 (2{,}377) \\
\bottomrule
\end{tabular}
\end{table}

\section{Lorenz system details}
\label{sec:si_lorenz}

\subsection{Equations, parameter domain, and numerical protocol}
We simulate
\begin{equation}
\dot x=\sigma(y-x),\qquad
\dot y=x(\rho-z)-y,\qquad
\dot z=xy-\beta z.
\end{equation}
The generators use logarithmic parameter coordinates
\begin{equation}
\log_{10}\sigma\in[0,3],\qquad
\log_{10}\rho\in[0,3],\qquad
\log_{10}\beta\in[-1,3],
\end{equation}
equivalently $\sigma\in[1,10^3]$, $\rho\in[1,10^3]$, and $\beta\in[10^{-1},10^3]$. Parameters are sampled uniformly in these logarithmic coordinates and linearly mapped to $[-1,1]^3$ for training. The classical parameter vector $(10,28,8/3)$ is used as a reference slice rather than as a training center. Each Lorenz conditional generator uses 200{,}000 aligned parameter--feature pairs and 100 shuffled training epochs.

For $\rho>1$, the nonzero equilibria are
\begin{equation}
C_\pm=\left(\pm\sqrt{\beta(\rho-1)},\ \pm\sqrt{\beta(\rho-1)},\ \rho-1\right).
\end{equation}
When $\sigma>\beta+1$, their Hopf loss of stability occurs on
\begin{equation}
\rho_H(\sigma,\beta)=\frac{\sigma(\sigma+\beta+3)}{\sigma-\beta-1}.
\end{equation}
Lorenz trajectories used to construct the simulation libraries were integrated over $t\in[0,150]$ with SciPy's \texttt{solve\_ivp}, using its default explicit RK45 method and tolerances $\mathrm{rtol}=10^{-3}$ and $\mathrm{atol}=10^{-6}$. The solver used adaptive internal time steps, while trajectories were requested at 15{,}000 equally spaced output times.

The distant-start library used the initial state $(1,0,0)$, whereas the near-equilibrium library used $1.001C_+$. The latter corresponds to the deterministic radial displacement $0.001C_+$, i.e., a $0.1\%$ perturbation directed away from the origin.

\subsection{Feature extraction and scalar-target manifolds}
Trajectory features are coordinate-wise standard deviations computed over the window $t\in[50,150]$, after discarding earlier transients. The scalar-manifold figures condition on $\log_{10}\operatorname{std}(x)$ or $\log_{10}\operatorname{std}(z)$. Most representative trajectories are initiated at $(1,0,0)$. The directly simulated cloud of $(\operatorname{std}(x),\operatorname{std}(y),\operatorname{std}(z))$ values serves as the three-dimensional feature object from which the scalar target sections in Fig.~2D are cut.

All Lorenz conditional generators, including the scalar $x$- and $z$-standard-deviation models, the joint $x$/$z$-standard-deviation model, and the transition-adjacent model, were trained on 200{,}000 aligned parameter--condition pairs for 100 epochs using shuffled minibatches of 256.
Validation histograms in Fig.~2E use 5{,}000 generated parameters per scalar target. Following the figure source, generated samples with $\log_{10}\mathrm{std}\le 0$ are excluded when estimating the scalar-target validation distributions. The excluded fractions are $0.0738$, $0.0364$, $0.0260$, $0.0298$, and $0.0580$ for the five $x$-std targets, and $0.0904$, $0.0468$, $0.0250$, $0.0180$, and $0.0220$ for the corresponding $z$-std targets.

\subsection{Transition-adjacent conditioning}
\label{sec:si_lorenz_transition}
The transition-adjacent Lorenz experiment uses two $x$-variability summaries computed from different initial-condition classes. One trajectory is initiated at $(1,0,0)$, and the other is initiated near the nonzero equilibria $C_\pm$ using $0.1\%$ perturbations around the stable fixed points. The target requires %
\begin{equation}
\log_{10}\operatorname{std}_x[(1,0,0)]\approx 1,
\qquad
\log_{10}\operatorname{std}_x[\text{near }C_+]\in[-1,0].
\end{equation}
This selects systems that still make large excursions from a distant start but collapse rapidly when started near the stable equilibria. Along the classical slice $\sigma=10$, $\beta=8/3$, the resulting cloud intersects the transient-chaos corridor between the homoclinic point $\rho\approx 13.926$ and the Hopf point $\rho_H\approx 24.74$. The upper boundary follows the analytic Hopf surface
\begin{equation}
\rho_H=\frac{\sigma(\sigma+\beta+3)}{\sigma-\beta-1},
\end{equation}
up to the visible thickening introduced by finite sampling and cloud expansion for visualization. We interpret this experiment as a transition-adjacent localization.

\subsection{Extraction of the transition-adjacent boundary}
The generated cloud was binned by occupied $(\log_{10}\sigma,\log_{10}\beta)$ columns. Before surface extraction, the occupied set was expanded by a radius of $0.06$ in log-parameter space, enclosed holes were filled, and the largest connected component was retained. This procedure produces an expanded outer envelope, so the reported boundary-to-Hopf separation may exceed that obtained directly from the unexpanded generated cloud. The point with the largest occupied $\log_{10}\rho$ was retained from each column so that densely sampled columns did not dominate the comparison. For each valid boundary point, we calculated the Euclidean distance in $(\log_{10}\sigma,\log_{10}\rho,\log_{10}\beta)$ space to the nearest point on the analytic Hopf surface. These distances had mean $0.20$, standard deviation $0.10$, median $0.20$, and interquartile range $0.12$--$0.27$. When the surfaces were instead compared at the same $(\sigma,\beta)$, the signed vertical offsets had mean $+0.36$, standard deviation $0.42$, median $0.29$, and interquartile range $0.16$--$0.40$ in $\log_{10}\rho$; the larger axis-aligned mean reflects regions where the Hopf surface is oblique to the $\log_{10}\rho$ axis.

\subsection{Lorenz rank and finite-cloud geometry}
For a regular scalar statistic, the exact inverse image in the three-dimensional parameter domain has codimension one. The displayed participation ratios between approximately 1.4 and 2.6 quantify the finite covariance shape of the generated sheets; they are not integer rank estimates. The fitted forward-Jacobian and local-PCA analyses provide separate local rank and finite-scale cloud summaries.

Table~\ref{tab:si_lorenz_scalar_rank_geometry} reports the target-wise diagnostics for the ten scalar Lorenz proposal clouds in Fig.~2, with 5,000 proposals per target. The global participation ratio $d_{\rm PR}^{\log}$ is computed in the plotted $(\log_{10}\sigma,\log_{10}\rho,\log_{10}\beta)$ coordinates and reproduces the Fig.~2 convention, whereas $d_{\rm PR}^{\rm sc}$ uses diffusion-scaled parameter coordinates. Brackets for the global PR columns are 95\% intervals from 1,000 within-cloud bootstrap resamples. The column $r_J/d_J$ gives the modal fitted-Jacobian rank/nullity under the fixed rule $s_i\geq0.05s_{\max}$. PCA95 is the modal 420-neighbor rank, with its modal center fraction in parentheses, and local PR is the median [IQR] over the same overlapping neighborhoods in diffusion-scaled coordinates.

\begin{table}[htbp]
\centering
\scriptsize
\caption{Target-wise Lorenz rank and finite-cloud geometry for scalar standard-deviation guides (5,000 proposals per target). Global-PR brackets are 95\% within-cloud bootstrap intervals; local-PR brackets report the median-neighborhood IQR.}
\label{tab:si_lorenz_scalar_rank_geometry}
\resizebox{\linewidth}{!}{%
\begin{tabular}{llccccc}
\toprule
Guide & Target & $d_{\rm PR}^{\log}$ & $d_{\rm PR}^{\rm sc}$ & $r_J/d_J$ & PCA95 mode & Local $d_{\rm PR}^{\rm sc}$ \\
\midrule
$\operatorname{std}(x)$ & 3   & 2.33 [2.28, 2.35] & 2.17 [2.09, 2.24] & 1/2 & 3 (94.0\%)  & 2.42 [2.22, 2.52] \\
$\operatorname{std}(x)$ & 5   & 1.99 [1.94, 2.04] & 1.86 [1.81, 1.90] & 1/2 & 3 (95.6\%)  & 2.34 [2.17, 2.40] \\
$\operatorname{std}(x)$ & 10  & 1.98 [1.97, 2.00] & 1.90 [1.87, 1.92] & 1/2 & 3 (85.5\%)  & 2.23 [2.12, 2.29] \\
$\operatorname{std}(x)$ & 30  & 1.81 [1.79, 1.83] & 1.67 [1.65, 1.69] & 1/2 & 3 (99.9\%)  & 2.33 [2.20, 2.39] \\
$\operatorname{std}(x)$ & 100 & 1.57 [1.55, 1.59] & 1.43 [1.42, 1.45] & 1/2 & 3 (100.0\%) & 2.61 [2.47, 2.69] \\
$\operatorname{std}(z)$ & 3   & 1.43 [1.40, 1.46] & 1.51 [1.47, 1.54] & 1/2 & 3 (77.9\%)  & 2.64 [2.11, 2.83] \\
$\operatorname{std}(z)$ & 5   & 2.55 [2.46, 2.65] & 2.61 [2.53, 2.66] & 1/2 & 3 (95.8\%)  & 2.68 [2.49, 2.80] \\
$\operatorname{std}(z)$ & 10  & 2.32 [2.29, 2.34] & 2.30 [2.27, 2.32] & 1/2 & 3 (98.4\%)  & 2.56 [2.35, 2.70] \\
$\operatorname{std}(z)$ & 30  & 2.18 [2.16, 2.19] & 2.08 [2.06, 2.11] & 1/2 & 3 (82.6\%)  & 2.45 [2.17, 2.64] \\
$\operatorname{std}(z)$ & 100 & 1.96 [1.94, 1.98] & 1.70 [1.67, 1.72] & 1/2 & 3 (95.4\%)  & 2.40 [2.15, 2.60] \\
\bottomrule
\end{tabular}%
}
\end{table}

As shown in Table~\ref{tab:si_lorenz_scalar_rank_geometry}, the fitted forward map had rank 1 and nullity 2 in all ten scalar-target clouds. The 95\% local-PCA rank had mode 3 in every cloud (modal fractions 77.9--100\%), whereas the median local participation ratios ranged from 2.23 to 2.68. Jacobian nullity is the local integer level-set dimension, while PCA95 and participation ratio describe finite-neighborhood thickness, curvature, truncation, and anisotropic sampling. Across the ten clouds, global participation ratios ranged from 1.43 to 2.55 in the plotted log-parameter coordinates and from 1.43 to 2.61 in diffusion-scaled coordinates.


\section{Izhikevich neuron details}
\label{sec:si_izh}

\subsection{Equations, parameter domain, and preprocessing}
The hybrid neuron model is
\begin{equation}
\dot v=0.04v^2+5v+140-u+I,\qquad
\dot u=a(bv-u),
\end{equation}
with reset $v\leftarrow c$, $u\leftarrow u+d$ whenever $v\ge30$ mV. The sampled box is
\begin{equation}
a\in[0.001,0.25],\quad b\in[-0.15,0.35],\quad
c\in[-75,-40],\quad d\in[-1,9].
\end{equation}
Parameter $a$ is log-transformed before normalization; $b,c,d$ are linearly normalized. All four normalized parameter coordinates are mapped to the range used by the diffusion model. Before generator training, $\log_{10}a$ was linearly mapped from $[-3,\log_{10}(0.25)]$ to $[-1,1]$, while $b$, $c$, and $d$ were linearly mapped from their respective sampled intervals to $[-1,1]$. Traces were simulated for 300 ms with BrainPy 2.6.0 using the Izhikevich neuron's default \texttt{exp\_auto} integration method and a fixed time step of $0.1$ ms. The neuron object was initialized before the sampled parameters were assigned, giving $(v_0,u_0)=(-70\ \mathrm{mV},-14)$. Within each numerical step, $(v,u)$ was first advanced over the full step; the threshold condition $v\geq30$ mV was then evaluated on the advanced voltage, and the reset $v\leftarrow c$, $u\leftarrow u+d$ was applied before the next step. The dynamics were deterministic, and parameter-library batch $k$ used NumPy seed $42+k$ for $k=0,\ldots,24$.

\subsection{Input protocol and response features}
The manifold-learning experiments use a fixed step-current protocol with baseline 0 and current strength $I=20$ during the interval 50--200 ms. Features are computed from spikes in that same 50--200 ms window after allowing the initial 50 ms to settle. The four response features are
\begin{equation}
\text{rate},\qquad \text{mean inter-spike interval (ISI)},\qquad \text{CV(ISI)},\qquad \text{Fano factor}.
\end{equation}
Firing rate is the spike count divided by the window duration $0.150$ s. Mean ISI and CV(ISI) are calculated from consecutive spike times using the population-standard-deviation convention; both are set to zero for traces with fewer than two spikes. The Fano factor is the population variance-to-mean ratio of spike counts in 15 nonoverlapping 10-ms bins and is set to zero when the mean bin count is zero.
Before conditioning, the four features are scaled as
\begin{equation}
\frac{\mathrm{rate}}{200\ \mathrm{Hz}},\qquad
\frac{\mathrm{mean\;ISI}}{0.05\ \mathrm{s}},\qquad
\frac{\mathrm{CV(ISI)}}{5},\qquad
\frac{\mathrm{Fano}}{5}.
\end{equation}

\subsection{Training and validation}
The Izhikevich generator was trained on 900{,}000 parameter--feature pairs for 100 epochs using shuffled minibatches of 4096 and a constant learning rate of $3\times10^{-3}$. A separate held-out test library contains 5{,}000 parameter--feature pairs that were not used during training. The examples displayed in Fig.~3 use conditioning guides and reference parameters from this held-out library. The feature distributions in Fig.~3 show 300 generated neurons for each of the two example targets. For broader quantitative validation, we selected 100 additional targets from the test library and generated and resimulated 1{,}000 parameter proposals per target. 

The 100 additional test targets provided broad coverage of firing rate, mean ISI, CV(ISI), and Fano factor. We generated and BrainPy-resimulated 1{,}000 new parameter proposals per target. A proposal was accepted when the largest absolute discrepancy across the four features, after normalization by the larger of the target magnitude and the corresponding feature floor, did not exceed 0.25. The feature floors were 10 Hz, 0.001 s, 0.05, and 0.10, respectively. Of the 100{,}000 resimulated proposals, 66{,}782 were accepted (66.78\%; 95\% target-bootstrap CI 62.15--71.32\%). In the four parameter coordinates normalized to $[-1,1]$, Euclidean distance from each reference parameter vector to its nearest generated proposal had a median of 0.0334 (IQR 0.0252--0.0432; range 0.0117--0.1487). A generated proposal lay within a descriptive radius of 0.10 in normalized coordinates for 98/100 targets, compared with 87/100 at radius 0.05 and 100/100 at radius 0.15. Across targets, the nearest-proposal feature error had a median of 0.0021 (range 0--0.0608), while the median of the within-target median errors was 0.1620 over all proposals and 0.1010 over accepted proposals. Generated feature means tracked their targets with Pearson correlations of 0.999, 0.999, 0.999, and 0.990 for firing rate, mean ISI, CV(ISI), and Fano factor, respectively; the corresponding median within-target population standard deviations were 6.058 Hz, 0.463 ms, 0.029, and 0.081.


\subsection{Mutual compensation of $(a,b,c,d)$ in the regular-spiking regime}
\label{sec:si_izh_compensation}

We analyze the regular-spiking regime under a constant input current during the active stimulation window. Let $v_{\mathrm{th}}=30$ mV denote the spike apex used to trigger reset. Immediately after the $n$th spike, the hybrid reset enforces
\begin{equation}
v(t_n^+)=c,
\qquad
u(t_n^+)=u_n^+.
\end{equation}
Because $v$ is reset to the fixed value $c$ after every spike, the spike-to-spike state is one-dimensional: once $u_n^+$ is specified, the next interspike interval and the next reset value are determined by $(a,b,c,d)$.

\paragraph{Exact spike-to-spike map.}
Between spikes, the recovery equation is linear:
\begin{equation}
\dot u + a u = a b v.
\end{equation}
Solving this equation over the interval $[t_n^+,t_n^+ + t]$ gives
\begin{equation}
u(t_n^+ + t)=e^{-at}u_n^+ + ab\int_0^t e^{-a(t-s)}v_n(s)\,\dd s,
\end{equation}
where $v_n(s)$ is the voltage trajectory during the $n$th interspike interval. If $T_n$ denotes the next hitting time of $v=30$, then after the reset jump $u\mapsto u+d$ the next post-spike value is
\begin{equation}
\label{eq:izh_return_map}
u_{n+1}^+ = P(u_n^+;a,b,c,d,I)
= d + e^{-aT_n}u_n^+ + ab\int_0^{T_n} e^{-a(T_n-s)}v_n(s)\,\dd s.
\end{equation}
Equation~\eqref{eq:izh_return_map} is an exact return map for the post-spike recovery variable. In the language of hybrid spiking dynamics, regular spiking corresponds to a stable fixed point of this adaptation map, whereas bursting and irregular spiking correspond to more complicated periodic or nonperiodic orbits of the map \citep{TouboulBrette2009,Nobukawa2018}.

Let $(u_*^+,T_*,v_*(t))$ denote a regular-spiking periodic orbit. Setting $u_{n+1}^+=u_n^+=u_*^+$ in Eq.~\eqref{eq:izh_return_map} yields
\begin{equation}
\label{eq:izh_fixedpoint_exact}
u_*^+\bigl(1-e^{-aT_*}\bigr)
=ab\int_0^{T_*} e^{-a(T_*-s)}v_*(s)\,\dd s + d.
\end{equation}
Introducing the exponentially weighted cycle-average voltage
\begin{equation}
\bar v_a(c,T_*,a)
:= \frac{a\int_0^{T_*} e^{-a(T_*-s)}v_*(s)\,\dd s}{1-e^{-aT_*}},
\end{equation}
Eq.~\eqref{eq:izh_fixedpoint_exact} becomes
\begin{equation}
\label{eq:izh_exact_balance}
u_*^+ = b\,\bar v_a(c,T_*,a) + \frac{d}{1-e^{-aT_*}}.
\end{equation}
This identity already exposes the mutual compensation mechanism. Parameter $d$ enters as the spike-triggered increment of the recovery variable, parameter $a$ controls how much of that increment survives until the next spike, parameter $b$ sets the strength of subthreshold coupling from $v$ to $u$, and parameter $c$ affects the full orbit $v_*(t)$ and therefore both $\bar v_a$ and $T_*$. The viable set is not a point because many parameter changes can preserve the balance in Eq.~\eqref{eq:izh_exact_balance}.

\paragraph{Slow-adaptation reduction in the regular-spiking regime.}
The canonical regular-spiking example has small $a$ \citep{Izhikevich2003}, and more generally regular-spiking responses are the regime in which $u$ evolves slowly compared with the fast voltage excursion between reset and spike. In that case the exponential kernel in Eq.~\eqref{eq:izh_exact_balance} is nearly flat over one interspike interval and
\begin{equation}
1-e^{-aT_*}=aT_*+O\bigl((aT_*)^2\bigr),
\qquad
\bar v_a = \bar v + O(aT_*),
\end{equation}
where
\begin{equation}
\bar v := \frac{1}{T_*}\int_0^{T_*} v_*(s)\,\dd s
\end{equation}
is the ordinary cycle-average voltage. Therefore
\begin{equation}
\label{eq:izh_slow_balance}
u_*^+ \approx b\,\bar v + \frac{d}{aT_*}
= b\,\bar v + \frac{r_*d}{a},
\qquad r_* = \frac{1}{T_*}.
\end{equation}
Equation~\eqref{eq:izh_slow_balance} is the simplest compensation law for regular firing. It says that the dominant spike-triggered adaptation load is controlled by the combination $r_*d/a$: each spike adds $d$ to $u$, and between spikes the variable forgets that increment on the timescale $1/a$. Thus larger $d$ can be offset by larger $a$ to preserve the same average adaptation load.

\paragraph{Interspike flight time and the role of $c$.}
In the same slow-adaptation regime, $u$ is approximately constant during a single interspike interval. Writing
\begin{equation}
F(v)=0.04v^2+5v+140,
\end{equation}
we obtain the scalar voltage equation
\begin{equation}
\dot v = F(v)+I-u_*^+,
\qquad v(0)=c,
\qquad v(T_*)=30.
\end{equation}
Hence the interspike interval is approximated by the flight-time integral
\begin{equation}
\label{eq:izh_flight_time}
T_* \approx \int_c^{30}\frac{\dd v}{F(v)+I-u_*^+}.
\end{equation}
Differentiating Eq.~\eqref{eq:izh_flight_time} gives
\begin{equation}
\label{eq:izh_T_derivs}
\frac{\partial T_*}{\partial c}
= -\frac{1}{F(c)+I-u_*^+} <0,
\qquad
\frac{\partial T_*}{\partial u_*^+}
= \int_c^{30}\frac{\dd v}{\bigl(F(v)+I-u_*^+\bigr)^2} >0.
\end{equation}
Thus a deeper reset ($c$ more negative) increases the period because it lengthens the voltage excursion to the next spike, while a larger effective recovery level $u_*^+$ also increases the period because it reduces the net depolarizing drive.

\paragraph{Local linear compensation law.}
Fix a regular-spiking target with asymptotic period $T_0$ (equivalently rate $r_0=1/T_0$) and linearize around a reference parameter vector
\begin{equation}
\theta_0=(a_0,b_0,c_0,d_0).
\end{equation}
From Eq.~\eqref{eq:izh_flight_time}, keeping $T_*$ fixed gives
\begin{equation}
0 \approx T_c\,\delta c + T_u\,\delta u_*^+,
\end{equation}
where $T_c$ and $T_u$ denote the derivatives in Eq.~\eqref{eq:izh_T_derivs} evaluated at $\theta_0$. At fixed $r_0$, the cycle-average voltage can be written more explicitly as
\begin{equation}
\bar v(u,c)=\frac{1}{T(u,c)}\int_c^{30} \frac{v\,\dd v}{F(v)+I-u},
\end{equation}
so $\bar v$ depends implicitly on both $u_*^+$ and $c$. Therefore the correct first-order variation of Eq.~\eqref{eq:izh_slow_balance} is
\begin{equation}
\delta u_*^+ \approx \bar v\,\delta b + b_0\bigl(\bar v_c\,\delta c + \bar v_u\,\delta u_*^+\bigr) + \frac{r_0}{a_0}\,\delta d - \frac{r_0d_0}{a_0^2}\,\delta a,
\end{equation}
where $\bar v_c := \partial \bar v/\partial c$ and $\bar v_u := \partial \bar v/\partial u$ are evaluated at $\theta_0$. Rearranging,
\begin{equation}
\delta u_*^+ \approx \frac{1}{1-b_0\bar v_u}
\left(
\bar v\,\delta b + b_0\bar v_c\,\delta c + \frac{r_0}{a_0}\,\delta d - \frac{r_0d_0}{a_0^2}\,\delta a
\right),
\end{equation}
provided $1-b_0\bar v_u \neq 0$. Substituting into the period constraint yields
\begin{equation}
\label{eq:izh_linear_comp}
A_a\,\delta a + A_b\,\delta b + A_c\,\delta c + A_d\,\delta d \approx 0,
\end{equation}
with
\begin{equation}
A_a=-\frac{T_u}{1-b_0\bar v_u}\frac{r_0d_0}{a_0^2},
\qquad
A_b=\frac{T_u}{1-b_0\bar v_u}\bar v,
\qquad
A_c=T_c+\frac{T_u}{1-b_0\bar v_u} b_0\bar v_c,
\qquad
A_d=\frac{T_u}{1-b_0\bar v_u}\frac{r_0}{a_0}.
\end{equation}
Equation~\eqref{eq:izh_linear_comp} is the detailed mutual-compensation relation sought in the main text. It makes the following parameter roles explicit. Parameter $d$ is the spike-triggered adaptation kick. Parameter $a$ is the recovery timescale that controls how quickly the kick is forgotten. Parameter $c$ changes the reset depth and therefore the geometric distance from reset to threshold. Parameter $b$ changes the subthreshold coupling and therefore the baseline around which the recovery variable evolves. Mutual compensation occurs because these parameters all enter the period constraint through the single scalar effective recovery load $u_*^+$, with the prefactor $(1-b_0\bar v_u)^{-1}$ accounting for the implicit dependence of the orbit average on $u_*^+$ itself.

\paragraph{Why the regular-spiking manifold is almost linear.}
A regular-spiking target is not determined by the period alone. In the present manuscript it also includes low values of ISI-CV and Fano factor. In the spike-map description those regularity statistics are controlled primarily by the stability of the fixed point of $P$. Let
\begin{equation}
\lambda = P'(u_*^+)
\end{equation}
denote the contraction factor of the return map at the fixed point. When $|\lambda|\ll 1$, perturbations in post-spike recovery decay rapidly from spike to spike, the transient adaptation dies out quickly, and the spike train is nearly periodic. When $|\lambda|$ approaches $1$, the spike train becomes more sensitive to small perturbations and the response is less regular \citep{TouboulBrette2009,ShlizermanHolmes2012}. Therefore a regular-spiking target can be represented locally as two smooth constraints,
\begin{equation}
G_1(a,b,c,d)=T(a,b,c,d)-T_0=0,
\qquad
G_2(a,b,c,d)=R(a,b,c,d)-R_0=0,
\end{equation}
where $R$ is any smooth local surrogate of regularity, such as $\lambda$, ISI-CV, or Fano factor in the small-noise/finite-window regime. If the Jacobian of $(G_1,G_2)$ has rank two, the implicit-function theorem gives a two-dimensional smooth manifold in the four-dimensional $(a,b,c,d)$ space.

The reason it looks \emph{almost linear} is then straightforward. Near a reference point $\theta_0$,
\begin{equation}
G(\theta_0+\delta\theta)
= G(\theta_0) + DG(\theta_0)\,\delta\theta + \frac12 D^2G(\theta_0)[\delta\theta,\delta\theta] + \cdots.
\end{equation}
The tangent space of the viable set is the null space of $DG(\theta_0)$, so to first order the manifold is an affine two-plane. Curvature is a second-order effect coming from $D^2G$. In the regular-spiking examples of the present manuscript, the generated clouds occupy only a small neighborhood of a stable periodic branch, so $DG$ changes little across the cloud and the quadratic term remains small. The resulting viable set therefore appears as a small, nearly planar chart in the displayed $(a,b,c,d)$ projections. This is exactly what one should expect from the local expansion above.

By contrast, irregular or burst-adjacent targets probe regions where the return map changes more rapidly, the reset nonlinearity matters more strongly, and nearby targets can sit closer to bifurcation or chaos-producing mechanisms of the hybrid reset system \citep{ShlizermanHolmes2012,Nobukawa2018}. In those regimes the coefficients in Eq.~\eqref{eq:izh_linear_comp} vary appreciably across the set, the second-order terms are no longer negligible, and the viable manifolds appear visibly curved rather than approximately affine.

\paragraph{Practical interpretation for model users.}
The derivation above suggests a simple workflow for users of the Izhikevich model. If a regular-spiking target is too slow because adaptation is too strong, the first quantity to inspect is the combination $d/a$: decreasing $d$ or increasing $a$ usually weakens adaptation in nearly equivalent ways. If the neuron fires too slowly even with appropriate adaptation, changing $c$ alters the geometric travel distance to the next spike. Parameter $b$ is then used to retune the subthreshold coupling because it shifts the baseline motion of $u$ between spikes. The main point is that these parameters should not be tuned independently in the regular regime: they enter the spike-period and regularity constraints through coupled balance laws, which is why the corresponding viable set is thin and close to linear.


\subsection{Projected curvature in regular and irregular firing}
\label{sec:si_izh_negcurv}

The regular-spiking reduction above explains why some targets produce nearly affine ribbons. The irregular targets in Fig.~3C require a different local reduction: the spike-to-spike map is no longer a strongly contracting fixed point, but the post-spike recovery variable still organizes the timing statistics. Write
\begin{equation}
u_n^+ = \bar u^+ + x_n,
\end{equation}
where $\bar u^+$ is the invariant mean of the post-spike recovery variable for the target-conditioned irregular response and $x_n$ is the spike-to-spike fluctuation. Linearizing the exact return map~\eqref{eq:izh_return_map} around the occupied part of this invariant set gives the weakly nonlinear stochastic recurrence
\begin{equation}
x_{n+1} = \lambda x_n + \frac12 \gamma x_n^2 + \xi_n,
\label{eq:izh_irreg_map}
\end{equation}
where $\lambda$ is the local memory factor of the recovery map, $\gamma$ is the first nonlinear correction, and $\xi_n$ collects residual fluctuations coming from the hybrid reset geometry, finite-window heterogeneity, and higher-order state dependence. This return-map viewpoint is standard for hybrid spiking models even when the response is not purely periodic \citep{TouboulBrette2009,ShlizermanHolmes2012}. In the irregular regime of interest, $|\lambda|$ is not small; it is the near-critical regime in which adaptation memory is retained over several spikes and the quadratic term is no longer negligible.

Let $T_n=\mathcal T(u_n^+,c)$ denote the $n$th interspike interval. Expanding around $\bar u^+$ gives
\begin{equation}
T_n = T_0 + T_u x_n + \frac12 T_{uu} x_n^2 + O(x_n^3),
\label{eq:izh_irreg_Texp}
\end{equation}
where $T_0=\mathcal T(\bar u^+,c)$ and $T_u>0$ by the same flight-time argument used in Eq.~\eqref{eq:izh_T_derivs}. If the quadratic term in Eq.~\eqref{eq:izh_irreg_map} is not dominant over a local patch, then the stationary variance of the linearized fluctuation is
\begin{equation}
\mathrm{Var}(x_n)\approx \frac{\sigma_\xi^2}{1-\lambda^2},
\qquad
\sigma_\xi^2:=\mathrm{Var}(\xi_n).
\label{eq:izh_xvar}
\end{equation}
Consequently,
\begin{equation}
\mathrm{Var}(T_n)\approx T_u^2\,\mathrm{Var}(x_n),
\qquad
\mathrm{CV(ISI)}^2 \approx \frac{T_u^2\,\sigma_\xi^2}{T_0^2(1-\lambda^2)}.
\label{eq:izh_cv_formula}
\end{equation}
Thus CV(ISI) fixes the fluctuation scale of the return map after the mean period has been fixed.

The temporal-bin Fano statistic adds a different piece of information because it is sensitive not only to interval variance but also to serial memory across intervals. The following correlated-renewal calculation is a mechanistic surrogate for this dependence, rather than an identity for the finite 15-bin dispersion estimator used in the simulations. For a binned count window with $m$ effective intervals, the surrogate gives
\begin{equation}
F \approx \mathrm{CV(ISI)}^2
\left[
1+2\sum_{k=1}^{m-1}\left(1-\frac{k}{m}\right)\rho_k
\right],
\label{eq:izh_fano_general}
\end{equation}
where $\rho_k$ is the lag-$k$ serial correlation of the interval sequence. Under the linearized map~\eqref{eq:izh_irreg_map}, $\rho_k\approx \lambda^k$, so for windows that are not extremely short,
\begin{equation}
F \approx \mathrm{CV(ISI)}^2\,\mathcal K_m(\lambda),
\qquad
\mathcal K_m'(\lambda)>0.
\label{eq:izh_fano_lambda}
\end{equation}
In particular, for long effective windows one recovers the familiar form $\mathcal K_m(\lambda)\approx (1+\lambda)/(1-\lambda)$. Therefore the four observables split into two active composite goals. Up to finite-window edge effects, rate and mean ISI mainly fix the mean timescale $T_0$, whereas CV(ISI) and Fano factor fix the fluctuation scale and the memory factor $(\sigma_\xi,\lambda)$ of the spike-to-spike map.

To translate those two goals back to $(a,b,c,d)$, extend the averaged balance law~\eqref{eq:izh_slow_balance} from a periodic orbit to the invariant irregular regime and define the effective recovery load
\begin{equation}
L := \ell_b\,b + r_0\frac{d}{a},
\qquad
r_0:=T_0^{-1},
\label{eq:izh_irreg_load}
\end{equation}
where $\ell_b$ is the invariant-measure coefficient multiplying $b$ (equal to $\bar v$ in the slow-adaptation periodic reduction). The mean-timescale goal is then
\begin{equation}
T(L,c)=T_0,
\label{eq:izh_mean_goal}
\end{equation}
with $T_L>0$ and $T_c<0$. By the implicit-function theorem,
\begin{equation}
L=\phi(c),
\qquad
\phi'(c)=-\frac{T_c}{T_L}>0.
\label{eq:izh_phi_curve}
\end{equation}
This is the first composite goal: shallower reset must be offset by a larger recovery load, whereas deeper reset must be offset by a smaller one, in order to preserve the mean spiking timescale.

The irregularity goal is different because the same load also affects the memory factor of the return map. A generic local form consistent with Eq.~\eqref{eq:izh_return_map} is
\begin{equation}
\lambda(a,L,c)\approx e^{-aT_0}\,\Lambda(L,c),
\label{eq:izh_lambda_factor}
\end{equation}
where $e^{-aT_0}$ is the direct decay of the recovery variable between spikes and $\Lambda(L,c)$ collects the positive feedback created by reset geometry and adaptation-dependent spike timing. The exact form of $\Lambda$ is model- and operating-point-specific, but the curvature argument only requires that $\Lambda$ vary smoothly with the same composite load that already preserves the mean statistics. Near a reference irregular target,
\begin{equation}
\Lambda(L,c)\approx \Lambda_0+\nu_L(L-L_0)+\nu_c(c-c_0),
\label{eq:izh_lambda_linear}
\end{equation}
so the pair $\bigl(\mathrm{CV(ISI)},F\bigr)$, which fixes a target memory factor $\lambda_0$ and a target fluctuation scale, imposes the second composite goal
\begin{equation}
\lambda\bigl(a,\phi(c),c\bigr)=\lambda_0.
\end{equation}
Solving Eq.~\eqref{eq:izh_lambda_factor} for $a$ gives
\begin{equation}
a(c)=\frac{1}{T_0}\log \Lambda_{\mathrm{eff}}(c)-\frac{1}{T_0}\log \lambda_0,
\qquad
\Lambda_{\mathrm{eff}}(c):=\Lambda\bigl(\phi(c),c\bigr).
\label{eq:izh_a_curve}
\end{equation}
Differentiating twice,
\begin{equation}
a''(c)=\frac{1}{T_0}
\left[
\frac{\Lambda_{\mathrm{eff}}''(c)}{\Lambda_{\mathrm{eff}}(c)}
-
\frac{\bigl(\Lambda_{\mathrm{eff}}'(c)\bigr)^2}{\Lambda_{\mathrm{eff}}(c)^2}
\right].
\label{eq:izh_a_curve_second}
\end{equation}
For the local affine model~\eqref{eq:izh_lambda_linear},
\begin{equation}
a''(c)=
\frac{\nu_L\phi''(c)}{T_0\,\Lambda_{\mathrm{eff}}(c)}
-
\frac{\bigl(\nu_L\phi'(c)+\nu_c\bigr)^2}{T_0\,\Lambda_{\mathrm{eff}}(c)^2}.
\label{eq:izh_a_curve_second_affine}
\end{equation}
Whenever the mean-goal curve $\phi(c)$ is not itself strongly convex, the second term dominates and $a''(c)<0$. This is the mathematical source of the negative bending: the irregularity target forces $a$ to depend \emph{log-concavely} on the same load--reset combination that already preserves the mean statistics.

The full 2D viable surface then follows from Eq.~\eqref{eq:izh_irreg_load}:
\begin{equation}
d(b,c)=\frac{a(c)}{r_0}\bigl[\phi(c)-\ell_b b\bigr].
\label{eq:izh_d_surface}
\end{equation}
Hence the surface is ruled in the $b$ direction but inherits the concavity of $a(c)$ in the $c$ direction. Any 3D projection that retains one of $(a,d)$ together with $c$ and one load coordinate therefore preserves the same antagonistic bending. In practice this includes the $(a,b,c)$, $(a,c,d)$, and $(b,c,d)$ views used in Fig.~3C. Other projections can change the apparent sign or magnitude, so the empirical projection-wise screen is needed rather than inferring a projection-invariant curvature sign from this reduction.

The important conceptual point is that the four observables do \emph{not} impose four unrelated conditions. Rate and mean ISI ask for a fixed average interspike timescale. CV(ISI) and Fano factor ask for a fixed level of spike-to-spike irregularity and memory. Those two goals are not aligned in parameter space: preserving the mean period is accomplished by one compensation law in the composite load $L$, while preserving irregularity requires a second constraint on the near-critical slope $\lambda$. Their intersection is therefore curved rather than affine. This reduced theory does not prove negative curvature in every coordinate chart or for every irregular target, but it explains why negative curvature is expected in the irregular manifolds and why it appears in several of the 3D projections of the generated clouds.

\subsection{Systematic curvature screen across regular and irregular targets}
\label{sec:si_izh_curvature_screen}
We evaluated 25 regular-spiking target manifolds spanning firing rate and 200 irregular target manifolds selected for broad coverage of firing rate, CV(ISI), and Fano factor. Each target contributed up to 2{,}200 generated points. Curvature was evaluated in the four three-parameter projections $(a,b,c)$, $(a,b,d)$, $(a,c,d)$, and $(b,c,d)$ using the estimator in Section~\ref{sec:si_geometry_estimators}. Table~\ref{tab:si_izh_curvature} reports the fraction of targets with negative median curvature and the median, across targets, of the target-level median signed-log curvature.

\begin{table}[htbp]
\centering
\small
\caption{Two-dimensional signed-curvature summary in three-parameter projections of Izhikevich viable sets.}
\label{tab:si_izh_curvature}
\begin{tabular}{llrrr}
\toprule
Target class & Projection & Targets / records & Fraction with negative median & Median of target medians \\
\midrule
Regular & $(a,b,c)$ & 25 & 0.800 & $-0.0312$ \\
Regular & $(a,b,d)$ & 25 & 0.280 & $\phantom{-}0.0243$ \\
Regular & $(a,c,d)$ & 25 & 0.720 & $-0.1222$ \\
Regular & $(b,c,d)$ & 25 & 1.000 & $-0.0290$ \\
Regular & all projections & 100 & 0.700 & $-0.0256$ \\
\midrule
Irregular & $(a,b,c)$ & 200 & 0.945 & $-0.1334$ \\
Irregular & $(a,b,d)$ & 200 & 0.935 & $-0.1317$ \\
Irregular & $(a,c,d)$ & 200 & 0.925 & $-0.3372$ \\
Irregular & $(b,c,d)$ & 200 & 0.790 & $-0.0878$ \\
Irregular & all projections & 800 & 0.8988 & $-0.1540$ \\
\bottomrule
\end{tabular}
\end{table}

For a single projection, the third column counts distinct targets. In the ``all projections'' rows it counts target--projection records (25 targets $\times$ 4 projections and 200 targets $\times$ 4 projections); inferential analyses across pooled projections therefore cluster by target.

The regular sweep is weakly curved and rate structured: its pooled median lies close to zero, while individual projections can have nonzero curvature. The irregular targets show a stronger and more projection-consistent negative shift. 
We therefore interpret negative signed-log curvature as a regime-level tendency across the four three-parameter projections.

To evaluate irregularity continuously without imposing a regular--irregular cutoff, we extended the analysis to $n=3{,}000$ targets selected for balanced coverage of CV(ISI). For each target and projection, we summarized curvature by the target-median signed-log value and examined its relationship with target CV(ISI). In the $(a,c,d)$ projection, the slope was $\beta=-0.1983$ (95\% CI, $-0.2052$ to $-0.1913$; $R^2=0.4981$). Main-text Fig.~3E shows the $(a,c,d)$ projection, whereas Fig.~\ref{fig:si-izh-curvature-cv} shows all four projections, with points colored by target Fano factor.

\setcounter{figure}{0}
\renewcommand{\thefigure}{S\arabic{figure}}

\begin{figure}[htbp]
\centering
\includegraphics[width=\textwidth]{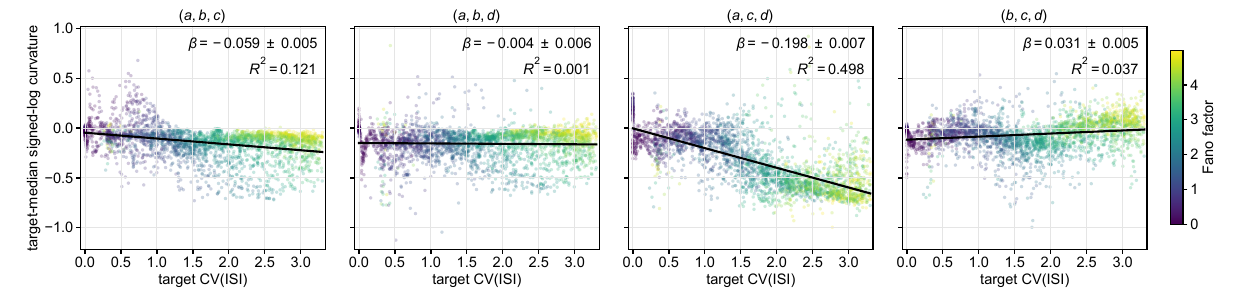}
\caption{Curvature--irregularity relation across three-parameter projections of Izhikevich viable sets. Target-median two-dimensional signed-log curvature versus target CV(ISI) in the $(a,b,c)$, $(a,b,d)$, $(a,c,d)$, and $(b,c,d)$ projections, with points colored by target Fano factor. Black lines and shading show projection-specific linear fits and 95\% target-bootstrap confidence intervals across $n=3{,}000$ CV-balanced targets. Each panel reports the slope $\beta$ with the half-width of its 95\% confidence interval and $R^2$.}
\label{fig:si-izh-curvature-cv}
\end{figure}

\section{dsODE details}
\label{sec:si_dsode}
\subsection{Model origin and simulation protocol}
The dsODE model is taken from \citet{Chang2025dsODE}, who derive a deterministic ODE description of finite homogeneous spiking networks from a coarse-grained Markov approximation. In that theory, the state variables include counts of neurons in voltage bins together with statistics of pending synaptic kicks. The model captures fixed points, oscillations, metastability, and finite-size effects in excitatory--inhibitory networks while remaining fast enough for systematic parameter exploration.

For the present manuscript, dsODE serves as a reduced but biologically meaningful testbed for viable-parameter-manifold learning. All dsODE simulations use $\Delta t=0.1$ ms, population sizes $n_E=300$ and $n_I=100$, total duration 5 s, and a 1 s burn-in period before feature extraction. In the 4D, 8D, and 10D main-text experiments the connection probability is fixed at $P=0.8$. The external drive was fixed at $J_E^{\mathrm{ext}}=J_I^{\mathrm{ext}}=7$ in the 4D, 8D, and 12D experiments, whereas $J_E^{\mathrm{ext}}$ and $J_I^{\mathrm{ext}}$ were varied independently in the 10D experiment. Excitatory and inhibitory mean rates and standard deviations are computed over the post-burn-in window and divided by 200 Hz before being used as target features. Parameters are linearly transformed to $[-1,1]$ for diffusion training.
Writing $t_b=1$ s and $t_f=5$ s, the implemented features are
\begin{equation}
\mu_Q=\frac{1}{t_f-t_b}\int_{t_b}^{t_f}r_Q(t)\,\dd t,
\qquad
\sigma_Q^2=\frac{1}{t_f-t_b}\int_{t_b}^{t_f}\bigl(r_Q(t)-\mu_Q\bigr)^2\,\dd t,
\quad Q\in\{E,I\}.
\end{equation}
\paragraph{dsODE implementation and initialization.}
For an equation-level specification, the dsODE is given by Eqs.~(15), (17), and (18) of \citet{Chang2025dsODE}, together with the recurrent-drive moments $u^{QR}$ and $D^{QR}$; the associated fluxes are defined in their Eq.~(16), and the efficient voltage-bin update is described in their Sec.~3.2.2 and Appendix~A.3. We used this direct voltage-bin update rather than a general-purpose ODE solver. Voltage was represented by 14 bins of width 10 spanning $[-40,100]$ in model voltage units, with firing threshold $M=100$. The inhibitory reversal level was $V=-66$ and was used to scale inhibitory synaptic effects according to postsynaptic voltage. At $t=0$, all excitatory and inhibitory population mass was placed in the zero-voltage bin; the refractory populations and pending-kick moments were zero, and the stored aggregate voltage was initialized to $10^{-4}$, an effectively zero value. Bin transfers were evaluated using the uniform within-bin approximation and Gaussian current convolution. Pending-kick moments were advanced by forward Euler.
\paragraph{Feature extraction and simulation handling.}
The means and standard deviations in the preceding display were computed directly from the discrete post-burn-in rate samples; the integrals provide the corresponding continuum notation, and no temporal interpolation or separate quadrature was used. When included, dominant frequency was the largest nonzero real-FFT component of the post-burn-in excitatory-rate trace at the native $10^{-4}$-s sampling interval, without sub-bin interpolation. Traces with sufficiently weak rate fluctuations to be classified as nonoscillatory, operationally defined by an excitatory-rate standard deviation of at most $5$ Hz, were assigned a dominant frequency of zero. 
\paragraph{Libraries and training.}
The 4D, 8D, 10D, and 12D training libraries contained 14{,}000, 10{,}146, 10{,}535, and 9{,}535 parameter--feature pairs, respectively. Targets were selected from the corresponding evaluation libraries. The 4D score models were trained for 500 epochs for the two-feature conditions and 1{,}000 epochs for the four-feature condition; the 8D, 10D, and 12D models were each trained for 5{,}000 epochs.
\subsection{Parameterizations and box constraints}
The manuscript uses several nested dsODE parameterizations.
\begin{enumerate}[leftmargin=*]
\item \textbf{4D synaptic strengths:}
\begin{equation}
(S_{EE},S_{IE},S_{EI},S_{II})
\in
[0.60,1.30]\times[0.90,1.60]\times[2.20,3.20]\times[1.10,3.00].
\end{equation}
The time constants were fixed at $(\tau_{EE},\tau_{IE},\tau_I,\tau_{\mathrm{ref}})=(1.4,1.2,4.5,4.0)$ ms.
\item \textbf{8D synaptic-plus-timescale variables:}
\begin{align}
&(\tau_{EE},\tau_{IE},\tau_I,\tau_{\mathrm{ref}},S_{EE},S_{IE},S_{EI},S_{II}) \nonumber
\\
&\in
[0.5,4.0]^2\times[3.0,6.0]\times[0.0,10.0] \nonumber\\
&\qquad\times[0.40,1.60]\times[0.60,1.80]\times[2.00,3.50]\times[1.00,3.00].
\end{align}
\item \textbf{10D extension with external drive:}
\begin{align}
&(J_E^{\mathrm{ext}},J_I^{\mathrm{ext}},\tau_{EE},\tau_{IE},\tau_I,\tau_{\mathrm{ref}},S_{EE},S_{IE},S_{EI},S_{II})
\in \nonumber 
\\
&
[0,20]\times[0,20]\times[0.5,4.0]\times[0.5,4.0]\times[3.0,6.0]\times[0.0,10.0] \nonumber 
\\
&\times[0.40,1.60]\times[0.60,1.80]\times[2.00,3.50]\times[1.00,3.00].
\end{align}
\item \textbf{12D extension with variable connection probabilities:}
The 12D parameter vector is ordered as
\begin{equation}
(\tau_{EE},\tau_{IE},\tau_I,\tau_{\mathrm{ref}},
P_{EE},P_{IE},P_{EI},P_{II},
S_{EE},S_{IE},S_{EI},S_{II}),
\end{equation}
with the componentwise parameter box
\begin{equation}
\begin{aligned}
&[0.5,3.5]\times[0.5,3.5]\times[3.0,6.0]\times[0.0,10.0]\\
&\quad\times[0.5,1.0]\times[0.5,1.0]\times[0.5,1.0]\times[0.5,1.0]\\
&\quad\times[0.60,1.30]\times[0.90,1.60]\times[2.30,3.10]\times[1.10,2.60].
\end{aligned}
\end{equation}
\end{enumerate}

\subsection{Target features and 4D background-regime manifolds}
\label{sec:si_dsode_4d}
Uncertainty and context dependence in recurrent coupling strengths, particularly for inhibitory efficacy, motivate varying all four recurrent strengths rather than holding the inhibitory channels fixed. Layer-4 measurements show substantial dependence on pathway, interneuron class, and short-term synaptic dynamics \citep{Stratford1996,Beierlein2003}. Earlier network calculations also show strong sensitivity of background firing-rate viability to small coupling perturbations. In the spontaneous-activity analysis of \citet[Table~1]{Xiao2021}, each coupling was perturbed separately around a reference parameter set with $(f_E^0,f_I^0)=(3.85,13.32)$ Hz, and viability was defined by $f_E/f_E^0\in(2/3,4/3)$ and $f_I/f_E\in(3,4.5)$. Some signed perturbations reported to the nearest percent were as small as $1\%$, motivating the joint variation of all four recurrent coupling strengths in our dsODE analysis.

Three dsODE target families appear in the manuscript:
\begin{enumerate}[leftmargin=*]
\item \textbf{4D activity moments:}
\begin{equation}
(\mu_E,\mu_I,\sigma_E,\sigma_I),
\end{equation}
where $\mu$ and $\sigma$ denote the mean and standard deviation of the E and I firing rates.
\item \textbf{2D or 4D subsets:}
mean-only or std-only guides for E and I, used to compare how different observables contract the same parameter space.
\item \textbf{7D augmented targets:}
\begin{equation}
(\mu_E,\mu_I,\sigma_E,\sigma_I,\mathrm{skew}_E,\mathrm{skew}_I,f_{\mathrm{dom}}),
\end{equation}
where $f_{\mathrm{dom}}$ is the dominant oscillation frequency.
\end{enumerate}

In the background regime inherited from \citet{Xiao2021}, we use $J_{\mathrm{ex}}=0.7$ in our dsODE setup, where excitatory and inhibitory mean rates remain tightly linked and a nominal two-rate guide is dominated by one constraint at coarse resolution in the 4D synaptic box. The corresponding mean-conditioned clouds have global participation ratios of $2.02$--$3.66$ in physical $S$ coordinates and cloud-median local participation ratios of $2.78$--$3.23$ in diffusion-scaled coordinates. Together with the fitted-rank sensitivity analysis, these results support an approximately codimension-1 organization of viable family in four parameters. In the more strongly driven regime emphasized in the main text, the E and I means decouple more strongly and mean-only conditioning becomes thinner. In the displayed examples, adding component-specific standard deviations separates nearby beat families that mean rates alone merge. This is why the 4D dsODE figures compare mean-only, std-only, and joint constraints. Further details are given in Sec.~\ref{sec:si_dsode_branch_classification}.

In the inhibition-plane coordinates
\begin{equation}
x_S=\frac{S_{IE}}{S_{II}},\qquad y_S=\frac{S_{EI}}{S_{EE}},
\label{eq:dsode-raw-inhibition-plane}
\end{equation}
the earlier balance organization is indexed by the product $x_Sy_S$. The generated clouds provide a two-dimensional view of relative recurrent couplings. The dimension diagnostics are computed in the full four-dimensional parameter space.

The background model was trained on approximately 13{,}000 examples using the same four synaptic-strength ranges specified above for the 8D setup. We selected 100 nonzero example targets with both mean rates at most 15 Hz, spanning $\mu_E=0.51$--$14.72$ Hz and $\mu_I=3.12$--$13.27$ Hz, with 5{,}000 generated parameter proposals per target.

The rank analysis supports an approximately codimension-1 organization for most targets at the coarser threshold, while a weaker second constraint remains detectable at the stricter threshold. Table~\ref{tab:si_dsode_background_summary} summarizes the rank sensitivity and participation-ratio distribution. 

\begin{table}[htbp]
\centering
\small
\caption{Background-regime proposal geometry ($J_{\mathrm{ex}}=0.7$) for 100 low-rate example targets. PR entries summarize medians and their between-target spread; local PR and $s_2/s_1$ are first summarized by their median within each cloud. Rank/nullity entries give within-cloud modes and target counts.}
\label{tab:si_dsode_background_summary}
\begin{tabular}{p{0.43\linewidth}p{0.49\linewidth}}
\toprule
Quantity & Result \\
\midrule
Global PR (physical $S$) & 2.53 (IQR 2.35--2.69) \\
Local PR (diffusion-scaled $S$) & 2.92 (IQR 2.86--2.97) \\
Relative second singular value $s_2/s_1$ & 0.065 (IQR 0.060--0.092) \\
Rank/nullity at $0.07s_{\max}$ & $1/3$ (55/100 targets); $2/2$ (45/100 targets) \\
Rank/nullity at $0.10s_{\max}$ & $1/3$ (77/100 targets); $2/2$ (23/100 targets) \\
\bottomrule
\end{tabular}
\end{table}


\subsection{Information in $(\mu_E,\mu_I,\sigma_E,\sigma_I)$ for periodic beat dynamics}
\label{sec:si_dsode_moments}

This subsection explains why the manuscript uses mean and standard deviation of the E and I rates as the baseline four-feature target in dsODE. The argument has three layers: an exact identity for any periodic trajectory, a pulse-train calculation for the beat-like trajectories seen in dsODE, and a local branch-separation condition whose empirical scope is evaluated by re-simulation.

\paragraph{Exact identity for any periodic trajectory.}
Let $r_Q(t)$, $Q\in\{E,I\}$, be a $T$-periodic firing-rate trajectory and define
\begin{equation}
\mu_Q=\frac{1}{T}\int_0^T r_Q(t)\,\dd t,
\qquad
\sigma_Q^2=\frac{1}{T}\int_0^T \bigl(r_Q(t)-\mu_Q\bigr)^2\,\dd t.
\end{equation}
Whenever $\sigma_Q>0$, the trajectory admits the exact normalization
\begin{equation}
r_Q(t)=\mu_Q+\sigma_Q\,q_Q(t/T),
\label{eq:dsode_periodic_normalization}
\end{equation}
where $q_Q$ is $1$-periodic and satisfies
\begin{equation}
\int_0^1 q_Q(s)\,\dd s=0,
\qquad
\int_0^1 q_Q(s)^2\,\dd s=1.
\end{equation}
Thus fixing $(\mu_Q,\sigma_Q)$ removes the affine degrees of freedom of the periodic signal: the DC offset and the overall oscillation scale are fixed exactly, and only the normalized shape $q_Q$ remains free. In Fourier form,
\begin{equation}
r_Q(t)=\mu_Q+\sum_{n\ge 1} a_{Q,n}\cos\frac{2\pi n t}{T}+b_{Q,n}\sin\frac{2\pi n t}{T},
\end{equation}
so Parseval gives
\begin{equation}
\sigma_Q^2=\frac12\sum_{n\ge 1}\bigl(a_{Q,n}^2+b_{Q,n}^2\bigr).
\label{eq:dsode_sigma_parseval}
\end{equation}
Equation~\eqref{eq:dsode_sigma_parseval} is exact: the standard deviation records the total non-DC oscillatory power, but not how that power is distributed across harmonics.

\paragraph{Rigorous pulse-train calculation for beat-like periodic trajectories.}
The periodic dsODE trajectories in the relevant regime are pulse-like rather than sinusoidal, so a more specific approximation is natural. Assume that within one beat family the population-$Q$ rate is well approximated by a non-overlapping pulse train on a common macro-period $T$,
\begin{equation}
r_Q(t)=b_Q+A_Q\sum_{j=1}^{k_Q} p_Q\!\left(\frac{t-t_{Q,j}}{w_Q}\right),
\label{eq:dsode_pulse_train}
\end{equation}
where $b_Q$ is the baseline rate, $A_Q$ is the pulse amplitude, $k_Q$ is the number of pulses per macro-period, $w_Q$ is the pulse width, and $p_Q$ is supported on $[0,1]$. Let
\begin{equation}
m_{1,Q}=\int_0^1 p_Q(s)\,\dd s,
\qquad
m_{2,Q}=\int_0^1 p_Q(s)^2\,\dd s,
\qquad
\delta_Q=\frac{k_Q w_Q}{T}.
\end{equation}
Because the pulses do not overlap, direct integration of Eq.~\eqref{eq:dsode_pulse_train} gives
\begin{equation}
\mu_Q=b_Q+A_Q\,\delta_Q m_{1,Q},
\label{eq:dsode_pulse_mu}
\end{equation}
and
\begin{equation}
\sigma_Q^2=A_Q^2\left(\delta_Q m_{2,Q}-\delta_Q^2 m_{1,Q}^2\right).
\label{eq:dsode_pulse_sigma}
\end{equation}
Therefore
\begin{equation}
\left(\frac{\sigma_Q}{\mu_Q-b_Q}\right)^2
=
\frac{m_{2,Q}}{\delta_Q m_{1,Q}^2}-1.
\label{eq:dsode_ratio_delta}
\end{equation}
Equation~\eqref{eq:dsode_ratio_delta} is rigorous under the pulse-train ansatz. For fixed baseline and pulse template, the ratio $\sigma_Q/(\mu_Q-b_Q)$ is a strictly decreasing function of the duty factor $\delta_Q$. Hence $(\mu_Q,\sigma_Q)$ determines the amplitude $A_Q$ and strongly constrains the duty factor. Since $\delta_Q=k_Qw_Q/T$, the beat count $k_Q$ enters explicitly.

This gives a precise conditional separation statement. If two beat families have disjoint duty-factor intervals,
\begin{equation}
\delta_Q\in \Delta_{Q,1}
\qquad\text{for one-beat,}
\qquad
\delta_Q\in \Delta_{Q,2}
\qquad\text{for two-beat,}
\end{equation}
with $\Delta_{Q,1}\cap\Delta_{Q,2}=\varnothing$, then the monotonicity of Eq.~\eqref{eq:dsode_ratio_delta} implies that their admissible intervals of $\sigma_Q/(\mu_Q-b_Q)$ are also disjoint. In that case, the pair $(\mu_Q,\sigma_Q)$ already separates one-beat from two-beat dynamics for that population. Using both populations only strengthens the constraint.

\paragraph{Local branch-separation condition.}
Under the pulse ansatz, the preceding separation statement is exact. On a smooth branch of periodic dsODE solutions with a fixed beat pattern, the baselines, pulse widths, and normalized pulse templates vary smoothly with parameters. The map
\begin{equation}
\thetaVec \mapsto (\mu_E,\mu_I,\sigma_E,\sigma_I)
\end{equation}
is therefore smooth on that branch. Persistent overlap of two branches over an open region of four-feature space would require four equalities between their smooth branch data, a nongeneric configuration unless imposed by symmetry or transition structure. This local transversality argument motivates the four-moment guide; empirical branch classification tests its adequacy in the sampled parameter box.

\paragraph{Information retained by the four moments.}
Equation~\eqref{eq:dsode_periodic_normalization} fixes the DC level and total marginal oscillatory power but leaves the normalized waveform free. Equation~\eqref{eq:dsode_sigma_parseval} makes the unresolved degrees of freedom explicit: frequency, E--I phase, harmonic allocation, and waveform can vary at fixed $(\mu_E,\mu_I,\sigma_E,\sigma_I)$. The empirical claim is that these four features separate the displayed one- and two-beat families within the explored box; other periodic branches can intersect in four-feature space.

\subsection{Beat classification and four-feature ablation}
\label{sec:si_dsode_branch_classification}
\paragraph{Beat classification.} Beat labels were assigned from the excitatory dsODE rate trace. Peak classification used the trace after 2~s to reduce residual transient effects and obtain a stable characterization. Peaks were detected with a minimum separation of 50 samples (5~ms at $\Delta t=0.1$~ms); the first and last detected peaks were discarded, and retained peaks had to exceed 0.1 in the raw dsODE population-rate representation to exclude near-zero peaks. The differences between successive retained peak heights were sorted and grouped with a group width of 20 in the same raw units. A trajectory was labeled low-variability (no-beat) when its post-burn-in E-rate standard deviation was at most 5~Hz or fewer than four peaks remained. Otherwise, one, two, or at least three peak-difference groups defined one-beat, two-beat, or multi-beat activity, respectively.
\paragraph{Nested-guide dimension ablation.}
Using the physical 4D synaptic-strength coordinates of main-text Figs.~5A and~7A, we compared E-mean, E/I-mean, and joint E/I-mean/standard-deviation guidance.  For ten example targets (5{,}000 proposals per target and guide), the mean global covariance participation ratio was 1.751, 1.686, and 1.582, respectively. Adding the I-rate mean and subsequently the two standard deviations reduced the covariance trace in all ten targets at each step, with median reductions of 45.8\% and 83.4\%, respectively. However, adding the standard deviations increased participation ratio in three targets. For example, at $(\mu_E,\mu_I)\simeq(35,64)$~Hz, participation ratio increased from 1.060 to 2.746 while the covariance trace decreased from 0.333 to 0.0128. Thus, adding conditioning features consistently contracted these proposal clouds and lowered participation ratio on average, but did not lower it uniformly across targets, as a smaller cloud can have a less anisotropic covariance spectrum.

In a normalized-coordinate analysis of 100 matched targets in the 8D parameter space (5{,}000 proposals per target and guide), the mean global participation ratios were 6.743, 5.739, and 4.518 for the one-, two-, and four-feature guides, respectively. Adding the I-rate mean reduced the covariance trace in all 100 targets, with a median reduction of 15.1\%; adding the two standard deviations reduced it in 99 of 100 targets, with a median reduction of 23.9\%. As the generated clouds retained variability in the 8D parameter space, adding conditioning features progressively constrained the parameter distributions, reducing their average PR dimension and contracting the clouds.

\subsection{Mechanistic explanation of $S$--$\tau$ compensation under fixed rate means and standard deviations}
\label{sec:si_dsode_stau}

To interpret the 8D viable manifolds, it is helpful to isolate one recurrent channel $Q\leftarrow R$. The full dsODE contains pending-kick variables rather than a single scalar synaptic filter, but near a fixed target and after projection onto one channel, a minimal local description consistent with that structure is
\begin{equation}
\tau_{QR}\dot h_{QR}(t)=-h_{QR}(t)+\kappa_{QR}P_{QR}\tau_{QR}r_R(t)+\xi_{QR}(t),
\qquad
I_{QR}(t)=\frac{S_{QR}}{\tau_{QR}}h_{QR}(t),
\label{eq:dsode_filter_model}
\end{equation}
where $r_R$ is the presynaptic population rate, $h_{QR}$ is a filtered drive or pending-kick variable, $P_{QR}$ is the connection probability, and $\xi_{QR}$ collects finite-size fluctuations or unresolved state dependence. Equation~\eqref{eq:dsode_filter_model} is a leading-order local channel reduction of the pending-kick dynamics; it isolates the gain role of $S_{QR}$ and the filtering role of $\tau_{QR}$.

Write $r_R(t)=\bar r_R+\delta r_R(t)$ and average Eq.~\eqref{eq:dsode_filter_model}. For a stationary target,
\begin{equation}
\bar h_{QR}=\kappa_{QR}P_{QR}\tau_{QR}\bar r_R,
\qquad
\bar I_{QR}=\kappa_{QR}P_{QR}S_{QR}\bar r_R.
\label{eq:dsode_mean_current}
\end{equation}
Thus, to first order, the \emph{mean} recurrent drive depends on the effective gain $P_{QR}S_{QR}$ and is only weakly sensitive to $\tau_{QR}$ separately. This explains why the original 4D mean-rate-conditioned manifolds are already organized in ratios of the synaptic strengths alone: the DC balance constraints primarily see products of connection probability and synaptic gain.

The \emph{standard deviation} of the rate or current depends differently on $\tau_{QR}$ because $\tau_{QR}$ filters temporal fluctuations. Taking the Fourier transform of Eq.~\eqref{eq:dsode_filter_model} gives
\begin{equation}
\widehat{\delta I}_{QR}(\omega)=\frac{\kappa_{QR}P_{QR}S_{QR}}{1+i\omega\tau_{QR}}\,\widehat{\delta r}_R(\omega)+\widehat{\zeta}_{QR}(\omega),
\label{eq:dsode_fourier_current}
\end{equation}
so the frequency response is low-pass with cutoff $1/\tau_{QR}$. Two useful asymptotic regimes follow.

\paragraph{Shot-noise-dominated fluctuations.}
If presynaptic events arrive approximately as a Poisson process with rate proportional to $P_{QR}\bar r_R$, then the contribution of one event can be represented by the exponential kernel
\begin{equation}
k_{QR}(t)=\frac{S_{QR}}{\tau_{QR}}e^{-t/\tau_{QR}}\mathbf 1_{t>0}.
\end{equation}
The corresponding variance contribution is
\begin{equation}
\mathrm{Var}(I_{QR})\propto P_{QR}\bar r_R\int_0^{\infty}k_{QR}(t)^2\,\dd t
= P_{QR}\bar r_R\frac{S_{QR}^2}{2\tau_{QR}}.
\label{eq:dsode_shotnoise_var}
\end{equation}
Hence fixed fluctuation level implies the compensation law
\begin{equation}
\frac{S_{QR}^2}{\tau_{QR}}\approx \mathrm{const},
\qquad\text{equivalently}\qquad
S_{QR}\propto \tau_{QR}^{1/2}.
\label{eq:dsode_shotnoise_comp}
\end{equation}

\paragraph{Beat-dominated fluctuations.}
If the temporal standard deviation is dominated instead by a recurrent oscillatory mode with dominant frequency $\omega_*$, then Eq.~\eqref{eq:dsode_fourier_current} implies the amplitude law
\begin{equation}
A_{QR}(\omega_*)\propto \frac{P_{QR}S_{QR}}{\sqrt{1+\omega_*^2\tau_{QR}^2}}.
\label{eq:dsode_beat_gain}
\end{equation}
Holding this amplitude fixed gives
\begin{equation}
P_{QR}S_{QR}\approx C_{QR}\sqrt{1+\omega_*^2\tau_{QR}^2}.
\label{eq:dsode_beat_comp}
\end{equation}
When $\omega_*\tau_{QR}\ll 1$, the timescale has only a weak effect and raw $S$-ratios remain informative. When $\omega_*\tau_{QR}\gg 1$, however, Eq.~\eqref{eq:dsode_beat_comp} reduces to
\begin{equation}
\frac{P_{QR}S_{QR}}{\tau_{QR}}\approx \mathrm{const},
\label{eq:dsode_highfreq_comp}
\end{equation}
which is exactly the regime in which a $S/\tau$ visualization becomes natural. Equations~\eqref{eq:dsode_shotnoise_comp} and \eqref{eq:dsode_highfreq_comp} therefore explain why positive $S$--$\tau$ correlations are expected under fixed mean/std targets, while also clarifying why the slope is target-dependent: shot-noise-like branches favor a square-root law, whereas beat-dominated branches favor an approximately linear $S\propto \tau$ or $S/\tau\approx\mathrm{const}$ relation.

\paragraph{Finite-frequency loop ratio.}
Let
\begin{equation}
K_{QR}(i\omega)=
\chi_Q(i\omega)\frac{P_{QR}S_{QR}}{1+i\omega\tau_{QR}}
\end{equation}
denote a local channel gain, where $\chi_Q$ is the postsynaptic population susceptibility. The magnitude ratio of the reciprocal E--I loop to the two within-population channels is
\begin{equation}
\mathcal R(\omega)=
\frac{|K_{IE}(i\omega)K_{EI}(i\omega)|}
{|K_{EE}(i\omega)K_{II}(i\omega)|}.
\label{eq:dsode_finite_frequency_ratio}
\end{equation}
When the connection probabilities and susceptibility factors are fixed or cancel to leading order and all relevant $\omega\tau_{QR}\gg1$, Eq.~\eqref{eq:dsode_finite_frequency_ratio} reduces to
\begin{equation}
\mathcal R_\tau=
\frac{(S_{IE}/\tau_{IE})(S_{EI}/\tau_I)}
{(S_{EE}/\tau_{EE})(S_{II}/\tau_I)}.
\label{eq:dsode_Rtau}
\end{equation}
For finite $\omega\tau$, the operative coordinate is $\mathcal R(\omega_*)$, evaluated at the target's dominant angular frequency $\omega_*$, rather than its high-frequency limit. 

\paragraph{Finite-window averaging and frequency blindness of moment targets.}
For an absolutely summable Fourier representation
\begin{equation}
r_Q(t)=\mu_Q+\sum_{k\ge1}A_{Qk}\cos(k\omega_*t+\phi_{Qk}),
\end{equation}
the mean over a window of length $T_w$ satisfies
\begin{equation}
\left|\widehat\mu_Q(T_w)-\mu_Q\right|
\le \frac{2}{\omega_*T_w}\sum_{k\ge1}\frac{|A_{Qk}|}{k}.
\label{eq:dsode_window_mean_bound}
\end{equation}
Applying the same integral bound to products of Fourier modes gives
\begin{equation}
\left|\widehat\sigma_Q^2(T_w)-\frac12\sum_{k\ge1}A_{Qk}^2\right|
\le \frac{C_Q}{\omega_*T_w},
\label{eq:dsode_window_variance_bound}
\end{equation}
where $C_Q$ is an explicit quadratic sum of Fourier amplitudes when the corresponding series is absolutely summable. Thus endpoint and initial-phase effects decay with the number of observed cycles. At higher frequency, the four marginal moments become less sensitive to phase and timing changes, even though frequency, E--I phase, and harmonic allocation remain dynamically distinct. This yields a mechanism for a positive frequency--width association: parameter changes that alter temporal organization while preserving baseline and total marginal power more readily remain inside a finite moment tolerance. It predicts preferential contraction of high-frequency clouds when frequency, E--I phase, cross-spectrum, or harmonic-power targets are added.

This mechanism can be stated locally in the plotted plane. Let $G=(\mu_E,\mu_I,\sigma_E,\sigma_I)$ and let $n_\tau$ denote a transverse parameter direction whose projection is normal to the ribbon in $(x_\tau,y_\tau)$. For a small displacement $s n_\tau$,
\begin{equation}
\|G(\thetaVec+s n_\tau)-G(\thetaVec)\|
=|s|\,\|DG(\thetaVec)n_\tau\|+o(|s|).
\label{eq:dsode_transverse_sensitivity}
\end{equation}
The feature tolerance therefore permits a transverse displacement of order $\varepsilon/\|DG(\thetaVec)n_\tau\|$. When derivatives of the Fourier amplitudes and phases remain bounded, Eqs.~\eqref{eq:dsode_window_mean_bound}--\eqref{eq:dsode_window_variance_bound} show that the endpoint- and phase-dependent contribution to this transverse sensitivity decays with the number of observed cycles. The resulting reduction in moment sensitivity widens the finite-tolerance ribbon even as the estimates themselves become less dependent on the initial phase.

\subsection{Why the $\tau$-adjusted plane yields thickened one-dimensional sets only for some targets}

Define the effective-gain coordinates
\begin{equation}
g_{QR}=\frac{S_{QR}}{\tau_{QR}},
\qquad
x=\frac{g_{IE}}{g_{II}},
\qquad
y=\frac{g_{EI}}{g_{EE}}.
\label{eq:dsode_adjusted_coords}
\end{equation}
Let $\Mset\subset\R^8$ be one target-conditioned viable set in the 8D parameter space, and let $\pi:\Mset\to\R^2$ be the projection $\pi(\thetaVec)=(x,y)$. If the chosen mean/std target depends mainly on one active combination of $(x,y)$, then the differential $D\pi$ restricted to $T_{\thetaVec}\Mset$ has local rank close to one. In that case the image $\pi(\Mset)$ is locally curve-like.

The projected ribbon records the active combination captured by $(x,y)$ and the unresolved degrees of freedom left by the four moment targets. In particular, a large $\mathcal R_\tau=xy$ places more weight on the reciprocal E--I loop relative to the within-population loops. Compensating changes in cross-loop gains and timescales can then redistribute dominant frequency, E--I phase, harmonic power, waveform, or beat timing while preserving marginal means and variances. More temporally distinct solutions can therefore occupy the same four-feature tolerance tube, producing a wider projected ribbon. Absolute timescales, refractory time, drive, and branch identity contribute additional residual width.


\subsection{Width calculation for dsODE inhibition-plane projections}
\label{sec:si_dsode_inhibition_plane_width}
The raw inhibition-plane projection was constructed from the four synaptic-strength components using the coordinates defined in Eq.~\eqref{eq:dsode-raw-inhibition-plane}. The same construction was used for the four-parameter $S$-only clouds and for the raw $S$ projection of the eight-dimensional clouds. We separately min--max normalized $x_S$ and $y_S$ using the fixed bounds $[0.20,1.80]$ and $[1.25,8.75]$, respectively, as determined from the prescribed parameter ranges. For each target cloud, 640 reproducibly selected generated samples served as patch centers, and the $k_n=64$ nearest neighbors were identified in the normalized plane. After centering each patch, let $\lambda_1\ge\lambda_2$ denote the eigenvalues of the resulting two-dimensional scatter matrix. The target-level transverse width was
\begin{equation}
\sigma_\perp=\operatorname{median}_{\text{patches}}
\sqrt{\frac{\lambda_2}{k_n-1}},
\qquad k_n=64,
\label{eq:dsode-transverse-width}
\end{equation}
which is the median local standard deviation along the minor principal axis of the normalized plane.

For the $\tau$-adjusted projection, the coordinates were
\begin{equation}
x_\tau=\frac{S^{\mathrm{IE}}/\tau^{\mathrm{IE}}}{S^{\mathrm{II}}/\tau^{\mathrm{I}}},
\qquad
y_\tau=\frac{S^{\mathrm{EI}}/\tau^{\mathrm{I}}}{S^{\mathrm{EE}}/\tau^{\mathrm{EE}}}.
\label{eq:dsode-tau-adjusted-inhibition-plane}
\end{equation}
Before neighborhood selection, $x_\tau$ and $y_\tau$ were separately min--max normalized using the fixed bounds $[0.15,21.60]$ and $[0.1042,11.6667]$, respectively, likewise determined from the prescribed parameter ranges. The transverse width of each $\tau$-adjusted cloud was then computed using the same patch construction and estimator in Eq.~\eqref{eq:dsode-transverse-width}. No target- or cloud-specific normalization was applied. For the paired raw-versus-$\tau$-adjusted comparison, the same 640 patch centers were used in both projections, whereas their nearest neighbors were selected separately in the corresponding normalized planes. In Fig.~6C, the product on the horizontal axis is the target value $\mathcal{R}_\tau=x_\tau y_\tau$ computed from each target's corresponding reference parameters, displayed on a logarithmic scale; the fixed normalization was used only for neighborhood selection and width calculation.

\begin{figure}[htbp]
\centering
\includegraphics[width=0.55\textwidth]{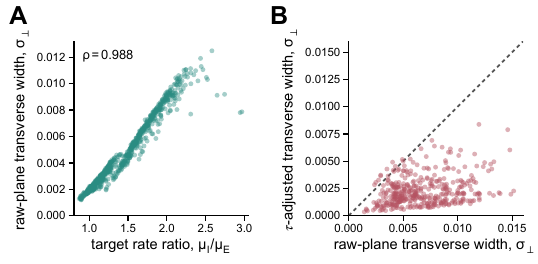}
\caption{Target-level transverse width of dsODE inhibition-plane projections. Each point represents one target-conditioned cloud. $\sigma_\perp$ is the median of the minor-axis standard deviation across 64-neighbor patches in fixed design-range-normalized coordinates. (A) Raw-plane transverse width versus $\mu_I/\mu_E$ for driven-regime targets ($n=725$). (B) Raw-plane versus $\tau$-adjusted transverse width for oscillatory targets ($n=400$). The dashed line denotes equality.}
\label{fig:si-dsODE-width}
\end{figure}

\subsection{Target-level width correlations and temporal diversity}
\label{sec:si_dsode_width_correlations}
The target-conditioned cloud was the independent statistical unit throughout. Across 725 driven-regime targets, each represented by 2,000 generated four-parameter synaptic-strength vectors, the raw inhibition-plane transverse width increased with $\mu_I/\mu_E$ (Fig.~\ref{fig:si-dsODE-width}A; two-sided Spearman rank test, $\rho=0.988$, 95\% target-bootstrap CI $[0.985,0.990]$, $P<10^{-300}$). The median $\sigma_\perp$ increased from $0.00201$ in the lowest $\mu_I/\mu_E$ quintile to $0.00905$ in the highest, a 4.49-fold difference. This association was stable across neighborhood sizes $k_n=16$--$256$ ($\rho=0.985$--$0.988$).

Across 400 oscillatory targets, each represented by 5,000 generated parameter vectors, the $\tau$-adjusted transverse width was positively associated with the dominant E-population frequency measured from the corresponding target trajectory (Fig.~6C, left; two-sided Spearman rank test, $\rho=0.383$, 95\% target-bootstrap CI $[0.291,0.469]$, $P=2.2\times10^{-15}$) and with the coordinate product $\mathcal{R}_\tau$ (Fig.~6C, right; two-sided Spearman rank test, $\rho=0.873$, 95\% target-bootstrap CI $[0.841,0.897]$, $P=3.9\times10^{-126}$). Both associations were stable across $k_n=16$--$512$: $\rho=0.382$--$0.387$ for frequency and $\rho=0.871$--$0.874$ for $\mathcal{R}_\tau$.

In the same 400-target cohort, replacing the raw $S$-ratio coordinates by the $S/\tau$ coordinates reduced the normalized transverse width in 388 targets (Fig.~\ref{fig:si-dsODE-width}B). The raw-plane and $\tau$-adjusted median widths were $0.00617$ and $0.00200$, respectively, and the median target-level adjusted-to-raw ratio was $0.322$ (95\% target-bootstrap CI $[0.294,0.338]$). The paired difference was significant (two-sided paired Wilcoxon signed-rank test, $P=8.6\times10^{-67}$; rank-biserial effect size $-0.996$, where negative values indicate smaller $\tau$-adjusted widths). The relative reduction was stable across $k_n=16$--$512$: the median adjusted-to-raw ratio remained between $0.318$ and $0.323$, and the $\tau$-adjusted width was smaller in 388--390 of 400 targets.


\subsection{External-drive, probability, and refractory-time compensation in 10D and 12D}
\label{sec:si_dsode_highdim}
The same effective-gain picture extends the 8D analysis to higher parameter dimensions. Let
\begin{equation}
w_{QR}^{\mathrm{eff}}=a_{QR}P_{QR}S_{QR}\Psi_{QR}(\tau_{QR})
\label{eq:dsode_effective_gain}
\end{equation}
be the effective recurrent gain relevant to the operating regime, where $\Psi_{QR}(\tau_{QR})$ is the appropriate timescale factor: $\Psi\equiv 1$ for the leading-order mean balance, $\Psi\propto \tau^{-1/2}$ for a shot-noise variance constraint, and $\Psi\propto (1+\omega_*^2\tau^2)^{-1/2}$ for beat-dominated gain constraints.

A local mean-rate balance model can then be written schematically as
\begin{align}
\bar r_E &= \Phi_E\!\left(J_E^{\mathrm{ext}}+w_{EE}^{\mathrm{eff}}\bar r_E-w_{EI}^{\mathrm{eff}}\bar r_I\right), \\
\bar r_I &= \Phi_I\!\left(J_I^{\mathrm{ext}}+w_{IE}^{\mathrm{eff}}\bar r_E-w_{II}^{\mathrm{eff}}\bar r_I\right),
\label{eq:dsode_balance_rates}
\end{align}
for suitable local gain functions $\Phi_E$ and $\Phi_I$. Linearizing Eq.~\eqref{eq:dsode_balance_rates} around a viable target with fixed $(\bar r_E,\bar r_I)$ yields
\begin{equation}
\delta J_E^{\mathrm{ext}}\approx -\bar r_E\,\delta w_{EE}^{\mathrm{eff}}+\bar r_I\,\delta w_{EI}^{\mathrm{eff}},
\qquad
\delta J_I^{\mathrm{ext}}\approx -\bar r_E\,\delta w_{IE}^{\mathrm{eff}}+\bar r_I\,\delta w_{II}^{\mathrm{eff}}.
\label{eq:dsode_external_comp}
\end{equation}
Equation~\eqref{eq:dsode_external_comp} predicts structured dependencies between external drive and recurrent couplings rather than independent variation, which is the qualitative pattern seen in the 10D pairwise density plots.

Probabilities enter in the same multiplicative manner as synaptic strengths. Consequently, in the 12D setting one expects compensation along surfaces of approximately constant $P_{QR}S_{QR}\Psi_{QR}(\tau_{QR})$, not along surfaces of constant $P_{QR}$ or $S_{QR}$ alone. This is why structured $P$--$S$ and $P$--$\tau$ relations are mechanistically plausible in the higher-dimensional generated clouds.

The refractory time $\tau_{\mathrm{ref}}$ plays a somewhat different role. A crude timing approximation is
\begin{equation}
r_Q\approx \frac{1}{\tau_{\mathrm{ref}}+T_Q^{\mathrm{drive}}},
\label{eq:dsode_refractory_rate}
\end{equation}
where $T_Q^{\mathrm{drive}}$ is the effective time needed for recurrent and external input to drive population $Q$ back to threshold. Increasing $\tau_{\mathrm{ref}}$ therefore lowers the rate unless it is offset by a decrease in $T_Q^{\mathrm{drive}}$, that is, by stronger excitation or larger external drive. At the same time, a larger refractory interval suppresses fast recurrent cycling and can move the dynamics from beat-associated branches toward no-beat branches. This dual role explains why $\tau_{\mathrm{ref}}$ often behaves less like a passive tangent coordinate and more like a branch-selection coordinate in the higher-dimensional examples.

\paragraph{Empirical high-dimensional summary.}
Consistent with the timescale-adjusted effective-gain interpretation, the normalized $S/\tau$ projection was narrower than the raw $S$-ratio projection in most targets ( Sec.~\ref{sec:si_dsode_width_correlations} and Fig.~\ref{fig:si-dsODE-width}B).




Because the standard-deviation guides are effectively inactive in the no-beat regime, we prefiltered the test sets using the raw guide values and selected 100 targets in each model--regime group. For no-beat targets the fitted forward map used the two rate means ($d_g=2$); for with-beat targets it used all four guide statistics ($d_g=4$). Local forward fits used training examples from the same regime. Table~\ref{tab:si_dsode_highdim_dimension_regime} reports the all-center analysis following Section~\ref{sec:si_geometry_estimators}. Brackets are 95\% target-bootstrap intervals for PR and interquartile ranges for the target-wise modal local-PCA rank.

\begin{table}[htbp]
\centering
\scriptsize
\setlength{\tabcolsep}{3.5pt}
\caption{Generated-cloud dimension in balanced dynamical-regime cohorts.}
\label{tab:si_dsode_highdim_dimension_regime}
\resizebox{\linewidth}{!}{%
\begin{tabular}{llccccc}
\toprule
Model & regime & $d_g$ & global PR & fitted $r_J/d_J$ & local-PCA$_{95}$ rank & local PR \\
\midrule
8D & no beat & 2 & 5.35 [5.24, 5.46] & 2/6 & 6 [6, 6] & 5.35 [5.29, 5.42] \\
8D & with beat & 4 & 3.64 [3.54, 3.73] & 4/4 & 5 [5, 6] & 4.40 [4.34, 4.47] \\
10D & no beat & 2 & 7.71 [7.58, 7.83] & 2/8 & 8 [8, 8] & 7.33 [7.23, 7.43] \\
10D & with beat & 4 & 5.48 [5.37, 5.58] & 4/6 & 6 [6, 7] & 5.78 [5.72, 5.85] \\
\bottomrule
\end{tabular}%
}
\end{table}

At the fixed $0.05s_{\max}$ rule, the target-wise modal fitted rank equaled the effective guide dimension throughout. No-beat targets had $r_J=2$, giving fitted nullities six in 8D and eight in 10D, whereas with-beat targets had $r_J=4$, giving fitted nullities four and six. The local-PCA and PR summaries were consistent with these regime-specific dimensions at finite scale. Local PR used the 420-nearest-neighbor covariance definition as in the Lorenz analysis, in diffusion-scaled parameter coordinates. Repeating the calculation with interior centers left the modal fitted rank/nullity unchanged and changed mean local PR a little.

\begin{figure}[t]
  \centering
  \includegraphics[width=0.98\linewidth]{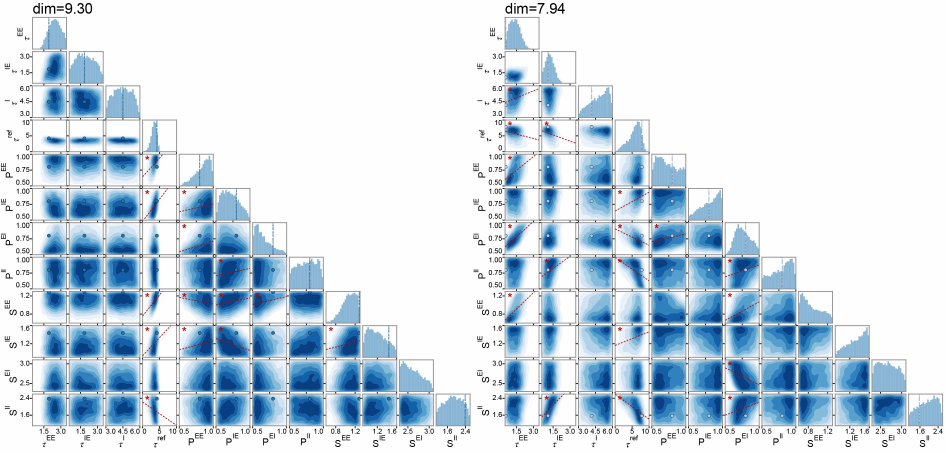}
  \caption{\textbf{Pairwise organization of 12D dsODE parameter generations.} The left matrix is conditioned on $[128.94,132.02,0,0]$ Hz and therefore represents a zero-standard-deviation, no-beat target. The right matrix is conditioned on $[31.20,51.02,99.22,105.37]$ Hz, where entries give the E/I firing-rate means and standard deviations. Each matrix contains 5,000 generated parameter vectors. Diagonal panels show marginal histograms; lower-triangle panels show log-compressed pairwise KDE densities. Colored points and dashed marginal lines mark constructed reference vectors. Red asterisks and dashed red fits denote screened pairwise associations defined in Sec.~\ref{sec:si-dsODE-pairwise-statistics}. The displayed \textnormal{dim} values are participation ratios in normalized 12D coordinates.}
  \label{fig:si-dsODE-12D}
\end{figure}

\paragraph{12D extension.}
We further extended the dsODE parameterization by allowing the four connection
probabilities to vary together with the eight strength and timescale
parameters. The conditional model was trained on 9,535 parameter--feature
pairs. Figure~\ref{fig:si-dsODE-12D} shows generations for two selected
four-feature targets, with 5,000 generated parameter vectors per target.

\subsection{Pairwise statistical annotations}
\label{sec:si-dsODE-pairwise-statistics}

The pairwise matrices in Fig.~6A, Fig.~6D, and Fig.~\ref{fig:si-dsODE-12D} were drawn using the same procedure. Diagonal panels show one-dimensional marginal histograms, whereas each lower-triangle panel shows a Gaussian kernel-density estimate of the corresponding two-parameter projection in the displayed physical units. The KDE used Scott's bandwidth rule and was evaluated on a $42\times42$ grid spanning the displayed parameter ranges. After applying $\log(1+\widehat p)$ for visual contrast, we drew nine equally spaced filled-contour levels spanning 8\%--100\% of the maximum transformed density within each panel. The fitted lines and statistical tests were calculated from the original generated parameter vectors.

For each target-conditioned cloud, we fitted an ordinary least-squares line to every lower-triangle parameter pair. A two-sided $t$ test of the fitted slope tested the null hypothesis of zero slope, equivalently zero Pearson correlation. One generated parameter vector was treated as the computational unit ($n=5{,}000$ per displayed cloud). Pearson $r$ quantified the direction and strength of each pairwise association. The $P$ values were adjusted by the Benjamini--Hochberg procedure separately within each matrix: 28 tests for the 8D cloud, 45 for the 10D cloud, and 66 for each 12D cloud. A red asterisk and dashed red fit were displayed only when $q<0.001$ and $|r|>0.2$, thereby requiring both statistical evidence and a minimum effect size.

Under this common rule, 11 of 28 associations were marked in Fig.~6A, 23 of 45 in Fig.~6D, and 14 of 66 and 18 of 66 in the two matrices of Fig.~\ref{fig:si-dsODE-12D}. These within-cloud tests describe organization of the generated conditional distributions.


\section{Cross-system guide ablations}
\label{sec:si_guide_ablations}
Let $G:\Theta\to\mathbb R^m$ be the current guide and let $g:\Theta\to\mathbb R$ be an added feature. At a regular point, imposing $g=g^*$ lowers the local fiber dimension by one precisely when
\begin{equation}
Dg(\thetaVec)\notin\operatorname{row}DG(\thetaVec).
\end{equation}
This local rank statement is separate from finite-cloud contraction: a tolerance, prior weighting, box truncation, and curvature can change participation ratio or projected spread without changing rank. Branch resolution is also separate. Two disconnected or dynamically distinct components may have the same $G$ value but different $g$ values; adding $g$ then selects or separates components even if the within-component rank change is small.

The matched one-, two-, and four-feature dsODE guide ablations, including the joint four-moment guide, are reported in the preceding subsection (Sec.~\ref{sec:si_dsode_branch_classification}). Table~\ref{tab:si_guide_ablations} gives the corresponding paired guide-subset comparisons for Lorenz and Izhikevich.


\begin{table}[htbp]
\centering
\scriptsize
\caption{Paired guide-ablation results for generated proposal clouds. Fitted rank denotes the numerical rank of the fitted local forward Jacobian $DI$ at the $0.05s_{\max}$ threshold; the corresponding nullity is $d_\theta-\operatorname{rank}DI$. For each target and guide, the reported rank is the median over local fits at all 5{,}000 generated centers. Contraction is the paired reduction $[1-\operatorname{tr}(C_{\mathrm{more}})/\operatorname{tr}(C_{\mathrm{fewer}})]$ in covariance trace in diffusion-normalized parameter coordinates, calculated from the complete generated clouds. Brackets give nonparametric target-bootstrap 95\% intervals for the paired mean; Lorenz uses the five matched standard-deviation target levels 3, 5, 10, 30, and 100, and Izhikevich uses $n=100$ targets.}
\label{tab:si_guide_ablations}
\begin{tabular}{p{0.11\linewidth}p{0.22\linewidth}p{0.26\linewidth}p{0.29\linewidth}}
\toprule
System & Guide comparisons & Incremental fitted rank & Finite-cloud contraction \\
\midrule
Lorenz & $x$-std $\to$ joint $x/z$-std; $z$-std $\to$ joint $x/z$-std & $\operatorname{rank}DI$ increased $1\to2$ at standard-deviation targets 10, 30, and 100 and remained $1\to1$ at targets 3 and 5 & Mean trace reductions $57.7\%$ [$31.9$, $71.4$] ($x$-std to joint) and $70.0\%$ [$60.2$, $81.7$] ($z$-std to joint); all five target levels contracted in both comparisons \\
Izhikevich & rate+mean ISI $\to$ full four features; CV+Fano $\to$ full four features & Median $\operatorname{rank}DI$ changed $1\to2$ and $2\to2$; mean paired increases were 1.27 [1.18, 1.36] and 0.35 [0.26, 0.44], respectively & Mean trace reductions $49.1\%$ [45.6, 52.5] and $29.5\%$ [25.6, 33.5], respectively; 99/100 targets contracted in both comparisons \\
\bottomrule
\end{tabular}
\end{table}


\subsection{Figure 7C tangent-plane comparison and estimator sensitivity}
\label{sec:si_fig7c_geometry}
At each center, let $K\in\mathbb R^{k\times q}$ be a column-orthonormal basis for the complete numerical nullspace of the fitted Jacobian for all feature components included in that cloud's guide, after the null vectors have been converted from locally standardized coordinates to diffusion-scaled parameter increments. Let $P_3\in\mathbb R^{3\times k}$ contain the first three local PCA loading rows. The singular value decomposition
\begin{equation}
P_3K=U\operatorname{diag}(g_j)V^\top, \qquad g_1\geq g_2\geq\cdots,
\end{equation}
defines the projected J-null modes. Figure~7C reports the acute out-of-plane angle
\begin{equation}
\alpha=\arcsin\!\left|e_3^\top u_1\right|,
\end{equation}
between leading projected mode~1 and the empirical PC1--PC2 plane. Because $u_1$ is a unit vector in PC1--PC2--PC3 coordinates, $|e_3^\top u_1|$ is its absolute component normal to the plane; hence the arcsine gives $0^\circ$ for an in-plane direction and $90^\circ$ for a plane-normal direction. The comparison plane is fixed at two dimensions and is not set to the estimated Jacobian nullity or the 95\% PCA rank. For the representative panels, the nullspace is mapped by the differential $D\Phi$ of the displayed-coordinate map, including display-axis scaling and, for dsODE, the quotient-rule derivatives of the plotted ratios $S^{IE}/S^{II}$ and $S^{EI}/S^{EE}$. If $\widetilde u_1$ is the leading left singular vector of $D\Phi K$ and $n_\Phi$ is the unit normal to the displayed local PCA plane, the corresponding panel angle is $\arcsin|n_\Phi^\top\widetilde u_1|$. The displayed J-null arrow lengths are proportional to the displayed projection gains relative to the leading gain. Their squared-gain fractions quantify projected squared displacement. The PC arrows have a common graphical length and do not encode these gains.

A center entered the all-eligible distribution only when its angle and fit statistic were finite, its numerical kernel was nonempty and smaller than the ambient parameter space, and the required PCA eigenvalue denominator was finite and positive. The solid curves further restrict to plane-resolved centers satisfying both $R_J^2\geq0.80$ and $\lambda_2/\lambda_3\geq3$, while the dashed curves show all eligible centers. Each density is normalized within its system and selection. The Jacobian diagnostic is the pooled Gaussian-weighted coefficient of determination of the standardized local parameter-to-feature regression,
\begin{equation}
R_J^2=1-\frac{\sum_i w_i\lVert z_i-\widehat z_i\rVert_2^2}{\sum_i w_i\lVert z_i-\bar z_w\rVert_2^2},
\end{equation}
where $z_i$ is the standardized feature vector and the sums include all fitted feature coordinates. Thus, $R_J^2$ measures how well the locally fitted linear parameter-to-feature map explains the weighted variation of the neighboring simulated features; a high value supports interpreting its numerical nullspace as a locally resolved compensation estimate. The eigengap uses decreasing local covariance eigenvalues and tests whether the retained PC1--PC2 plane is separated from PC3.

For Lorenz, Izhikevich, and dsODE, respectively, the all-eligible/plane-resolved distributions contained 2{,}432/243, 4{,}850/2{,}387, and 84{,}283/5{,}758 centers. In the plane-resolved subsets, the median [IQR] angles were $1.23[0.63,2.58]^\circ$, $2.12[0.89,5.00]^\circ$, and $4.97[2.24,9.76]^\circ$; the corresponding means $\pm$ standard deviations were $2.60^\circ\pm4.56^\circ$, $4.06^\circ\pm5.15^\circ$, and $7.72^\circ\pm9.59^\circ$. The means exceed the medians because the distributions have right tails. The all-eligible medians were $12.28^\circ$, $2.90^\circ$, and $18.74^\circ$, showing that the local fit and plane-resolution diagnostics help exclude locally ill-resolved centers with either a poorly fitted Jacobian or insufficient separation between PC2 and PC3, including centers near the edge of the sampled cloud when boundary truncation degrades these diagnostics. Such cases would otherwise broaden the angle distributions and obscure the local tangent comparison.

Sensitivity was evaluated at the same 128 deterministic centers from each of 778 clouds, changing one setting at a time. We compared 1{,}000, 2{,}000, and 3{,}000 training neighbors; 160, 420, and 800 generated-cloud neighbors for PCA; and numerical-rank cutoffs $0.02s_{\max}$, $0.05s_{\max}$, and $0.10s_{\max}$. The plane-resolution screen was recomputed under each setting. Across the neighborhood choices, plane-resolved median angles remained low, ranging from $1.08^\circ$ to $2.34^\circ$ for Lorenz, $1.77^\circ$ to $3.11^\circ$ for Izhikevich, and $4.31^\circ$ to $7.30^\circ$ for dsODE, although PCA neighborhood size also changed the number of centers passing the eigengap screen. In dsODE, increasing the rank cutoff from $0.05s_{\max}$ to $0.10s_{\max}$ changed the fitted rank at 2{,}266 of 5{,}758 baseline plane-resolved centers and increased the median angle from $4.97^\circ$ to $7.09^\circ$. The plane-resolution criteria do not require separation between the leading projected gains $g_1$ and $g_2$. When these gains are nearly equal, the labeled mode~1 can rotate within the leading projected null subspace. This occurs for the plane-resolved Lorenz centers, but the median basis-invariant fraction of total projected nullspace gain contained in the PC1--PC2 plane is 99.91\%, indicating that the Jacobian-null plane and empirical PCA plane agree closely.

\section{Computational-cost scaling for viable-manifold recovery}
\label{sec:si_cost_scaling}

\label{sec:cost_scaling}
This section decomposes the computational budget for viable-set recovery into simulator calls, neural-network optimization, generator sampling, and local linear algebra. Geometry contributes separate target-localization and tangent-coverage factors; global conditional learning can be amortized across targets. One \emph{parameter evaluation} means one simulation followed by feature extraction.

\subsection{Empirical proposal and search comparison}
\label{sec:si_empirical_method_comparison}
We compared diffusion generation with a conditional flow-matching generator, uniform random sampling and rejection, local continuation, and random-walk MCMC under the target-specific acceptance criteria in Section~\ref{sec:si_sampling_validation}. The reported rows aggregate the corresponding Lorenz and Izhikevich target sets and representative 4D, 8D, and 10D dsODE subsets.

Diffusion used the retained annealed-Langevin sampler with 1{,}024 score-network evaluations per proposal. For the aligned flow-matching baseline, each scaled training parameter $x_1$ and guide $g$ was paired with an independent $x_0$ drawn uniformly from $[-1,1]^d$ and $t$ drawn uniformly from $[0,1]$; the vector field was trained at $x_t=(1-t)x_0+tx_1$ against velocity $x_1-x_0$. Its input combined $x_t$, $g$, and a 64-dimensional Fourier embedding of $t$, followed by four 256-unit SiLU hidden layers. The model was optimized with AdamW for 20{,}000 minibatch updates, using minibatches of 4{,}096 and a learning rate of $3\times10^{-4}$. Samples were transported from the uniform base distribution using 64 forward-Euler steps. The flow-matching runs used the same training arrays and parameter and guide scaling as the corresponding diffusion runs. Neither generator called the dynamical-system simulator during post-training generation. Simulations marked in these methods were used to validate generated parameters

For Lorenz and Izhikevich, uniform rejection simulated 1{,}000 independent normalized-box samples per target, local continuation used 50 random starts, 32 simulated neighbors per local Jacobian estimate, perturbation size 0.05, and at most 30 iterations, and MCMC used 50 random-start chains with at most 1{,}000 proposals per chain. Because each dsODE simulation is substantially more expensive, its baseline subset used 100 uniform samples per target, four local-continuation starts with 16 neighbors and at most 20 iterations, and ten MCMC starts with at most 30 proposals per chain.

\begin{table}[htbp]
\centering
\scriptsize
\caption{Aggregate empirical comparison. Entries are success fraction / counted simulator calls per accepted parameter. For diffusion and flow matching, the second quantity counts post-generation validation simulations. For the three simulator-driven methods, it counts simulations used during search. A dash indicates that no accepted point was observed.}
\label{tab:si_empirical_baselines}
\resizebox{\textwidth}{!}{%
\begin{tabular}{lccccc}
\toprule
Target family & Diffusion & Flow matching & Uniform rejection & Local continuation & MCMC \\
\midrule
Lorenz $x$-std & 0.8864 / 1.13 & 0.9668 / 1.03 & 0.0198 / 50.5 & 0.5520 / 511 & 0.9280 / 132 \\
Lorenz $z$-std & 0.7060 / 1.42 & 0.8856 / 1.13 & 0.0208 / 48.1 & 0.5800 / 555 & 0.8760 / 219 \\
Lorenz joint $x$/$z$-std & 0.2178 / 4.59 & 0.2282 / 4.38 & 0.0034 / 294 & 0.4640 / 818 & 0.7520 / 440 \\
Izhikevich four-feature targets & 0.6256 / 1.60 & 0.6136 / 1.63 & 0.00467 / 214 & 0.5607 / 606 & 0.8587 / 369 \\
Izhikevich regular-firing targets & 0.5363 / 1.86 & 0.7837 / 1.28 & 0.00133 / 750 & 0.5533 / 746 & 0.8467 / 554 \\
dsODE 4D $S$ mean-rate subset & 0.9525 / 1.05 & 0.9338 / 1.07 & 0.0450 / 22.2 & 0.8125 / 161 & 0.4500 / 48.9 \\
dsODE 8D mean/std subset & 0.4113 / 2.43 & 0.4375 / 2.29 & 0 / -- & 0.6875 / 314 & 0.1000 / 291 \\
dsODE 10D mean/std subset & 0.2838 / 3.52 & 0.3050 / 3.28 & 0 / -- & 0.4375 / 605 & 0 / -- \\
\bottomrule
\end{tabular}}
\end{table}


The two neural generators showed broadly comparable target-matching performance across the Lorenz, Izhikevich, and dsODE targets. Flow matching sometimes produced broader accepted clouds, but the result depended on the target and metric: among 100 valid target-level comparisons, it had larger covariance log-volume, participation dimension, and median pairwise distance in 60, 56, and 70 cases, respectively. The two objectives yielded comparably effective conditional generators with target-dependent differences in accepted-cloud geometry.

The dsODE subsets extend the simulator-search comparison to a more expensive dynamical model. Uniform rejection accepted 18/400 points in the 4D subset and none of 400 points in either the 8D or 10D subset. Local continuation reached the criterion from a substantial fraction of starts (13/16, 11/16, and 7/16 in the 4D, 8D, and 10D subsets, respectively) but required 161, 314, and 605 simulations per accepted terminal point in the corresponding subsets. MCMC succeeded from a smaller fraction of starts (18/40, 4/40, and 0/40 in the 4D, 8D, and 10D subsets, respectively), while also requiring repeated simulator evaluations. These measurements support the practical value of amortized conditional generation when many accepted parameters are required across many targets: after construction of the common simulation library and model training, both diffusion and flow matching produce new target-conditioned proposals without further simulator calls.

The reported costs provide an operational comparison rather than an end-to-end wall-clock benchmark. Diffusion and flow matching use simulations only for post-generation validation, whereas random rejection, local continuation, and MCMC require simulations during target-specific search. Continuation and MCMC success is also defined per random start rather than per proposal. Training-library construction and neural optimization are reusable one-time costs, so the comparison primarily characterizes simulator use after training. 

\subsection{Regular level-set setup and cost bookkeeping}
Let $\Theta\subset\R^k$ be a bounded parameter domain and let
\begin{equation}
\IMap:\Theta\rightarrow\R^c
\end{equation}
be the feature map used for conditioning, where $c$ is the local codimension of the target constraint. For a regular value $y$, define
\begin{equation}
\Mset_y=\{\thetaVec\in\Theta:\IMap(\thetaVec)=y\},
\qquad m=k-c,
\end{equation}
so that $m$ is the intrinsic dimension of the viable manifold. Assume that $\IMap$ is $C^2$ in a tubular neighborhood of $\Mset_y$, that the smallest nonzero singular value of $D\IMap$ is bounded away from zero there, and that $\Mset_y$ has positive reach. These are the standard regularity conditions under which the viable-set picture is meaningful.

We separate the basic unit costs as
\begin{align}
C_{\mathrm{sim}} &:= \text{one simulator run + feature extraction},\\
C_{\mathrm{gen}} &:= \text{one amortized generator draw},\\
C_{\mathrm{opt}}(n) &:= \text{optimization cost for training a neural generator on }n\text{ pairs},\\
C_{\mathrm{ret}} &:= \text{library retrieval / kernel evaluation cost per query},\\
C_{\mathrm{la}}(k,c) &:= \text{local linear-algebra overhead in }k\text{ parameters and codimension }c.
\end{align}
In the Lorenz, Izhikevich, and dsODE settings considered here, $C_{\mathrm{sim}}$ is typically the dominant term for non-amortized search, but keeping the other terms explicit clarifies why diffusion becomes advantageous only after reuse across many targets.

For \(E\) training epochs, minibatch size \(b\), and per-update cost \(C_{\mathrm{step}}\), optimizer bookkeeping gives
\begin{equation}
N_{\mathrm{step}}=E\left\lceil\frac{n}{b}\right\rceil,
\qquad
C_{\mathrm{opt}}(n)=N_{\mathrm{step}}C_{\mathrm{step}}.
\label{eq:si_optimizer_bookkeeping}
\end{equation}
At fixed architecture, epoch count, and batch size, this contribution is \(O(n)\).

By the coarea formula, the volume of the $\varepsilon$-tube
\begin{equation}
\Mset_{y,\varepsilon}=\{\thetaVec:\norm{\IMap(\thetaVec)-y}\le \varepsilon\}
\end{equation}
obeys
\begin{equation}
\operatorname{Vol}_k(\Mset_{y,\varepsilon})
=\nu_c\varepsilon^c\int_{\Mset_y}\frac{1}{J_{\IMap}(\thetaVec)}\,d\mathcal H^m(\thetaVec)
+O(\varepsilon^{c+1}),
\label{eq:tube_scaling_revised}
\end{equation}
with
\begin{equation}
J_{\IMap}(\thetaVec)=\sqrt{\det\big(D\IMap(\thetaVec)D\IMap(\thetaVec)^T\big)}.
\end{equation}
Hence, for approximately uniform sampling in $\Theta$,
\begin{equation}
p_\varepsilon
:= \mathbb P\{\norm{\IMap(\thetaVec)-y}\le \varepsilon\}
= C_y\varepsilon^c+O(\varepsilon^{c+1}).
\label{eq:accept_prob_revised}
\end{equation}
This is the codimension penalty. Independently, resolving the manifold at parameter-space scale $\eta$ requires a covering number
\begin{equation}
M_\eta:=N_{\Mset}(\eta)\asymp C_{\Mset}\eta^{-m}.
\label{eq:covering_scaling_revised}
\end{equation}
Thus any method that returns a cloud resolving $\Mset_y$ must pay for two separate tasks: entering an $\varepsilon$-thin tube in the $c$ normal directions, and covering the $m$ tangent directions at scale $\eta$. The first creates an $\varepsilon^{-c}$ rare-event factor; the second creates an $\eta^{-m}$ coverage factor.

\subsection{Full cost of direct-search and local geometric methods}

We compare four baselines that do not learn a global conditional generator. Fix a target value \(y\). Each baseline must either return \(M_\eta\) accepted parameter points, or build an atlas with \(M_\eta\) local charts, where
\begin{equation}
M_\eta \asymp \eta^{-m}
\end{equation}
is sufficient to resolve \(\Mset_y\) at parameter-space scale \(\eta\). Let \(p_0\) denote the proposal measure on \(\Theta\) (uniform on \(\Theta\) unless otherwise stated), and define
\begin{equation}
p_\varepsilon(y)
:=
\mathbb P_{\thetaVec\sim p_0}\!\left\{\norm{\IMap(\thetaVec)-y}\le \varepsilon\right\}
\asymp C_y \varepsilon^c .
\end{equation}
The factor \(p_\varepsilon(y)^{-1}\) is the normal-direction rarity, whereas \(M_\eta\) is the tangent-direction coverage requirement.

\paragraph{Random search plus acceptance.}
This baseline draws \(\thetaVec_i \stackrel{\mathrm{i.i.d.}}{\sim} p_0\), computes \(z_i=\IMap(\thetaVec_i)\), and accepts \(\thetaVec_i\) whenever \(\norm{z_i-y}\le \varepsilon\). The expected number of simulator calls needed to obtain \(M_\eta\) accepted points is therefore
\begin{equation}
\mathcal C_{\mathrm{rand}}^{(T)}(\eta,\varepsilon)
\asymp
T\,M_\eta\,p_\varepsilon(y)^{-1}\,C_{\mathrm{sim}}
\asymp
T\,C_{\mathrm{sim}}\,\eta^{-m}\varepsilon^{-c},
\label{eq:random_full_cost}
\end{equation}
for \(T\) target values. This is the clearest expression of the codimension penalty.

\paragraph{Feature-space library retrieval / kernel conditioning.}
Here one first precomputes a static library
\begin{equation}
\mathcal L_{N_{\mathrm{lib}}}
=
\{(\thetaVec_i,\IMap(\thetaVec_i))\}_{i=1}^{N_{\mathrm{lib}}},
\qquad
\thetaVec_i \stackrel{\mathrm{i.i.d.}}{\sim} p_0 .
\end{equation}
At query time one either performs hard retrieval,
\begin{equation}
A_y^\varepsilon
=
\{i:\norm{\IMap(\thetaVec_i)-y}\le \varepsilon\},
\end{equation}
or kernel conditioning with weights
\begin{equation}
w_i(y)\propto K_\tau\!\big(\IMap(\thetaVec_i)-y\big),
\qquad
N_{\mathrm{eff}}(y)
:=
\frac{(\sum_i w_i(y))^2}{\sum_i w_i(y)^2}.
\end{equation}
To resolve \(\Mset_y\), one needs \(N_{\mathrm{eff}}(y)\gtrsim M_\eta\). Under approximately uniform proposal sampling in \(\Theta\) and locally bounded distortion of \(\IMap\),
\begin{equation}
\mathbb E\!\left[N_{\mathrm{eff}}(y)\right]\lesssim N_{\mathrm{lib}}\,p_\varepsilon(y),
\end{equation}
so one must have
\begin{equation}
N_{\mathrm{lib}}
\gtrsim
M_\eta\,p_\varepsilon(y)^{-1}
\asymp
\eta^{-m}\varepsilon^{-c}
\label{eq:library_size_revised}
\end{equation}
for a single target. For target neighborhoods \(A_1,\ldots,A_T\) with masses \(p_j\), a single i.i.d.\ library produces multinomial occupancies with means \(N_{\mathrm{lib}}p_j\). Simultaneous coverage therefore requires
\begin{equation}
N_{\mathrm{lib}}
\gtrsim
\max_{1\le j\le T}
\frac{M_\eta+\log(T/\delta)+\sqrt{M_\eta\log(T/\delta)}}{p_j}
\label{eq:library_size_many_targets}
\end{equation}
up to universal concentration constants to obtain at least \(M_\eta\) points near every target with probability \(1-\delta\). The dependence on \(T\) is thus logarithmic at fixed target masses; any additional scaling enters through the smallest \(p_j\), which itself can decrease as more separated target regions partition a fixed feature distribution. The full cost is therefore
\begin{equation}
\mathcal C_{\mathrm{lib}}^{(T)}(\eta,\varepsilon)
\gtrsim
N_{\mathrm{lib}}\,C_{\mathrm{sim}}
+
T\,C_{\mathrm{ret}}(N_{\mathrm{lib}}),
\label{eq:library_full_cost_one}
\end{equation}
with
\begin{equation}
C_{\mathrm{ret}}(N_{\mathrm{lib}})
=
\begin{cases}
O(N_{\mathrm{lib}}\,c), & \text{naive full scan},\\[1mm]
O(\log N_{\mathrm{lib}}+M_\eta), & \text{indexed retrieval in low feature dimension}.
\end{cases}
\end{equation}
Thus a static library removes online simulator calls only by prepaying the same \(\eta^{-m}\varepsilon^{-c}\) rarity.

\paragraph{SMC / ABC targeting a softened loss.}
Sequential Monte Carlo and approximate Bayesian computation replace hard acceptance by a softened target
\begin{equation}
\pi_\tau(\thetaVec\mid y)
\propto
p_0(\thetaVec)\,K_\tau\!\big(\IMap(\thetaVec)-y\big),
\qquad
K_\tau(z)=\exp\!\left[-\frac{\norm{z}^2}{2\tau}\right]
\quad\text{or}\quad
K_\tau(z)=\mathbf 1\{\norm{z}\le \varepsilon\},
\end{equation}
with terminal scale \(\tau_L\asymp \varepsilon^2\). A prior-start SMC/ABC scheme uses a sequence \(\tau_0>\tau_1>\cdots>\tau_L\), and reweights, resamples, and mutates particles until they approximate \(\pi_{\tau_L}(\cdot\mid y)\). Because only a fraction \(p_\varepsilon(y)\) of prior mass reaches the terminal tube, the simulator cost to obtain \(M_\eta\) effective terminal particles satisfies
\begin{equation}
\mathcal C_{\mathrm{SMC}}^{(T)}(\eta,\varepsilon)
\gtrsim
T\,\kappa_{\mathrm{SMC}}\,M_\eta\,p_\varepsilon(y)^{-1}\,C_{\mathrm{sim}},
\label{eq:smc_full_cost}
\end{equation}
where \(\kappa_{\mathrm{SMC}}\ge 1\) collects the number of tempering levels, resampling loss, and move-step inefficiency. SMC/ABC can improve constants, but it does not remove the codimension barrier when initialized from the prior.

\paragraph{MCMC targeting the same softened loss.}
A Metropolis--Hastings chain targeting \(\pi_\tau(\cdot\mid y)\) with local proposal \(q(\thetaVec'\mid\thetaVec)\) faces two distinct regimes. If it is initialized from the prior, then burn-in inherits the same \(p_\varepsilon(y)^{-1}\) rarity as rejection search. If it is initialized already near \(\Mset_y\), then the posterior thickness in the \(c\) normal directions is \(O(\varepsilon)\), so an isotropic random-walk proposal must use step size \(O(\varepsilon)\) to maintain \(O(1)\) acceptance. Tangential exploration is then diffusive: moving an \(O(1)\) geodesic distance along the manifold requires \(O(\varepsilon^{-2})\) accepted steps. A local lower bound is therefore
\begin{equation}
\mathcal C_{\mathrm{RW\text{-}MCMC}}^{(T)}(\eta,\varepsilon)
\gtrsim
T\,M_\eta\,
\kappa_{\mathrm{mix}}(m,\text{curvature},\text{components})\,
\varepsilon^{-2}\,C_{\mathrm{sim}},
\label{eq:rwmcmc_full_cost}
\end{equation}
with an additional \(p_\varepsilon(y)^{-1}\) burn-in factor if no near-manifold initialization is available. Disconnected viable components further increase the cost through multiple chains or tempering.

\paragraph{Local linear / nullspace continuation.}
Continuation is intrinsically a \emph{local} method. It assumes a seed \(\thetaVec^{(0)}\) on \(\Mset_y\), or at least within the corrector convergence radius \(O(h^2)\), and extends an atlas outward from that seed. At chart center \(\thetaVec^{(j)}\), one estimates the Jacobian \(J_j=D\IMap(\thetaVec^{(j)})\), computes a basis for \(\ker J_j\), takes predictor steps of size \(h\) in tangent coordinates, and applies a Newton or Gauss--Newton corrector back to \(\Mset_y\). If the viable set has several connected components, then one needs at least one seed per component; otherwise unseen components remain invisible.

When sensitivities are hard to obtain, one Jacobian refresh costs \(c_J k\) simulator evaluations, with \(c_J=1\) for forward differences and \(c_J=2\) for central differences. If each chart uses \(q_{\mathrm{pred}}\) predictor attempts, \(q_{\mathrm{cor}}\) corrector iterations, and a fraction \(\rho_J\in[0,1]\) of corrector iterations refreshes the Jacobian, then the simulator overhead per chart is
\begin{equation}
c_C(k)
:=
q_{\mathrm{pred}}
+
q_{\mathrm{cor}}\big(1+\rho_J c_J k\big),
\end{equation}
so \(c_C(k)=O(k)\) for finite-difference sensitivities and \(c_C(k)=O(1)\) when exact or adjoint sensitivities are available. The non-simulation linear-algebra cost per chart is
\begin{equation}
C_{\mathrm{la}}(k,c)=O(kc^2+c^3+km^2),
\end{equation}
coming from QR/SVD nullspace extraction and the solution of \(c\times c\) Newton systems. At comparable resolution \(h\sim \eta\), the full continuation cost is therefore
\begin{equation}
\mathcal C_{\mathrm{cont}}^{(T)}(h)
\asymp
T\,M_h
\Big[
\big(c_J k + c_C(k)\big)C_{\mathrm{sim}}
+
C_{\mathrm{la}}(k,c)
\Big],
\qquad
M_h\asymp h^{-m}.
\label{eq:continuation_full_cost}
\end{equation}
Continuation removes the rare-event factor \(p_\varepsilon(y)^{-1}\), but pays instead for repeated local Jacobian estimation and chart construction, target by target and component by component.

\subsection{Ambient conditional learning, oracle fiber estimation, and geometric recovery}
\label{sec:si_three_level_theory}
The manuscript's learning framework begins with ambient samples
\begin{equation}
\theta_i\sim\pi\quad\text{on }\Theta\subset\mathbb R^k,
\qquad Y_i=\IMap(\theta_i),
\qquad i=1,\dots,N,
\end{equation}
without prior knowledge of the location of a selected fiber. Its theory therefore separates three statistical objects.

\paragraph{Level I: global joint or conditional-law learning.}
The first task is to learn the joint law of $(\Theta,Y)$ or the Markov kernel $y\mapsto\mu_y=\mathcal L(\Theta\mid Y=y)$ over a target region. This task depends on the ambient parameter dimension $k$, the effective feature-image dimension $r$, the regularity of $\IMap$ and $\pi$, the rarity of the target, and the regularity of the family $y\mapsto\mu_y$. Recent conditional-distribution results address variants of this global problem under uniform smoothness, reach, density, and lower-mass assumptions and typically control a risk averaged over $Y$ \citep{TangLinYang2025,TangYang2025DistributionRegression}. Those results are relevant benchmarks, but applying them to a simulator-induced inverse family requires verification of their assumptions.

\paragraph{Level II: pointwise inverse-measure estimation.}
For a specified regular target $y^*$, the relevant population law is the coarea-weighted inverse measure $\mu_{y^*}$ in Eq.~\eqref{eq:coarea_main_revised}. Let $h$ be a feature-space localization bandwidth. If $p_Y$ is continuous and positive at $y^*$ in $r_{y^*}$ independent feature coordinates, then
\begin{equation}
N_{\mathrm{eff}}(y^*,h)
\asymp N\,v_{r_{y^*}}p_Y(y^*)h^{r_{y^*}}
\label{eq:si_target_effective_sample}
\end{equation}
library points are expected near the target. A schematic pointwise error decomposition is
\begin{equation}
\mathbb E W_1(\widehat\mu_{y^*},\mu_{y^*})
\ \lesssim\
C_{\mathrm{loc}}h^{\alpha_y}
+\mathcal E_{\mathrm{oracle}}\!\left(N_{\mathrm{eff}}(y^*,h);d,\alpha_\theta,\beta\right)
+\mathcal E_{\mathrm{model}}+\mathcal E_{\mathrm{opt}},
\label{eq:si_pointwise_decomposition}
\end{equation}
where $d=k-r_{y^*}$ is the local fiber dimension, $\alpha_y$ controls variation of the conditional law across targets, $\alpha_\theta$ and $\beta$ describe within-fiber density and support regularity, and the final terms represent approximation and optimization errors. Equation~\eqref{eq:si_pointwise_decomposition} is a bookkeeping decomposition, not a proved minimax rate for the present estimator.

Tang and Yang's unconditional manifold-supported diffusion theory \citep{TangYang2024} supplies an oracle benchmark for $\mathcal E_{\mathrm{oracle}}$: it assumes that the observations are already sampled from the target-supported law. Under their assumptions, the $W_1$ error scales, up to logarithmic factors, as
\begin{equation}
\mathcal E_{\mathrm{oracle}}(n;d,\alpha_\theta,\beta)
=\widetilde O\!\left(n^{-1/2}\vee n^{-(\alpha_\theta+1)/(2\alpha_\theta+d)}\right)
\label{eq:si_oracle_TY_rate}
\end{equation}
under the corresponding smoothness compatibility condition. In the present setting, $n$ in Eq.~\eqref{eq:si_oracle_TY_rate} cannot be replaced by the full ambient library size $N$: the target-localization factor in Eq.~\eqref{eq:si_target_effective_sample}, information shared across neighboring targets, and conditional-model approximation must first be analyzed. No end-to-end minimax rate is claimed here.

\paragraph{Level III: support and geometry recovery.}
A small $W_1$ error controls average transport but does not by itself guarantee Hausdorff support recovery, discovery of low-mass components, correct relative component mass, tangent or normal accuracy, curvature accuracy, or topology. Such conclusions require additional lower-mass, reach, separation, and derivative assumptions together with a metric or estimator suited to support geometry. Pointwise and uniform rates for simulator-induced inverse measures, followed by error propagation to support, components, tangents, and curvature, remain open for this framework.

\paragraph{End-to-end cost accounting.}
Let $N_{\mathrm{lib}}$ be the actual ambient library size used for global training. Without an end-to-end theorem, the diffusion cost is recorded directly as
\begin{equation}
a_y(\varepsilon)
:=
\Pr_{\tilde\thetaVec\sim\widehat\mu_y}
\left\{\|\IMap(\tilde\thetaVec)-y\|\le\varepsilon\right\},
\label{eq:si_diffusion_acceptance}
\end{equation}
the empirically measured probability that a proposal from the trained conditional generator passes direct re-simulation at tolerance \(\varepsilon\). The total cost for \(T\) targets is
\begin{equation}
\mathcal C_{\mathrm{diff}}^{(T)}
=N_{\mathrm{lib}}C_{\mathrm{sim}}
+C_{\mathrm{opt}}(N_{\mathrm{lib}})
+T\frac{M_\eta}{a_y(\varepsilon)}
\bigl(C_{\mathrm{gen}}+C_{\mathrm{sim}}\bigr),
\label{eq:si_diffusion_cost_corrected}
\end{equation}
where $a_y(\varepsilon)$ is measured by re-simulation and $M_\eta$ is the desired accepted-cloud or covering budget. The first two terms are paid once and amortized across targets; the final term is target specific. Equation~\eqref{eq:si_diffusion_cost_corrected} is cost bookkeeping, not a statistical guarantee that $M_\eta$ points cover every component.

Equivalently, the amortized cost per target is
\begin{equation}
\frac{\mathcal C_{\mathrm{diff}}^{(T)}}{T}
=
\frac{N_{\mathrm{lib}}C_{\mathrm{sim}}+C_{\mathrm{opt}}(N_{\mathrm{lib}})}{T}
+
\frac{M_\eta}{a_y(\varepsilon)}
\bigl(C_{\mathrm{gen}}+C_{\mathrm{sim}}\bigr).
\label{eq:si_diffusion_amortized_cost}
\end{equation}

\subsection{Direct comparison across methods}
Equations~\eqref{eq:random_full_cost}, \eqref{eq:smc_full_cost}, \eqref{eq:rwmcmc_full_cost}, \eqref{eq:continuation_full_cost}, and \eqref{eq:si_diffusion_cost_corrected} separate the dominant costs.

Rejection search, prior-initialized SMC/ABC, and static random libraries pay the codimension factor $\varepsilon^{-c}$ before resolving tangent geometry. Continuation replaces this factor with seed acquisition, repeated Jacobian/corrector work, and an atlas of size $M_h\asymp h^{-m}$. Conditional diffusion pays the measured ambient-library and optimization costs once and then supplies target-specific proposals with empirical validation probability $a_y(\varepsilon)$.

The crossover conditions are explicit. Diffusion is preferable to rejection whenever
\begin{equation}
\frac{N_{\mathrm{lib}}\,C_{\mathrm{sim}}+C_{\mathrm{opt}}(N_{\mathrm{lib}})}{T}
+
\frac{M_\eta}{a_y(\varepsilon)}\big(C_{\mathrm{gen}}+C_{\mathrm{sim}}\big)
\ll
M_\eta \varepsilon^{-c} C_{\mathrm{sim}},
\label{eq:diff_beats_random}
\end{equation}
that is, when the one-time library is amortized over many targets and the learned proposal has acceptance $a_y(\varepsilon)\gg p_\varepsilon$. Diffusion is preferable to local continuation whenever
\begin{equation}
\frac{N_{\mathrm{lib}}\,C_{\mathrm{sim}}+C_{\mathrm{opt}}(N_{\mathrm{lib}})}{T}
+
\frac{M_\eta}{a_y(\varepsilon)}\big(C_{\mathrm{gen}}+C_{\mathrm{sim}}\big)
\ll
M_\eta \Big[(c_J k + c_C(k))C_{\mathrm{sim}} + C_{\mathrm{la}}(k,c)\Big].
\label{eq:diff_beats_cont}
\end{equation}
The left-hand side decreases per target as the same trained model is reused. For a single low-dimensional target with a known seed and reliable sensitivities, continuation avoids the ambient library and training terms.

\subsection{Summary table}
\begin{table}[H]
\centering
\small
\caption{Cost bookkeeping for viable-set sampling at parameter resolution $\eta$ and feature tolerance $\varepsilon$. Here $m=k-c$ is the local fiber dimension, $c$ is the codimension, $M_\eta\asymp \eta^{-m}$ is a nominal accepted-cloud or chart budget, and $T$ is the number of targets. The diffusion row uses the actual ambient library size; no end-to-end sample-complexity rate is asserted.}
\begin{tabular}{p{0.22\linewidth}p{0.44\linewidth}p{0.24\linewidth}}
\toprule
Method & Full cost for $T$ targets & Dominant bottleneck \\
\midrule
Random search / rejection &
$T\,M_\eta\,p_\varepsilon^{-1} C_{\mathrm{sim}}
\asymp T\,C_{\mathrm{sim}}\,\eta^{-m}\varepsilon^{-c}$ &
Codimension rarity $\varepsilon^{-c}$ plus tangent coverage $\eta^{-m}$ \\
\addlinespace
Static random library + NN/KDE conditioning &
$N_{\mathrm{lib}} C_{\mathrm{sim}} + T C_{\mathrm{ret}}$, with $N_{\mathrm{lib}}\gtrsim M_\eta p_\varepsilon^{-1}$ for one target and larger for many separated targets &
The same rare-event geometry is prepaid in the library \\
\addlinespace
Prior-initialized SMC / ABC &
$\gtrsim T\,C_{\mathrm{sim}}\,\eta^{-m}\varepsilon^{-c}$ &
Particle collapse into thin feature tubes \\
\addlinespace
Random-walk MCMC near manifold &
$\gtrsim T\,M_\eta\,\varepsilon^{-2} C_{\mathrm{sim}}$ &
Diffusive tangent mixing under normal-direction stiffness \\
\addlinespace
Local linear / nullspace continuation &
$T\,M_h\Big[(c_J k + c_C(k))C_{\mathrm{sim}} + C_{\mathrm{la}}(k,c)\Big]$, with $M_h\asymp h^{-m}$ and $h\sim\eta$ &
Atlas size $h^{-m}$, repeated Jacobian/corrector work, and a seed on each target component \\
\addlinespace
Conditional diffusion &
$N_{\mathrm{lib}} C_{\mathrm{sim}} + C_{\mathrm{opt}}(N_{\mathrm{lib}}) + T\,\dfrac{M_\eta}{a_y(\varepsilon)}(C_{\mathrm{gen}}+C_{\mathrm{sim}})$ &
Ambient library and global conditional learning, followed by target-specific proposal validation \\
\bottomrule
\end{tabular}
\end{table}

The comparison identifies distinct operating regimes. Rejection-style methods and ambient random libraries pay target rarity. Continuation pays for seeds and repeated local geometry. A conditional generator pays for one ambient library and training run, then amortizes proposal generation across targets. Tang and Yang \citep{TangYang2024} characterize the oracle problem in which samples from one manifold-supported target law are already available; their rate does not determine the ambient library size in the present pipeline.



\bibliographystyle{unsrtnat}
\bibliography{pnas_viable_manifolds_refs_updated_v3}